\documentclass[reqno,10pt, centertags, draft]{amsart}
\usepackage{amsmath,amsthm,amscd,amssymb,latexsym,upref}

\def\theequation{\@arabic\c@equation}

\newcommand{\bbN}{{\mathbb{N}}}
\newcommand{\bbR}{{\mathbb{R}}}

\newcommand{\bbC}{{\mathbb{C}}}

\newcommand{\cB}{{\mathcal B}}
\newcommand{\cC}{{\mathcal C}}

\newcommand{\cH}{{\mathcal H}}

\newcommand{\cJ}{{\mathcal J}}
\newcommand{\cK}{{\mathcal K}}

\newcommand{\cV}{{\mathcal V}}

\newcommand{\no}{\nonumber}
\newcommand{\lb}{\label}
\newcommand{\f}{\frac}

\newcommand{\ol}{\overline}
\newcommand{\ti}{\tilde  }
\newcommand{\wti}{\widetilde}
\newcommand{\hatt}{\widehat}

\newcommand{\loc}{\text{\rm{loc}}}

\newcommand{\ran}{\text{\rm{ran}}}
\newcommand{\nul}{\text{\rm{nul}}}

\newcommand{\dom}{\text{\rm{dom}}}

\newcommand{\supp}{\text{\rm{supp}}}

\newcommand{\AC}{\text{\rm{AC}}}
\newcommand{\bi}{\bibitem}
\renewcommand{\Re}{\text{\rm Re}}
\renewcommand{\Im}{\text{\rm Im}}

\DeclareMathOperator{\AKNS}{AKNS}
\DeclareMathOperator{\NS}{NLS}

\DeclareMathOperator{\wlim}{w-lim}

\newcommand{\Span}{\operatorname{span}}

\renewcommand{\bar}[1]{\overline{#1}}

\numberwithin{equation}{section}

\newtheorem{theorem}{Theorem}[section]
\newtheorem{lemma}[theorem]{Lemma}
\newtheorem{corollary}[theorem]{Corollary}
\newtheorem{hypothesis}[theorem]{Hypothesis}
\newtheorem{remark}[theorem]{Remark}

\begin{document}

\title[Spectral Analysis for $\NS_-$ Darboux Transformations]{ Spectral
Analysis of Darboux Transformations for the Focusing 
$\boldsymbol{\NS}$ Hierarchy} 
\author[R.\ C.\ Cascaval]{Radu C.\ Cascaval}
\address{Department of Mathematics,
University of Missouri, Columbia, MO 65211, USA}
\email{radu@math.missouri.edu}
\author[F.\ Gesztesy]{Fritz Gesztesy}
\address{Department of Mathematics,
University of Missouri, Columbia, MO 65211, USA}
\email{fritz@math.missouri.edu}
\urladdr{http://www.math.missouri.edu/people/fgesztesy.html}
\author[H.\ Holden]{Helge Holden}
\address{Department of Mathematical Sciences,
Norwegian University of
Science and Technology, NO--7491 Trondheim, Norway,
\textit{and}
Simula Research Laboratory, PO Box 134,
NO--1325 Lysaker, Norway}
\email{holden@math.ntnu.no}
\urladdr{http://www.math.ntnu.no/\~{}holden/}
\author[Y.\ Latushkin]{Yuri Latushkin}
\address{Department of Mathematics,
University of Missouri, Columbia, MO 65211, USA}
\email{yuri@math.missouri.edu}
\date{October, 2002.}
\subjclass{Primary: 34L05, 34L40, 35Q51, 35Q55. Secondary: 34B20, 47A10.}
\keywords{Dirac operator, focusing nonlinear Schr\"odinger equation,
transformation operators, Darboux transformations, J-self-adjointness,
Weyl--Titchmarsh solutions.}

\begin{abstract}
We study Darboux-type transformations associated with the focusing
nonlinear Schr\"odinger equation ($\NS_-$) and their effect on spectral
properties of the underlying Lax operator. The latter is a formally 
$\cJ$-self-adjoint (but non-self-adjoint) Dirac-type differential
expression of the form  
$$M(q)=i\begin{pmatrix} \frac{d}{dx} & -q \\ -\bar q
&-\frac{d}{dx}\end{pmatrix},$$ satisfying $\cJ M(q)\cJ =M(q)^*$,
where $\cJ $ is defined by
$\cJ =\left(\begin{smallmatrix} 0 & 1\\ 1&
0\end{smallmatrix}\right)\cC$, and $\cC$ denotes the antilinear 
conjugation map in 
$\bbC^2$, $\cC (a, b)^\top = (\ol a, \ol b)^\top$, $a,b\in\bbC$. As
one of our principal results we prove that under the most general 
hypothesis $q\in L^1_{\loc}(\bbR)$ on $q$, the maximally defined operator
$D(q)$ generated by $M(q)$ is actually $\cJ$-self-adjoint in
$L^2(\bbR)^2$. Moreover, we establish the existence of
Weyl--Titchmarsh-type solutions
$\Psi_+(z,\cdot) \in L^2([R,\infty))^2$ and $\Psi_-(z,\cdot) \in 
L^2((-\infty,R])$ for all $R\in\bbR$ of
$M(q)\Psi_\pm (z)=z\Psi_\pm (z)$ for $z$ in the resolvent set of $D$.

The Darboux transformations considered in this paper are the analog of
the double commutation procedure familiar in the KdV and Schr\"odinger
operator contexts. As in the corresponding case of Schr\"odinger
operators, the Darboux transformations in question guarantee
that the resulting potentials $q$ are locally nonsingular. Moreover,
we prove that the construction of $N$-soliton
$\NS_-$ potentials $q^{(N)}$ with respect to a general $\NS_-$ background
potential $q\in L^1_{\loc}(\bbR)$, associated with the Dirac-type
operators $D\big(q^{(N)}\big)$ and $D(q)$, respectively, amounts to the
insertion of $N$ complex conjugate pairs of $L^2(\bbR)^2$-eigenvalues
$\{z_1,\bar z_1,\dots,z_N,\bar z_N\}$ into the spectrum 
$\sigma(D(q))$ of $D(q)$, leaving the rest of the spectrum 
(especially, the essential spectrum $\sigma_{\rm e}(D(q))$) invariant, 
that is,
\begin{align} 
\sigma \big(D\big(q^{(N)}\big)\big)&=\sigma
(D(q))\cup\{z_1,\bar z_1,\dots,z_N,\bar z_N\}, \no \\
\sigma_{\rm e} \big(D\big(q^{(N)}\big)\big)
&=\sigma_{\rm e} (D(q)). \no
\end{align} 
These results are obtained by establishing the existence of bounded
transformation operators which intertwine the background Dirac operator
$D(q)$ and the Dirac operator $D\big(q^{(N)}\big)$ obtained after $N$
Darboux-type transformations. 
\end{abstract}

\maketitle

\section{Introduction} \lb{s1}

Various methods of inserting eigenvalues in spectral gaps
of one-dimensional Schr\"odinger operators $H(q)$ associated
with  differential expressions of the type 
\begin{equation}
L(q)=-\frac{d^2}{dx^2} + q \lb{1.0}
\end{equation}
in $L^2(\bbR)$ (or in $L^2((a,\infty))$, $a \ge -\infty$), with $q$
real-valued and locally integrable, have attracted an enormous amount of
attention. This is due to their prominent role in diverse fields such as
the inverse scattering approach, supersymmetric  quantum mechanics,
level  comparison theorems, as a tool to construct soliton solutions of
the Korteweg-de Vries (KdV) hierarchy  relative to (general) KdV 
background solutions, and in connection with B\"acklund transformations
for the KdV  hierarchy. The literature on this subject is too extensive
to go into details here, but we refer to the detailed accounts given in
\cite{Ge93}, \cite{GH00},
\cite[App.\ G]{GH02}, \cite{GST96}, \cite{GT96} and the references 
cited therein. Historically, these methods of inserting eigenvalues go
back to Jacobi \cite{Ja46} and Darboux
\cite{Da82} with decisive later contributions by Crum \cite{Cr55},
Schmincke \cite{Sc78}, and, especially, Deift \cite{De78}. 

Two particular methods turned out to be of special importance: The single
commutation method, also called the Crum-Darboux  method \cite{Cr55},
\cite{Da82} (actually going back at least to Jacobi \cite{Ja46}) and the
double commutation method, to be found, for instance, in the seminal work
of Gel'fand and Levitan \cite{GL51}. (The latter can be obtained by a
composition of two separate single commutation steps, explaining the name
double commutation.)

The single commutation method, although very simply implemented, has the
distinct disadvantage of relying on positivity properties of  certain
solutions $\psi$ of $H(q) \psi = \lambda \psi$, which confines its
applicability  to the insertion of eigenvalues below the spectrum of
$H(q)$ (assuming $H(q)$ to be bounded from below). A complete spectral
characterization of this method has been  provided by Deift
\cite{De78} (see also \cite{Sc78}) on the basis of unitary equivalence
of $A^*A|_{\ker(A)^\perp}$ and $AA^*|_{\ker(A^*)^\perp}$ for a  densely
defined closed linear operator $A$ in a (complex, separable) Hilbert
space.

The double commutation method on the other hand, allows one to insert
eigenvalues into {\em any} spectral gap of $H(q)$. Although relatively 
simply implemented also, a complete spectral characterization of the
double  commutation method for Schr\"odinger-type operators was more
recently achieved  in \cite{Ge93} on the basis of Weyl--Titchmarsh
$m$-function techniques and subsequently in \cite{GT96} (for general
Sturm--Liouville operators on arbitrary intervals) using a functional
analytic approach based on the notion of (intertwining) transformation
operators.

In this paper we concentrate on the analog of the double commutation
method for Dirac-type operators associated with the Lax operator for the
focusing nonlinear Schr\"odinger ($\NS_-$) hierarchy. Assuming
$q$ to be locally integrable, the Dirac-type operator corresponding to
the Lax differential expression in the $\NS_-$ case is associated with
the $2\times 2$ matrix-valued differential expression
\begin{equation}
M(q)=i\begin{pmatrix} \frac{d}{dx} & -q \\ -\bar q
&-\frac{d}{dx}\end{pmatrix} \lb{1.1}
\end{equation}
for $x\in\bbR$ (cf., e.g., \cite[Part I, Sect.\ I.2]{FT87}, \cite{ZS72},
and \cite{ZS73}). The maximally defined Dirac-type operator associated
with $M(q)$ in the (two-component) Hilbert space $L^2(\bbR)^2$ will then
be denoted by $D(q)$. By way of contrast, the corresponding (formally
self-adjoint) Lax differential expression for the defocusing $\NS_+$ case
is given by 
\begin{equation}
i\begin{pmatrix} \frac{d}{dx} & -q \\ \bar q
&-\frac{d}{dx}\end{pmatrix}. \lb{1.1a}
\end{equation}

As it turns out there is no natural analog of the single commutation
method for the Dirac operators associated with the focusing and
defocusing nonlinear Schr\"odinger hierarchies ($\NS_\pm$). However, the
complexified version of the $\NS_\pm$ hierarchies, the
Ablowitz--Kaup--Newell--Segur ($\AKNS$) hierarchy, supports two natural
analogs of the single commutation method. In order to briefly describe
them, we recall that the Dirac-type Lax differential expression
associated with the $\AKNS$ hierarchy is given by
\begin{equation}
M(p,q)=i\begin{pmatrix} \frac{d}{dx} & -q \\ p
&-\frac{d}{dx}\end{pmatrix} \lb{1.2}
\end{equation}
(cf.\, e.g., \cite{AKNS74} and \cite[Ch.\ 3]{GH02}) in terms of two
locally integrable coefficients $p,q$ on $\bbR$. The focusing ($\NS_-$)
and defocusing ($\NS_+$) nonlinear Schr\"odinger hierarchies are then
associated with the constraints 
\begin{equation}
\text{$\NS_\pm$: } p(x)=\pm\bar{q(x)}, \lb{1.3a}
\end{equation}
respectively. In this paper we will concentrate on the focusing $\NS_-$
case only. 

The two analogs of the single commutation method for the $\AKNS$ case,
which are usually called elementary Darboux transformations, can then be
described as follows. Suppose 
\begin{equation}
M(p,q)\Psi(z_1,x)=z_1\Psi(z_1,x), \quad
\Psi(z_1,x)=(\psi_1(z_1,x),\psi_2(z_1,x))^\top, \;
(z_1,x)\in\bbC\times\bbR, \lb{1.3}
\end{equation}
and 
\begin{equation}
M(p,q)\wti\Psi(\tilde z_1,x)=\ti z_1\wti\Psi(\ti z_1,x), \quad
\wti\Psi(\ti z_1,x)=(\ti\psi_1(\ti z_1,x),\ti\psi_2(\ti z_1,x))^\top, \;
(\ti z_1,x)\in\bbC\times\bbR. \lb{1.4}
\end{equation}
Then the two elementary Darboux transformations in the $\AKNS$ context
are given by (cf.\ \cite{Ko82}, \cite{KR92})
\begin{equation}
(p,q)\mapsto (\hat p_{z_1}, \hat q_{z_1}), \lb{1.5}
\end{equation}
where
\begin{align}
\begin{split}
\hat p_{z_1}(x)&=-\psi_2(z_1,x)/\psi_1(z_1,x), \\
\hat q_{z_1}(x)&=q'(x)-\psi_2(z_1,x)/\psi_1(z_1,x)q(x)^2+2iz_1q(x),  
\lb{1.6}
\end{split}
\end{align}
and 
\begin{equation}
(p,q)\mapsto (\check{p}_{\ti z_1},\check{q}_{\ti z_1}), \lb{1.7}
\end{equation}
where
\begin{align}
\begin{split}
\check{p}_{\ti z_1}(x)&=-p'(x)
+\ti\psi_1(\ti z_1,x)/\ti\psi_2(\ti z_1,x)p(x)^2+2i\ti z_1p(x), \\
\check{q}_{\ti z_1}(x)&=\ti\psi_1(\ti z_1,x)/\ti\psi_2(\ti z_1,x).  
\lb{1.8}
\end{split}
\end{align}
Similar to the case of Schr\"odinger operators, the analog of the
double commutation method for Dirac-type operators associated with
\eqref{1.2} is then obtained by an appropriate composition of the two
elementary Darboux transformations \eqref{1.6} and \eqref{1.8}. This 
two-step procedure is denoted by  
\begin{equation}
(p,q)\mapsto(p^{(1)}_{z_1,\ti z_1},q^{(1)}_{z_1,\ti z_1}) \lb{1.9}
\end{equation}
and leads to (cf., e.g., \cite{Ko82}, \cite{KR92}, \cite[Sect.\
4.2]{MS91}, \cite{SZ87})
\begin{align}\begin{split}
p^{(1)}_{z_1,\ti z_1}(x)&=p(x)-2i(\ti z_1-z_1)\psi_2(z_1,x)\ti\psi_2(\ti
z_1,x) /W(\Psi (z_1,x),\ti\Psi (\ti z_1,x)), \\
q^{(1)}_{z_1,\ti z_1}(x)&=q(x)-2i(\ti z_1-z_1)\psi_1(z_1,x)\ti\psi_1(\ti
z_1,x) /W(\Psi (z_1,x),\ti\Psi (\ti z_1,x)). \lb{1.10}
\end{split}
\end{align}
(Here $W(F,G)$ denotes the Wronskian of $F,G\in\bbC^2$). In contrast to
\eqref{1.5}, \eqref{1.6} and \eqref{1.7}, \eqref{1.8}, the
two-step procedure \eqref{1.9}, \eqref{1.10} with $\ti z_1=\bar{z_1}$, is
compatible with the $\NS_\pm$ cases and one explicitly obtains
the following Darboux-type transformation in the $\NS_-$ case,
\begin{equation}
q(x) \mapsto q^{(1)}_{z_1}(x)=q(x)+4 \, \Im(z_1)\f{\psi_1(z_1,x)
\bar{\psi_2(z_1,x)}}{|\psi_1 (z_1,x)|^2+|\psi_2(z_1,x)|^2}. \lb{1.11}
\end{equation} 
The transformation \eqref{1.11} for Dirac-type operators assoiated with 
\eqref{1.1} in the $\NS_-$ context represents the analog of the double
commutation method for Schr\"odinger operators and leads to locally
nonsingular $\NS_-$  potentials $q^{(1)}_{z_1}$, assuming $q$ to be free
of local singularities.

By analogy to the KdV and Schr\"odinger operator case, one
expects the $\NS_-$ potential $q^{(1)}_{z_1}(x)$ to produce an 
eigenvalue at the spectral point $z_1$ for the associated Dirac operator
$D\big(q^{(1)}_{z_1}\big)$, assuming $z_1$ to be a point in
the resolvent set of the ``background'' operator $D(q)$. Actually, by a
simple symmetry consideration, one expects a pair of eigenvalues
$(z_1,\bar{z_1})$ in the point spectrum of $D\big(q^{(1)}_{z_1}\big)$.
To prove this fact and to show that the remaining spectral
characteristics (especially, the essential spectrum of $D(q)$) remain
invariant under the Darboux-type transformation \eqref{1.11}, is the
principal purpose of this paper. More precisely, if we denote by
$q^{(N)}_{z_1,\dots,z_N}$ the $\NS_-$ potential obtained after an
$N$-fold iteration of the Darboux-type transformation and by
$D\big(q^{(N)}_{z_1,\dots,z_N}\big)$ the resulting Dirac operator, we
will prove that 
\begin{align} 
\sigma \big(D\big(q^{(N)}_{z_1,\dots,z_N}\big)\big)&=\sigma
(D(q))\cup\{z_1,\bar z_1,\dots,z_N,\bar z_N\}, \lb{1.12} \\
\sigma_{\rm p} \big(D\big(q^{(N)}_{z_1,\dots,z_N}\big)\big)
&=\sigma_{\rm p} (D(q))\cup\{z_1,\bar z_1,\dots,z_N,\bar z_N\}, \lb{1.13}
\\ 
\sigma_{\rm e} \big(D\big(q^{(N)}_{z_1,\dots,z_N}\big)\big)
&=\sigma_{\rm e} (D(q)). \lb{1.14}
\end{align} 
(Here $\sigma(S)$, $\sigma_{\rm p}(S)$, and $\sigma_{\rm e}(S)$ denote
the spectrum, point spectrum, and essential spectrum of a densely
defined closed operator $S$ in a complex separable Hilbert space $\cH$, 
cf.\ Section \ref{s5} for more details on spectra, etc.) Actually, we
will go a step beyond \eqref{1.12}--\eqref{1.14} and establish the
existence of bounded transformation operators which intertwine
$D\big(q^{(N)}_{z_1,\dots,z_N}\big)$ and $D(q)$. 

When trying to embark on proving results of the type
\eqref{1.12}--\eqref{1.14} for Dirac-type operators associated
with the differential expression \eqref{1.1} in the $\NS_-$ context,
one finds oneself at a distinct disadvantage when compared to the case
of Schr\"odinger operators with real-valued potentials: While $L$ in
\eqref{1.0} is formally self-adjoint for $q$ real-valued, $M(q)$ in
\eqref{1.1} is never self-adjoint (except, in the trivial case $q=0$).
As a consequence, the original approach to a complete spectral
characterization of the double commutation method for Schr\"odinger
operators in \cite{Ge93}, based on Weyl--Titchmarsh theory and hence on
spectral theory, is doomed from the start as there simply is no spectral
theory for non-self-adjoint Dirac-type operators associated with
\eqref{1.1} under our general hypothesis $q\in L^1_\loc(\bbR)$. That
leaves one with only one possible line of attack, the analog of the
transformation operator approach developed in the general
Sturm--Liouville context in \cite{GT96}. As it will turn out in Section
\ref{s6}, this approach is indeed successful although it requires
more sophisticated and elaborate arguments  compared to those in
\cite{GT96}.

While the differential expression $M(q)$ in \eqref{1.1} is never formally
self-adjoint (if $q\neq 0$), it is, however, formally
$\cJ$-self-adjoint, that is,
\begin{equation}
\cJ M(q)\cJ =M(q)^*. \lb{1.17}
\end{equation} 
Here $\cJ $ is defined by
\begin{equation}
\cJ =\begin{pmatrix} 0 & 1\\ 1 & 0\end{pmatrix}\cC, \lb{1.18}
\end{equation}
and $\cC$ denotes the antilinear  conjugation map in $\bbC^2$, 
\begin{equation}
\cC (a, b)^\top = (\ol a, \ol b)^\top, \quad a,b\in\bbC. \lb{1.19} 
\end{equation}
As one of our principal results in this paper we will prove in
Section \ref{s3} that under the most general hypothesis $q\in
L^1_{\loc}(\bbR)$, the maximally defined Dirac operator $D(q)$ associated
with $M(q)$ is in fact $\cJ$-self-adjoint,
\begin{equation}
\cJ D(q)\cJ =D(q)^*\; (=D(-q)). \lb{1.20}
\end{equation} 
As an aside we should mention that the corresponding maximally defined
Lax operator associated with the defocusing $\NS_+$ differential
expression \eqref{1.1a} is in fact self-adjoint assuming $q\in
L^1_{\loc}(\bbR)$ only (this is proved in the references mentioned in
Section \ref{s3}). This should be contrasted with the case of
one-dimensional Schr\"odinger differential expressions
$L(q)$ in
\eqref{1.0} (the Lax differential expression associated with the KdV
hierarchy). If $q\in L^1_{\loc}(\bbR)$ in \eqref{1.0} is real-valued,
then $L(q)$ is formally self-adjoint but the maximally defined operator
$H(q)$ in $L^2(\bbR)$ associated with $L(q)$ may not be self-adjoint.
The latter situation occurs precisely when $L(q)$ is in the limit circle
case (as opposed to the limit point case) at $+\infty$ and/or $-\infty$ 
(cf.\ \cite[Ch.\ 9]{CL85}).
This is in sharp contrast to the focusing (respectively, defocusing)
$\NS$ case where $D(q)$ is always $\cJ$-self-adjoint (respectively,
self-adjoint).

Summarizing, we derive the following principal new results in this
paper, assuming the optimal condition $q\in L^1_{\loc}(\bbR)$:

$-$ $\cJ$-self-adjointness of $D(q)$.

$-$ The existence of Weyl--Titchmarsh-type solutions of $M(q)\Psi=z\Psi$
for all \\
\indent \indent $z\in\rho(D(q))$.

$-$ The existence and boundedness of transformation operators
intertwining the \\
\indent \indent operators $D\big(q^{(N)}\big)$ and $D(q)$.

$-$ A spectral analysis of $\NS_-$ Darboux transformations (cf.\
\eqref{1.12}--\eqref{1.14}).
\smallskip

Finally we briefly describe the content of each section. Section
\ref{s2} first introduces our main notation and then proceeds to a
review of Darboux transformations for $\AKNS$ and $\NS_-$ systems.
Section \ref{s3} is devoted to a proof of the $\cJ$-self-adjointness
property \eqref{1.20} of $D(q)$, assuming $q\in L^1_\loc(\bbR)$ only.
Section \ref{s4} constructs eigenvalues of $D\big(q^{(1)}_{z_1}\big)$ at
pairs $z_1,\bar{z_1}$, $z_1\in\rho(D(q))$ and associated
$L^2(\bbR)^2$-eigenfunctions. Section \ref{s5} derives some basic
spectral properties of general Dirac-type operators $D(q)$ and 
establishes the existence of Weyl--Titchmarsh-type solutions associated
with $M(q)$. This shows a remarkable similarity to self-adjoint systems
and appears to be without precedent in this non-self-adjoint context. Our
final Section \ref{s6} establishes the existence of bounded
transformation operators intertwining $D\big(q^{(1)}_{z_1}\big)$ and
$D(q)$ and then employs these transformation operators to prove the
spectral properties
\eqref{1.12}--\eqref{1.14}.   

All results in the principal part of this paper, Sections
\ref{s3}--\ref{s6}, are proved under the optimal condition $q\in
L^1_\loc(\bbR)$. Moreover, practically all results in
Sections \ref{s3}--\ref{s6} are new as long as one goes beyond bounded or
periodic potentials $q$. In particular, Theorem \ref{RT4.1}
(characterizing transformation operators) and Theorem \ref{RT4.2}
(proving \eqref{1.12}--\eqref{1.14}) appear to be the first of their
kind under any assumptions on $q$.

In this paper we confine ourselves to a stationary (i.e.,
time-independent) approach only. Applications to the
time-dependent focusing $\NS_-$ equation and to nonlinear optics will be
made in a subsequent paper
\cite{CGHL03}. \\

\section{Darboux-type Transformations for $\AKNS$ and $\NS_-$ Systems}
\lb{s2}

In this section we take a close look at Darboux-type transformations for
non-self-adjoint Dirac-type differential expressions $M(q)$ (cf.\
\eqref{Y1.1}) applicable to $\AKNS$ systems, with special emphasis on the
case of the focusing nonlinear Schr\"odinger equation $\NS_-$ (cf.\
\eqref{2.5}). 

Throughout this paper we use the following notation: $'=d/dx$; for a
matrix $A$ with complex-valued entries, $A^\top$ denotes the transposed
matrix, $\bar{A}$ the matrix  with complex conjugate entries, 
and $A^*=\bar{A}^\top=\bar{A^\top}$ the adjoint matrix. Occasionally, we
use the following $2\times 2$ matrices,
\begin{align}
\begin{split}
I_2&=\begin{pmatrix}1 & 0\\0 & 1 \end{pmatrix}, \quad 
\sigma_1=\begin{pmatrix} 0 & 1\\ 1& 0\end{pmatrix}, \quad 
\sigma_3=\begin{pmatrix}1 & 0\\0 & -1\end{pmatrix}, \quad \\
\sigma_4&=\begin{pmatrix}0 & 1\\-1 & 0\end{pmatrix}, \quad 
\sigma_5=\begin{pmatrix}1 & 0 \\0 & 0\end{pmatrix}, \quad 
\sigma_6=\begin{pmatrix}0 & 0\\0 & 1\end{pmatrix}.
\end{split}
\end{align} 
If $A=(a_1,a_2)^\top$ and $B=(b_1,b_2)^\top$ are $2\times 1$
column-vectors, then $(A,B)_{\bbC^2}
=B^*A=a_1\bar{b_1} +a_2\bar{b_2}$ denotes the usual scalar product
in $\bbC^2$, $\|A\|_{\bbC^2} =(|a_1|^2+|a_2|^2)^{1/2}$ the associated
norm, and $A^\bot =(a_2,-a_1)=A^\top\sigma_4$ the $1\times 2$
row-vector  perpendicular to $A$ (in the sense that $A^\bot A=0$). We
also use the notation 
\begin{equation}
W(A,B)=B^\bot A=-A^\bot B=a_1b_2-a_2b_1
=\big(A,\bar{B^\bot}\big)_{\bbC^2}
\end{equation}
for the Wronskian of $A$ and $B$. The space of $2\times 2$
matrices with entries in $\bbC$ is denoted by $\bbC^{2\times 2}$ and
the operator norm of a $2\times 2$ matrix $A$ induced by the
usual norm in $\bbC^2$ is denoted by $\|A\|_{\bbC^{2\times 2}}$. In
the following, $\Omega\subseteq\bbR$ denotes an open subset of
$\bbR$ and $\AC_\loc(\Omega)^{2\times m}$, $m=1,2$, denotes
the set  of $2\times m$ matrices with locally absolutely continuous
entries on $\Omega$ (for $2\times 1$ columns we use the
corresponding notation $\AC_\loc(\Omega)^2$). We define\footnote{For
brevity we write $L^p(\Omega)$ for $L^p(\Omega;dx)$ and suppress the
Lebesgue measure $dx$ whenever possible.}
$L^p(\Omega)^{2\times m}$ and $L^p_{\loc}(\Omega)^{2\times m}$,
$m=1,2$, to consist of $2\times m$ matrices with entries in
$L^p(\Omega)$ and $L^p_{\loc}(\Omega)$, respectively. 
In the special case $\Omega=\bbR$ and 
$F,G \in L^2(\bbR)^2$, the scalar product of $F$ and $G$ is 
denoted by $(F,G)_{L^2}=\int_{\bbR} dx\,(F(x),G(x))_{\bbC^2}$
with associated norm of $F$ given by
$\|F\|_{L^2}=\left(\int_\bbR
dx\,\|F(x)\|_{\bbC^2}^2\right)^{1/2}$. Finally, the
open complex upper (respectively, lower) half-plane is denoted by
$\bbC_+$ (respectively, $\bbC_-$); the domain, range, and kernel (null
space) of a linear operator $T$ are denoted by $\dom(T)$, $\ran(T)$, and
$\ker(T)$, respectively.

\begin{hypothesis} \lb{h2.1}
Let $\Omega\subseteq\bbR$ open and assume $p,q\in L^1_\loc(\Omega)$.
\end{hypothesis}

Assuming Hypothesis~\ref{h2.1} and $z\in \mathbb{C}$, we introduce the 
$2\times 2$ matrix $U(z,p,q)$ and the $2\times 2$ matrix-valued
differential expression $M(p,q)$ by
\begin{equation}
U(z,p,q)=\begin{pmatrix}-iz & q \\ p & iz\\\end{pmatrix},
\quad M(p,q)=i\begin{pmatrix} \frac{d}{dx} &
-q\\p&-\frac{d}{dx}\end{pmatrix}. \label{Y1.1}
\end{equation}

The functions $p$ and $q$ in \eqref{Y1.1} are 
referred to as $\AKNS$ {\it potentials} due to the fact that $M(p,q)$ is
the Lax differential expression associated with the $\AKNS$ hierarchy
(see, e.g., \cite{AKNS74} and \cite[Ch.\ 3]{GH02}). The particularly
important special case $p=-\bar{q}$ will be referred to as the $\NS_-$
case (due to the obvious connection of
\eqref{Y1.1} with the zero curvature representation and the Lax operator
for the focusing nonlinear Schr\"odinger equation, see, e.g., \cite[Part
1, Sect.\ I.2]{FT87}, \cite{ZS72}, and \cite{ZS73}), and then $q$ is
called an $\NS_-$ {\it potential}. For given $z\in\mathbb{C}$,
$\Omega\subseteq\bbR$, and $\AKNS$ potentials $(p,q)$, a function $\Psi
(z,\cdot)\in\AC_\loc(\Omega)^2$ is called a {\em $z$-wave function
associated with $(p,q)$ on $\Omega$} if
$\Psi'(z,x)=U(z,p,q)\Psi(z,x)$ holds for a.e.\ $x\in \Omega$, that is,
if $\Psi =(\psi_1,\psi_2)^\top$ satisfies the following first-order
system of differential equations
\begin{align}
&\psi'_1(z,x)=-iz\psi_1(z,x)+q(x)\psi_2(z,x),\quad
\psi'_2(z,x)=iz\psi_2(z,x)+p(x)\psi_1(z,x) \label{Y1.2} 
\end{align}
for a.e.\ $x\in\Omega$. Equivalently,
$\Psi(z)=(\psi_1(z),\psi_2(z))^\top\in\AC_\loc(\Omega)^2$ is a $z$-wave
function associated with $(p,q)$ on $\Omega$ if and only if
$M(p,q)\Psi(z)=z\Psi(z)$ on $\Omega$ in the distributional sense. If
for some $z\in\bbC$, $\Psi(z)$ and $\Phi(z)$ are $z$-wave functions
associated with $(p,q)$ on $\Omega$, their Wronskian is well-known to be
constant with respect to $x\in\Omega$,
\begin{equation}
\f{d}{dx}W(\Psi(z,x),\Phi(z,x))=0, \quad x\in\Omega. \lb{2.4}
\end{equation}
More generally, if $\Psi(z_1)$ and $\Phi(z_2)$ are $z_1$- and $z_2$-wave
functions associated with $(p,q)$ on $\Omega$, then
\begin{align}
&\f{d}{dx}W(\Psi(z_1,x),\Phi(z_2,x))
=i(z_2-z_1)[\psi_1(z_1,x)\phi_2(z_2,x)
+\psi_2(z_1,x)\phi_1(z_2,x)] \no \\
& \hspace*{3.7cm} =i(z_2-z_1)\Psi(z_1,x)^\top \sigma_1 \Phi(z_2,x), 
\lb{2.6} \\
& \Psi(z_1)=(\psi_1(z_1),\psi_2(z_1))^\top, \; 
\Phi(z_2)=(\phi_1(z_2),\phi_2(z_2))^\top, \; z_1,z_2\in\bbC, \quad
x\in\Omega. \no
\end{align}
In the $\NS_-$ case $p=-\bar q$ we use the notation
\begin{equation}
U(z,q)=\begin{pmatrix}-iz & q(x) \\ -\bar{q(x)} & iz\\\end{pmatrix},
\quad M(q)=i\begin{pmatrix} \frac{d}{dx} &
-q \\ -\bar q &-\frac{d}{dx}\end{pmatrix} \label{2.5}
\end{equation}
instead of \eqref{Y1.1}, and we then call any distributional
solution $\Psi(z)$ of $M(q)\Psi(z)=z\Psi(z)$ an $\NS_-$ $z$-wave function
associated with $q$.
 
If $\Psi(z)$ is a $z$-wave function associated with  $(p,q)$, and 
$\Psi(z,x_0)=0$ for some $x_0\in \Omega$, then $\Psi (z,x)$ vanishes
identically for all $x$ in an open neighborhood of $x_0$ in $\Omega$ by
the unique solvability of the Cauchy problem for \eqref{Y1.2}. Therefore,
we will always assume that $\Psi (z,x)\neq 0$ for all $x\in \Omega$. 

If $q=0$ (and analogously if $p=0$) a.e.\ on $\Omega$, then the system
\eqref{Y1.2} decomposes and yields 
\begin{align}
\begin{split}
\psi_1(z,x)&=C_1 \exp(-izx), \\
\psi_2(z,x)&=C_1 \bigg(\int^x_{x_0}
dx'\, p(x')\exp(-2izx')\bigg)\exp(izx)+ C_2 \exp(izx)
\end{split}
\end{align}  
for some constants $C_1, C_2\in\bbC$, $x_0\in\Omega$. 

For further use we collect some simple consequences of \eqref{Y1.2}.
First we introduce the antilinear (i.e., conjugate linear) operator
$\cJ$ defined by 
\begin{equation}
\cJ=\sigma_1\cC= \begin{pmatrix} 0 & 1\\ 1& 0\end{pmatrix}\cC, 
\quad \cJ^2=I_2, \lb{2.16}
\end{equation} 
with $\cC$ the antilinear conjugation map 
\begin{equation}
\cC (a, b)^\top =( \ol a, \ol b)^\top, \quad
a,b\in\bbC. \lb{2.12}
\end{equation}
Moreover, we introduce the antilinear involution $\cK$ defined by
\begin{equation}
\cK =\sigma_4 \cC= \begin{pmatrix}0 & 1\\-1 & 0\end{pmatrix}\cC,  
\quad \cK^2=-I_2.
\end{equation}
We also note that 
\begin{equation}
\| \cJ F\|_{\bbC^2}=\|F\|_{\bbC^2}, \quad 
\| \cK F\|_{\bbC^2}=\|F\|_{\bbC^2}, \quad F\in\bbC^2. \lb{2.10}
\end{equation}
\begin{lemma}\label{YL1} Assume Hypothesis \ref{h2.1}, $z\in
\mathbb{C}$, and suppose $\Psi(z) =(\psi_1(z),\psi_2(z))^\top$ is
a $z$-wave function associated with $(p,q)$ on $\Omega$. Then the
following assertions hold. \\
$(i)$ $\phi (z,x)=-\psi_2(z,x)/\psi_1(z,x)$ satisfies the Riccati-type
equation 
\begin{equation}
-\phi'(z,x)+q(x)\phi(z,x)^2 +2iz\phi(z,x)-p(x)=0
\end{equation}
on the set $\{x\in\Omega\,|\, \psi_1(z,x)\neq 0\}$. \\
$(ii)$ $\varphi (z,x)=\psi_1(z,x)/\psi_2(z,x)$ satisfies the
Riccati-type equation 
\begin{equation}
-\varphi'(z,x)-p(x)\varphi(z,x)^2-2iz\varphi(z,x)+q(x)=0 
\end{equation}
on the set $\{x\in\Omega\,|\, \psi_2(z,x)\neq 0\}$. \\
Assume in addition the $\NS_-$ case $p=-\bar q$.  Then the
following assertions hold. \\
$(iii)$ If $\Psi(z)=(\psi_1(z),\psi_2(z))^\top$ is a $z$-wave function
associated with $q$ then 
\begin{equation}
\cK \Psi (z)=\sigma_4 \cC \Psi(z)=\big(\overline{\psi_2(z)},
-\overline{\psi_1(z)}\big)^\top=(\Psi(z)^\bot)^* \lb{2.9}
\end{equation}
is a $\bar{z}$-wave function associated with  $q$ and 
\begin{equation}
M(q)=-\cK M(q)\cK . \lb{2.11}
\end{equation}
$(iv)$ The following identity holds
\begin{equation}
\big(\| \Psi (z,x)\|_{\bbC^2}^2\big)'=
2\Im(z)\big[|\psi_1(z,x)|^2-|\psi_2(z,x)|^2\big]. \lb{2.13}
\end{equation}
$(v)$ $M(q)$ and $M(q)^*$ are formally unitarily equivalent in the 
sense that 
\begin{equation}
M(q)^*=\sigma_3M(q)\sigma_3 =M(-q). \lb{2.14}
\end{equation}
In addition, $M(q)$ is formally $\cJ$-self-adjoint in the following
sense
\begin{equation}
\cJ M(q) \cJ = M(q)^*. \lb{2.15}
\end{equation}
\end{lemma}

Consider $\AKNS$ potentials $p,q\in L^1_\loc(\Omega)$. Fix $z,z_1,\ti
z_1\in \mathbb{C}$, a $z_1$-wave function $\Psi (z_1)$, and a $\ti
z_1$-wave function $\ti\Psi (\ti z_1)$ associated with $(p,q)$. Our
objective is to construct new potentials $p^{(1)},q^{(1)}\in
L^1_\loc(\Omega^{(1)})$ for some open set $\Omega^{(1)}\subseteq\Omega$,
and the corresponding 
$z$-wave functions associated with $(p^{(1)},q^{(1)})$ on $\Omega^{(1)}$.
In the $\NS_-$ case $p=-\bar q$ we choose $\ti z_1=\bar{z_1}$ and 
$\ti\Psi (\bar{z_1})=\cK\Psi (z_1)$, see  Lemma~\ref{YL1}\,(iii).

\begin{remark}\label{LR1} Let $\Gamma\in \AC_\loc(\Omega)^{2\times
2}$, $A,B\in L^1(\Omega)^{2\times 2}$ for some open subset
$\Omega\subseteq\bbR$, suppose that the following identity holds
\begin{equation}\Gamma '(x)+\Gamma (x)A(x)-B(x)\Gamma (x)=0 
\, \text{ for a.e.\ $x\in \Omega$},\label{Y3.1}
\end{equation}
and assume that $\Phi\in\AC_\loc(\Omega)^2$ satisfies the first-order
system $\Phi'=A\Phi$ on $\Omega$.  Then the function $\Phi^{(1)}$,
defined by $\Phi^{(1)}=\Gamma \Phi$, satisfies
$\Phi^{(1)}\in\AC_\loc(\Omega)^2$ and the first-order system
$(\Phi^{(1)})'=B\Phi^{(1)}$ on $\Omega$.
\end{remark}

\begin{lemma}\label{YL2} Assume Hypothesis~\ref{h2.1} and $z,z_1\in
\mathbb{C}$. In addition, suppose 
$\Psi(z_1)=(\psi_1(z_1),\psi_2(z_1))^\top$ is a $z_1$-wave function
associated with $(p,q)$ on $\Omega$ and introduce
\begin{equation}
\hat\Omega_{z_1}=\{x\in\Omega \,|\, \psi_1(z_1,x)\neq 0\}. 
\end{equation}
Define $\hat p_{z_1}$, $\hat q_{z_1}$,
$\hat{\Gamma}_0(q,\hat p_{z_1})$, and $\hat{\Gamma} (z,q,\hat p_{z_1})$ 
on $\hat\Omega_{z_1}$ by
\begin{align} 
&\hat p_{z_1}(x)=-{\psi_2(z_1,x)}/{\psi_1(z_1,x)},\quad
 \hat q_{z_1}(x)=q'(x)+\hat{p}_{z_1}(x)q(x)^2+2iz_1q(x), \label{Y3.2} \\
&\hat{\Gamma}_0(x,q,\hat p_{z_1})= -\frac{1}{2} \begin{pmatrix}
q(x)\hat p_{z_1}(x) & q(x) \\ \hat p_{z_1}(x) & 1\end{pmatrix}, \\
& \hat{\Gamma} (z,x,q,\hat p_{z_1})=i(z-z_1)\sigma_5
+\hat{\Gamma}_0(x,q,\hat p_{z_1}). \label{Y3.3}
\end{align}
Then $\Gamma=\hat{\Gamma}$ satisfies \eqref{Y3.1} on 
$\hat\Omega_{z_1}$ with $A$ and $B$ given by 
\begin{equation}
A(z,x)=U(z,p,q), \quad B(z,x,z_1)=U(z,\hat{p}_{z_1},\hat{q}_{z_1}), 
\quad x\in\hat\Omega_{z_1}.
\end{equation}
\end{lemma}
\begin{proof} Using Lemma~\ref{YL1}\,(i) one verifies that 
$\Gamma =\hat{\Gamma}_0$ satisfies \eqref{Y3.1} with
$A=U(z_1,p,q)$ and $B=U (z_1,\hat{p}_{z_1},\hat{q}_{z_1})$. 
Since $U (z)=U(z_1)-i(z-z_1)\sigma_3$, the  conclusion follows.
\end{proof}

\noindent Similarly, by Lemma~\ref{YL1}\,(ii), one has the following
analogous result.

\begin{lemma}\label{YL3} Assume Hypothesis~\ref{h2.1} and let $z,\ti
z_1\in \mathbb{C}$. In addition, suppose that $\ti\Psi(\ti z_1)=(\tilde
\psi_1(\ti z_1),\tilde \psi_2(\ti z_1))^\top$ is a $\ti z_1$-wave 
function associated with $(p,q)$ on $\Omega$ and introduce
\begin{equation} 
\check\Omega_{\ti z_1}=\{x\in\Omega \,|\, \ti\psi_2(\ti z_1,x)\neq 0\}. 
\end{equation}
Define $\check p_{\ti z_1}$, $\check q_{\ti z_1}$, 
$\check{\Gamma}_0(p,\check q_{\ti z_1})$, and 
$\check{\Gamma} (z,p,\check q_{\ti z_1})$ on 
$\check\Omega_{\ti z_1}$ by
\begin{align}
&\check{p}_{\ti z_1}(x)=-p'(x)+\check{q}_{\ti z_1}(x)p(x)^2
+2i\ti z_1p(x),
\quad \check{q}_{\ti z_1}(x)={\ti\psi_1(\ti z_1,x)}/\ti\psi_2(\ti z_1,x), 
\label{Y4.1} \\
&\check{\Gamma}_0(x,p,\check q_{\ti z_1})=\frac{1}{2}\begin{pmatrix}-1 &
\check{q}_{\ti z_1}(x) \\ p(x)  & -p(x)\check{q}_{\ti
z_1}(x)\end{pmatrix}, \\ 
&\check{\Gamma}(z,x,p,\check q_{\ti z_1})=i(z-\ti z_1)\sigma_6
 +\check{\Gamma}_0(x,p,\check q_{\ti z_1}).\label{Y4.2}
\end{align}
Then $\Gamma=\check{\Gamma}$ satisfies \eqref{Y3.1} on 
$\check\Omega_{\ti z_1}$ with $A$ and $B$ given by 
\begin{equation}
A(z,x)=U (z,p,q), \quad B(z,x,\ti z_1)=U(z,\check{p}_{\ti z_1},
\check{q}_{\ti z_1}), \quad x\in\check\Omega_{\ti z_1}.
\end{equation} 
\end{lemma}

The Darboux-type transformations characterized by \eqref{Y3.3} and
\eqref{Y4.2} are also called elementary Darboux transformations. They
have been discussed, for instance, in \cite{Ko82} and \cite{KR92}. In the
special context of algebro-geometric $\AKNS$ solutions, the effect of
elementary Darboux transformations on the underlying compact
hyperelliptic curve (in connection with the insertion and deletion of
eigenvalues as well as the isospectral case) was studied in detail in
\cite{GH00}, \cite[App.\ G]{GH02} (see also
\cite{GW98}, \cite{GW98a}).  

Next, we will construct the transformation matrix 
$\Gamma(z,\Psi(z_1),\ti\Psi(\ti z_1))$ that satisfies equation
\eqref{Y3.1} with $A(z,x)=U(z,p,q)$ and 
$B(z,x)=U \big(z,p^{(1)}_{z_1,\ti z_1},q^{(1)}_{z_1,\ti z_1}\big)$
as the product of $\hat{\Gamma}(z,\hat p_{z_1},\check q_{\ti z_1})$ and 
$\check{\Gamma}(z,q,\hat{p}_{z_1})$. Since we will choose 
$\ti z_1=\bar{z_1}$ in the $\NS_-$ context, we omit the
$\ti z_1$-dependence in $p^{(1)}$, $q^{(1)}$, $\Omega^{(1)}$,
$\Phi^{(1)}$, etc., in the $\NS_-$ case in the following.

\begin{theorem}\label{YT1} 
Assume Hypothesis~\ref{h2.1}. \\
$(i)$ Suppose $z,z_1,\ti z_1\in \mathbb{C}$,
and assume that $\Psi (z_1)=(\psi_1(z_1),\psi_2(z_1))^\top$ and 
$\ti\Psi (\ti z_1)=(\ti\psi_1(\ti z_1),\ti\psi_2(\ti z_1))^\top$ are
$z_1$- and $\ti z_1$-wave functions, respectively, associated 
with $(p,q)$ on $\Omega$. In addition, introduce 
\begin{equation}
\Omega^{(1)}_{z_1,\ti z_1}=\{x\in\Omega \,|\,
W(\Psi (z_1,x),\ti\Psi(\ti z_1,x))\neq 0\}. 
\end{equation}
Define
$p^{(1)}_{z_1,\ti z_1}$, $q^{(1)}_{z_1,\ti z_1}$, and 
$\Gamma(z,\Psi(z_1),\ti\Psi(\ti z_1))$ on $\Omega^{(1)}_{z_1,\ti z_1}$ by
\begin{align}
&p^{(1)}_{z_1,\ti z_1}(x)=p(x)-2i(\ti z_1-z_1)\psi_2(z_1,x)\ti\psi_2(\ti
z_1,x) /W(\Psi (z_1,x),\ti\Psi (\ti z_1,x)),\label{Y4.4} \\
&q^{(1)}_{z_1,\ti z_1}(x)=q(x)-2i(\ti z_1-z_1)\psi_1(z_1,x)\ti\psi_1(\ti
z_1,x) /W(\Psi (z_1,x),\ti\Psi (\ti z_1,x)),\label{Y4.3}\\
&\Gamma(z,x,\Psi(z_1),\ti\Psi(\ti z_1))=-\frac{i}{2}zI_2
-\frac{i}{2}W(\Psi (z_1,x),\ti\Psi (\ti z_1,x))^{-1} \no \\
& \hspace*{3.5cm} \times \big(\ti z_1\ti\Psi (\ti z_1,x)\Psi
(z_1,x)^\bot  -z_1\Psi (z_1,x)\ti\Psi(\ti z_1,x)^\bot \big). \label{Y4.5}
\end{align}
Then  $\Gamma$ satisfies the first-order system 
\begin{align} 
&\Gamma'(z,x,\Psi(z_1),\ti\Psi(\ti z_1))
+\Gamma(z,x,\Psi(z_1),\ti\Psi(\ti z_1)) U(z,p,q) \no \\
&\quad -U \big(z,p^{(1)}_{z_1,\ti z_1},q^{(1)}_{z_1,\ti z_1}\big)
\Gamma(z,x,\Psi(z_1),\ti\Psi(\ti z_1)) =0 \label{Y5.1}
\end{align}
a.e.\ on $\Omega^{(1)}_{z_1,\ti z_1}$. Thus, if $\Upsilon (z)$ is a 
$z$-wave function associated with $(p,q)$ on $\Omega$, then 
$\Upsilon^{(1)}_{z_1,\ti z_1}(z)$, defined by  
\begin{equation}
\Upsilon ^{(1)}_{z_1,\ti z_1}(z,x)=\Gamma(z,x,\Psi(z_1),\ti\Psi(\ti z_1))
\Upsilon(z,x), 
\end{equation}
is a $z$-wave function associated with  $(p^{(1)}_{z_1,\ti
z_1},q^{(1)}_{z_1,\ti z_1})$ on $\Omega^{(1)}_{z_1,\ti z_1}$. \\
$(ii)$ Assume the $\NS_-$ case $p=-\bar{q}$ and $z,z_1\in \mathbb{C}$. 
Then, for $\ti z_1=\bar{z_1}$, $\ti\Psi(\ti z_1)=\cK \Psi(z_1)
=\big(\bar{\psi_2(z_1)},-\bar{\psi_1(z_1)}\big)^\top$, and 
\begin{equation}
\Omega^{(1)}_{z_1}=\{x\in\Omega \,|\,
W(\Psi (z_1,x),\cK \Psi(z_1,x))\neq 0\},
\end{equation}
formulas
\eqref{Y4.3}--\eqref{Y4.5}  simplify to 
\begin{align} 
q^{(1)}_{z_1}(x)&=q(x)+4\Im(z_1)\psi_1(z_1,x)\bar{\psi_2(z_1,x)}
\|\Psi (z_1,x)\|_{\bbC^2}^{-2}, \label{Y5.2} \\ 
p^{(1)}_{z_1}(x)&=p(x)-4\Im(z_1)\psi_2(z_1,x)
\bar{\psi_1(z_1,x)} \|\Psi (z_1,x)\|_{\bbC^2}^{-2} \no \\
&=-\bar{q^{(1)}_{z_1}(x)}, \label{Y5.3} \\
& \hspace*{-1cm} 
\Gamma (z,x,\Psi(z_1),\cK \Psi(z_1))=-(i/2)(z-z_1)I_2 \no \\
& \hspace*{2.75cm} +\Im(z_1) \| \Psi (z_1,x)\|_{\bbC^2}^{-2}
\cK \Psi(z_1,x)\Psi (z_1,x)^\bot 
\label{Y5.4}
\end{align}
for $x\in \Omega^{(1)}_{z_1}$. In particular, if $\Upsilon (z)$ is a
$z$-wave function associated with $q$ on $\Omega$, then 
$\Upsilon^{(1)}_{z_1}(z)$, defined by  
\begin{equation}
\Upsilon ^{(1)}_{z_1}(z,x)=\Gamma(z,x,\Psi(z_1),
\cK \Psi(z_1))\Upsilon(z,x), 
\end{equation}
is a $z$-wave function associated with  $q^{(1)}_{z_1}$ on
$\Omega^{(1)}_{z_1}$ $($which may vanish identically w.r.t. 
$x\in\Omega^{(1)}_{z_1}$, cf.\  Remark \ref{YR2b}$)$.
\end{theorem}
\begin{proof} First, we use Lemma~\ref{YL2} for $z=\ti z_1$ and 
$\Psi (z_1)$ to construct $\hat{p}_{z_1}$, $\hat{q}_{z_1}$, 
$\hat{\Gamma}(\ti z_1,q,\hat p_{z_1})$ as in \eqref{Y3.2}--\eqref{Y3.3}.
Define
$\hat{\Psi}_{z_1} (\ti z_1)=\hat{\Gamma}(\ti z_1,q,\hat p_{z_1})\Psi
(\ti z_1)$. By  Lemma~\ref{YL2} and Remark~\ref{LR1}  we conclude that
$\hat{\Psi}_{z_1}(\ti z_1)$ is a $\ti z_1$-wave function associated with 
$(\hat{p}_{z_1},\hat{q}_{z_1})$. Moreover,
\begin{align}
\hat{\Psi}_{z_1}(\ti z_1,x)&=i(\ti z_1-z_1)\psi_1(\ti z_1,x)
\begin{pmatrix}1\\0\end{pmatrix} \no \\ & \quad
-\frac{1}{2}\begin{pmatrix}  q(x) & 0 \\ 0 & 1\end{pmatrix} \big(\hat
p_{z_1}(x)\psi_1(\ti z_1,x)  +\psi_2(\ti z_1,x)\big)\begin{pmatrix} 1 \\
1 \end{pmatrix},
\end{align}
where $\hat{p}_{z_1}$ is defined in \eqref{Y3.2}. 
We now apply Lemma~\ref{YL3} replacing $\ti{\Psi}(\ti z_1)$ by
$\hat\Psi_{z_1}(\ti z_1)$ and $(p,q)$ by $(\hat p_{z_1},\hat
q_{z_1})$. Then $\check{q}_{z_1,\ti z_1}=\hat{\psi}_{1,z_1}(\ti
z_1)/\hat{\psi}_{2,z_1}(\ti z_1)$, as required by \eqref{Y4.1},
coincides with $q^{(1)}_{z_1,\ti z_1}$,  as defined in \eqref{Y4.3},
\begin{equation}
\check{q}_{z_1,\ti z_1}(x)=q^{(1)}_{z_1,\ti z_1}(x). 
\end{equation}
By formula \eqref{Y4.1} for $\check{p}_{\ti z_1,z_1}$ and 
Lemma~\ref{YL1}\,(i) for $\hat{p}_{z_1}=-\psi_2(z_1)/\psi_1(z_1)$ one
infers 
\begin{align}
\check{p}_{\ti z_1,z_1}(x)&=-\hat{p}'_{z_1}(x)
+q^{(1)}_{z_1,\ti z_1}(x)\hat p_{z_1}(x)^2
+2i\ti z_1\hat p_{z_1}(x) \no \\
&=-\big(q(x)\hat p_{z_1}(x)^2
+2iz_1\hat p_{z_1}(x)-p(x)\big)
+q^{(1)}_{z_1,\ti z_1}(x)\hat p_{z_1}(x)^2  
 +2i\ti z_1\hat p_{z_1}(x) \no \\
&=p^{(1)}_{z_1,\ti z_1}(x).
\end{align}
Using \eqref{Y4.2} and \eqref{Y3.2}--\eqref{Y3.3} one computes
\begin{align}
& \check{\Gamma} (z,\hat p_{z_1},\check
q_{z_1,\ti z_1})\hat{\Gamma} (z,q,\hat p_{z_1}) \no \\ 
& =-\bigg( (z-\ti z_1)\sigma_6+
\frac{i}{2}\begin{pmatrix} 1 & 0\\0& -\hat{p}_{z_1}\end{pmatrix} 
\begin{pmatrix} 1 & -\check q_{z_1,\ti z_1} \\ 
1 & -\check q_{z_1,\ti z_1}
\end{pmatrix}\bigg) \no \\
& \quad \times \bigg((z-z_1)\sigma_5 +
\frac{i}{2}\begin{pmatrix} q & 0\\0 & 1\end{pmatrix}
\begin{pmatrix} \hat{p}_{z_1} &
1\\\hat{p}_{z_1} & 1\end{pmatrix}\bigg) \no \\ 
& =-(i/2)zI_2-(i/2){W(\Psi (z_1),\ti \Psi (\ti z_1))}^{-1} \no \\
&\quad\times\begin{pmatrix}
\ti z_1\psi_2(z_1)\ti\psi_1(\ti z_1)-z_1\psi_1(z_1)\ti\psi_2(\ti z_1) &
-(\ti z_1-z_1)\psi_1(z_1)\ti\psi_1(\ti z_1)\\ 
(\ti z_1-z_1)\psi_2(z_1)\ti\psi_2(\ti z_1)  &
-\ti z_1\psi_1(z_1)\ti\psi_2(\ti z_1)+z_1\psi_2(z_1)\ti\psi_1(\ti z_1)
\end{pmatrix} \no \\ 
& =\Gamma(z,\Psi(z_1),\ti\Psi(\ti z_1)). 
\end{align}
To check \eqref{Y5.1} one uses Lemmas~\ref{YL2} and \ref{YL3},
\begin{align}
& \Gamma' (z,\Psi(z_1),\ti\Psi(\ti z_1)) \no \\
& =\check{\Gamma}' (z,\hat p_{z_1},\check q_{z_1,\ti z_1})
\hat{\Gamma} (z,q, \hat p_{z_1})+
\check{\Gamma}(z,\hat p_{z_1},\check q_{z_1,\ti z_1})
\hat{\Gamma}'(z,q,\hat p_{z_1}) \no \\ 
&=[-\check{\Gamma}(z,\hat p_{z_1},\check q_{z_1,\ti z_1}) 
U(z,\hat{p}_{z_1},\hat{q}_{z_1}) \no \\
& \quad \;\;\; +U(z,\check p_{z_1,\ti z_1},\check q_{z_1,\ti z_1}) 
\check{\Gamma}(z,\hat p_{z_1},\check q_{z_1,\ti z_1})]
\hat{\Gamma}(z,q, \hat p_{z_1}) \no \\
&\quad +\check{\Gamma}(z,\hat p_{z_1},\check q_{z_1,\ti z_1})
[-\hat{\Gamma}(z,q,\hat p_{z_1}) U(z,p,q) 
+U (z,\hat{p}_{z_1},\hat{q}_{z_1})\hat{\Gamma}(z,q, \hat p_{z_1})]
\no \\  
&= U(z,\check p_{z_1,\ti z_1},\check q_{z_1,\ti z_1})
\Gamma(z,\Psi(z_1),\ti\Psi(\ti z_1)) 
-\Gamma(z,\Psi(z_1),\ti\Psi(\ti z_1)) U(z,p,q).
\end{align}
Formulas \eqref{Y5.3}--\eqref{Y5.4} follow from
\eqref{Y4.4}--\eqref{Y4.5} since by Lemma~\ref{YL1}\,(iii),
\begin{equation} 
W(\Psi (z_1),\ti\Psi (\bar{z_1}))=W(\Psi (z_1),\cK \Psi (z_1))=-\|
\Psi (z_1)\|_{\bbC^2}^2.
\end{equation}
\end{proof}

For a general treatment of B\"acklund (Darboux) and gauge transformations
and their interrelations we refer, for instance, to \cite[Sect.\
4.1]{MS91}  and \cite{SZ87}.

Finally we add a few more facts valid in the $\NS_-$ case.

\begin{remark}\label{YR2a} 
Assume the $\NS_-$ case $p=-\bar q$. If $\Upsilon(z)
=(\upsilon_1(z),\upsilon_2(z))^\top$  is a $z$-wave function 
associated with $q$, then $\cK \Upsilon(z,x)$ is a $\bar z$-wave
function associated with $q$ and 
\begin{align}
\Upsilon^{(1)}_{z_1}(\bar{z},x)
&=\Gamma(\bar{z},x,\Psi(z_1),\cK \Psi(z_1))\Upsilon(\bar{z},x) \no \\
&=\Gamma(\bar{z},x,\Psi(z_1),\cK \Psi(z_1))\cK \Upsilon(z,x) \no \\
&=\Big(\overline{\upsilon^{(1)}_{2,z_1}(z,x)},
-\overline{\upsilon^{(1)}_{1,z_1}(z,x)}\Big)^\top \no \\
&=\cK \Upsilon^{(1)}_{z_1}(z,x)
\end{align}
is a $\bar{z}$-wave function associated with $q_{z_1}^{(1)}$ $($cf.\
Lemma~\ref{YL1}\,$(iii))$.
\end{remark}
\begin{remark}\label{YR2b}
Assume the $\NS_-$ case $p=-\bar q$. \\
$(i)$ Take $z=z_1$ and $\Upsilon (z_1)=\Psi(z_1)$ in Theorem~\ref{YT1}. 
Then 
\begin{equation}
\Psi^{(1)}_{z_1}(z_1,x)=\Gamma (z_1,x,\Psi(z_1),\cK \Psi(z_1))\Psi
(z_1,x)=0
\end{equation}
since $\Psi (z_1)^\bot \Psi ({z}_1)=0$. \\
$(ii)$ Take $z=\bar{z_1}$ and
$\cK \Upsilon(z_1)=\cK \Psi(z_1)$ in Theorem~\ref{YT1}. Then 
\begin{equation}
\cK \Psi^{(1)}_{z_1}(z_1,x)
=\Gamma(\bar{z_1},x,\Psi(z_1),\cK \Psi(z_1))\cK \Psi(z_1,x)=0 
\end{equation}
since $\Psi (z_1)^\bot\cK \Psi(z_1)=\| \Psi (z_1)\|_{\bbC^2}^2$ by
Lemma~\ref{YL1}\,$(iii)$.
\end{remark}

In the $\NS_-$ case, Theorem \ref{YT1} also shows that $q^{(1)}_{z_1}$ 
is locally nonsingular whenever $q$ is. More precisely, one has the
following result.

\begin{corollary}\label{YCor1} 
Assume the $\NS_-$ case $p=-\bar q$ and suppose
$z_1\in\bbC$. Then, if $q\in L^p_\loc(\bbR)$ for some
$p\in [1,\infty)\cup\{\infty\}$ $($respectively, if $q\in C^k(\bbR)$ for
some $k\in \bbN_0$$)$, the $\NS_-$ potential $q^{(1)}_{z_1}$ given by
\eqref{Y5.2} also satisfies $q^{(1)}_{z_1}\in L^p_\loc(\bbR)$
$($respectively, $q^{(1)}_{z_1}\in C^k(\bbR)$$)$.
\end{corollary}
\begin{proof}
Since $\big|\psi_1(z_1,x)\bar{\psi_2(z_1,x)}\big|
\|\Psi (z_1,x)\|_{\bbC^2}^{-2}\leq 1/2$, one concludes from
\eqref{Y5.2} that $(q_{z_1}^{(1)}-q)\in L^\infty(\bbR)$. Again by
\eqref{Y5.2}, $q$ and $q^{(1)}_{z_1}$ share the same
$L^p_{\loc}$ and $C^k$ properties since by \eqref{Y1.2}
(with $p=-\bar q$) one has $\partial^{m}_x \psi_j(z_1,\cdot)\in
\AC_{\loc}(\bbR)$, $j=1,2$, whenever $\partial^{m}_x q\in
L^1_\loc(\bbR)$. 
\end{proof}

\section{$\cJ$-Self-Adjointness of $\NS_-$ Dirac-Type Operators}
\lb{s3}

It is a known fact that the Dirac-type Lax differential expression in
the defocusing $\NS_+$ case is always in the limit point case at
$\pm\infty$. Put differently, the maximally defined Dirac-type operator
corresponding to the defocusing $\NS_+$ case (cf.\ \eqref{1.1a}) is
always self-adjoint. Classical references in this context are
\cite[Sect.\ 8.6]{LS91},
\cite{We71}, which use some additional conditions (such as
real-valuedness and/or regularity) of the coefficient $q$. A simple
proof of this fact under most general conditions on $q$ was recently
communicated to us by  Hinton \cite{Hi99} (cf.\ also \cite{Cl94},
\cite{CG02} and \cite{LM02} for matrix-valued extensions of this
result).  In this section we show that the analogous result holds for
Dirac-type differential expressions
$M(q)$ in \eqref{2.5} in the focusing $\NS_-$ case, when
self-adjointness is replaced by $\cJ$-self-adjointness. 

First, we recall some basic facts about $\cJ $-symmetric 
and $\cJ $-self-adjoint operators in a complex Hilbert space $\cH$ 
(see, e.g., \cite[Sect.\ III.5]{EE89} and \cite[p.\ 76]{Gl65}) with
scalar product denoted by $(\cdot ,\cdot)_\cH$ (linear in the first 
and antilinear in the second place) and corresponding norm denoted by
$\|\cdot\|_\cH$. Let $\cJ$ be a conjugation operator in $\cH$, that is,
$\cJ $ is an antilinear involution satisfying
\begin{equation}
	(\cJ u,v)_\cH=(\cJ v,u)_\cH \text{ for all }u,v\in \cH,\quad \cJ^2=I.
\lb{35.1}
\end{equation}
In particular,
\begin{equation}
	(\cJ u,\cJ v)_\cH=(v,u)_\cH, \quad u,v\in \cH. \lb{35.1a}
\end{equation}
A densely defined linear operator $S$ in $\cH$ is called 
$\cJ      $-symmetric if 
\begin{equation}
S\subseteq \cJ  S^*\cJ \, \text{ (equivalently, if  
$\cJ S\cJ\subseteq S^*$).}  \lb{35.1b}  
\end{equation}
Clearly, \eqref{35.1b} is equivalent to
\begin{equation}
	(\cJ u,Sv)_\cH=(\cJ Su,v)_\cH, \quad u,v\in \dom(S). \lb{35.2}
\end{equation}
Here $S^*$ denotes the adjoint operator of $S$ in $\cH$. If $S$ is
$\cJ$-symmetric, so is its closure $\ol{S}$. The operator
$S$ is called $\cJ $-self-adjoint if  
\begin{equation}
	S=\cJ S^*\cJ \, \text{ (equivalently, if $\cJ S\cJ=S^*$).}
\end{equation}
Finally, a densely defined, closable operator $T$ is called essentially 
$\cJ$-self-adjoint if its closure $\ol{T}$ is $\cJ $-self adjoint,
that is, if 
\begin{equation}
\ol{T}=\cJ T^* \cJ. 
\end{equation}
Next, assuming $S$ to be $\cJ$-symmetric, one introduces the following
inner product $(\cdot,\cdot)_*$ on $\dom(\cJ S^*\cJ )=\dom(S^*\cJ)$
according to
\cite{Kn81} (see also \cite{Ra85}),
\begin{equation}
	(u,v)_*=(\cJ u,\cJ v)_\cH+(S^*\cJ u,S^*\cJ v)_\cH, 
\quad u,v\in \dom(\cJ S^*\cJ),
\end{equation}
which renders $\dom(\cJ S^*\cJ )$ a Hilbert space. Then the
following theorem holds ($I$ denotes the identity operator in $\cH$).
\begin{theorem}[Race  \cite{Ra85}] 
Let $S$ be a densely defined closed $\cJ $-symmetric operator. Then
\begin{equation}
	\dom (\cJ S^*\cJ )=\dom (S) \oplus_* \ker((S^*\cJ )^2+I), \lb{35.3}
\end{equation}
where $\oplus_*$ means the orthogonal direct sum with respect to the inner
product $(\cdot,\cdot)_*$. In particular, a densely defined closed $\cJ
$-symmetric operator $S$ is $\cJ $-self-adjoint if and only if 
\begin{equation}
	\ker((S^*\cJ )^2+I)=\{ 0\} .\lb{35.4}
\end{equation}
\end{theorem}

We will apply \eqref{35.4} to (maximally defined) Dirac-type operators
associated with the differential expression $M(q)$ in \eqref{2.5}
relevant to the focusing $\NS_-$ hierarchy and prove the fundamental 
fact that such Dirac operators are always
$\cJ$-self-adjoint under most general conditions on the coefficient
$q$ (cf.\ Theorem \ref{RT5.1}). 

It will be convenient to make the following $\NS_-$ assumption throughout
the remainder of this section.

\begin{hypothesis}\label{h3.0} 
Suppose $q\in L^1_\loc(\bbR)$ and assume the $\NS_-$ case $p=-\bar q$. 
\end{hypothesis}

Given Hypothesis~\ref{h3.0}, we now introduce the following maximal and
minimal Dirac-type operators in $L^2(\bbR)^2$ associated with the
differential expression $M(q)$,
\begin{align}
&D_{\max}(q)F=M(q)F, \lb{3.1} \\
&F\in\dom(D_{\max}(q))=\big\{G\in L^2(\bbR)^2\,\big|\, G\in
AC_\loc(\bbR)^2; \; M(q)G\in L^2(\bbR)^2\big\}, \no \\
&D_{\min}(q)F=M(q)F,  \lb{3.2} \\
&F\in\dom(D_{\min}(q))=\{G\in\dom(D_{\max}(q)) \,|\,
\supp(G) \text{ is compact}\}. \no
\end{align}
It follows by standard techniques (see, e.g., \cite[Ch.\ 8]{LS91} and
\cite{We71}) that under Hypothesis~\ref{h3.0},
$D_{\min}(q)$ is densely defined and closable in $L^2(\bbR)^2$ and
$D_{\max}(q)$ is a densely defined closed operator in $L^2(\bbR)^2$. 
Moreover (cf.\ \eqref{2.14}), one infers (see, e.g., \cite[Lemma
8.6.2]{LS91} and \cite{We71} in the analogous case of symmetric Dirac
operators)
\begin{equation}
\ol{D_{\min}(q)}=D_{\max}(-q)^*,  \, \text{ or equivalently, }\,
D_{\min}(q)^*=D_{\max}(-q). \lb{3.3}
\end{equation}

The following result will be the crucial ingredient in the proof of
Theorem \ref{RT5.1}, the principal result of this section.
\begin{theorem}
Assume Hypothesis \ref{h3.0}. 
Let $N(q)$ be the following $($formally self-adjoint\,$)$ 
differential expression
\begin{equation}
	N(q)=i\begin{pmatrix} \frac{d}{dx} &
	-q\\\bar{q}&\frac{d}{dx}\end{pmatrix} \lb{35.6}
\end{equation} 
and denote by $\wti D_{\max}(q)$ the maximally defined Dirac-type operator
in $L^2(\bbR)^2$ associated with $N(q)$,  
\begin{align}
&\wti D_{\max}(q)F=N(q)F, \lb{35.6a} \\
&F\in\dom(\wti D_{\max}(q))=\big\{G\in L^2(\bbR)^2\,\big|\, G\in
AC_\loc(\bbR)^2; \; N(q)G\in L^2(\bbR)^2\big\}. \no 
\end{align}
Then one infers: \\
(i) The following identity holds
\begin{equation}
	M(-q)M(q)=N(q)^2.\lb{35.7}
\end{equation}
(ii)
Let $U_q=U_q(x)$ satisfy the initial value problem
\begin{equation}
	U_q'= \begin{pmatrix} 0 & q\\ -\bar{q}& 0 \end{pmatrix} U_q, 
\quad U_q(0)=I_2. \lb{35.8}
\end{equation}
Then $\{U_q(x)\}_{x\in \bbR}$ is a family of unitary matrices in $\bbC^2$
with entries in $AC_{\loc}(\bbR)\cap L^\infty(\bbR)$ satisfying
\begin{equation}
	U_q^{-1}N(q)U_q=i\frac{d}{dx}I_2.\lb{35.9}
\end{equation}
(iii) Let $\mathcal{U}_q$ denote the multiplication operator with $
U_q(\cdot )$ on $L^2(\bbR)^2$. Then $\wti D_{\max}(q)$ is unitarily
equivalent to the maximally defined operator in $L^2(\bbR)^2$ 
associated with the differential expression $i\frac{d}{dx}I_2$,
\begin{align}
&	\mathcal{U}_q^{-1}\wti D_{\max}(q)\mathcal{U}_q=\bigg(
	i\frac{d}{dx}I_2\bigg)_{\max}, \lb{35.9a} \\
&\dom \bigg(\bigg( i\frac{d}{dx}I_2\bigg)_{\max}\bigg)= H^{1,2}(\bbR)^2
=\big\{ F\in L^2(\bbR)^2\, \big| \, F\in AC_{\loc}(\bbR)^2; \; F'\in
L^2(\bbR)^2\big\}. \no
\end{align}
Moreover,
\begin{align}
&\mathcal{U}_q^{-1}D_{\max}(-q)D_{\max}(q)\mathcal{U}_q =
	\bigg( -\frac{d^2}{dx^2}I_2\bigg)_{\max}, \lb{35.10} \\
&\dom \bigg(\bigg( -\frac{d^2}{dx^2}I_2\bigg)_{\max}\bigg)
= H^{2,2}(\bbR)^2 \no \\ 
& \hspace*{3.63cm}
=\big\{F\in L^2(\bbR)^2\, \big|\, F,F'\in AC_{\loc}(\bbR)^2; \; 
F', F''\in L^2(\bbR)^2\big\}. \no
\end{align}
\end{theorem}
\begin{proof}
That $N(q)$ is formally self-adjoint and $M(-q)M(q)=N(q)^2$ as stated in
$(i)$ is an elementary matrix calculation. \\
To prove $(ii)$, we note that the initial value problem
\eqref{35.8} is well-posed in the sense of Carath\'eodory since $q\in
L^1_{\loc}(\bbR)$ (cf., e.g., \cite[p.\ 45--46]{Ha82})  with a solution
matrix $U_q$ with entries in $AC_\loc(\bbR)$. Moreover, for each
$x\in\bbR$, $U_q(x)$ is a unitary matrix in $\bbC^2$, since
$U_q'=-B(q)U_q$, with
$B(q)=\left(\begin{smallmatrix} 0 & -q \\ \ol q & 0
\end{smallmatrix}\right)$ skew-adjoint. Thus, the entries $U_{q,j,k}$,
$1\leq j,k \leq 2$ of $U_q$ (as well as those of $U_q^{-1}$) actually
satisfy 
\begin{equation}
U_{q,j,k}\in AC_{\loc}(\bbR)\cap L^\infty(\bbR), \quad 1 \leq j,k \leq 2.
\lb{35.10a}
\end{equation}
Next, fix
$F\in AC_{\loc}(\bbR)^2$, such that $\mathcal{U}_q^{-1}F\in
H^{1,2}(\bbR)^2$. Then 
\begin{align}
U_q\bigg( i\frac{d}{dx}I_2\bigg) U_q^{-1}F&=i\frac{d}{dx}F+ iU_q
\frac{d}{dx}(U_q^{-1})F \no\\
&=i\frac{d}{dx}F+iU_q(U_q^{-1}B(q)^*)F \no\\
&=N(q)F, \lb{35.11}
\end{align}
where we used the fact that  $(U_q^{-1})'=U_q^{-1}B(q)^*$. 
Thus, $(ii)$ follows. \\
Moreover, by \eqref{35.11} one concludes
$\dom(\wti D_{\max}(q))= \mathcal{U}_q H^{1,2}(\bbR)^2$ by \eqref{35.10a}
and the fact that $U_q$ is unitary in $\bbC^2$. This proves 
\eqref{35.9a}. \\ 
Clearly $(i)$ and $(ii)$ yield the relation 
\begin{equation}
U_q^{-1}M(-q)M(q)U_q = -\frac{d^2}{dx^2}I_2. 
\end{equation}
Thus, \eqref{35.10} will follow once we prove the following facts:
\begin{align}
& (i) \; U_q F\in L^2(\bbR)^2 \text{ if and only if } F\in L^2(\bbR)^2,
\lb{35.12} \\
& (ii) \; U_q F\in AC_\loc (\bbR)^2 \text{ if and only if } F\in
AC_\loc(\bbR)^2, \lb{35.13} \\
& (iii) \; M(q)U_q F\in L^2(\bbR)^2 \text{ if and only if } 
F'\in L^2(\bbR)^2, \lb{35.14} \\
& (iv) \; M(q)U_q F\in AC_\loc (\bbR)^2 \text{ if and only if } 
F'\in AC_\loc(\bbR)^2, \lb{35.15} \\
& (v) \; M(-q)M(q)U_q F\in L^2(\bbR)^2 \text{ if and only if } 
F''\in L^2(\bbR)^2. \lb{35.16}  
\end{align}
Clearly \eqref{35.12} and \eqref{35.16} hold since $U_q$ is unitary in
$\bbC^2$. \eqref{35.13} is valid since $U_{q,j,k}, U^{-1}_{q,j,k}\in
AC_{\loc}(\bbR)\cap L^\infty(\bbR)$, $j,k=1,2$. Next, an explicit
computation yields
\begin{equation}
M(q)U_qF=i\begin{pmatrix} U_{q,1,1}f_1'+U_{q,1,2}f_2' \\ 
-U_{q,2,1}f_1'-U_{q,2,2}f_2' \end{pmatrix}, \quad F=(f_1, f_2)^\top.
\lb{35.17}
\end{equation}
Introducing 
\begin{equation}
V_q=\sigma_3 U_q\sigma_3
=\begin{pmatrix} U_{q,1,1} & -U_{q,1,2} \\ -U_{q,2,1} & U_{q,2,2}
\end{pmatrix}, \lb{35.18}
\end{equation}
one infers $V_{q,j,k}, V^{-1}_{q,j,k}\in
AC_{\loc}(\bbR)\cap L^\infty(\bbR)$, $j,k=1,2$ and 
\begin{equation}
V_q^{-1}M(q)U_qF=i(f_1', -f_2')^\top, \lb{35.19}
\end{equation}
and hence \eqref{35.14} and \eqref{35.15} hold. This proves 
\eqref{35.10}.
\end{proof}

\begin{remark}
We note that by \eqref{35.9a}, $\wti D_{\max}(q)$, the maximally defined,
self-adjoint operator in $L^2(\bbR)^2$ associated with the
$2\times 2$ matrix-valued differential expression $N(q)$ has a purely
absolutely continuous spectrum that equals $\bbR$,
\begin{equation}
\sigma(\wti D_{\max}(q))=\sigma_{\rm{ac}}(\wti D_{\max}(q))=\bbR.
\end{equation}
$($We refer to Section \ref{s5} for a discussion of various spectral
decompositions. In the present context we just note that $\sigma (T)$ and
$\sigma_{\rm{ac}}(T)$ denote the spectrum and absolutely continuous
spectrum of a self-adjoint operator $T$ in a separable complex Hilbert
space $\cH$.$)$ 
\end{remark}

The principal result of this section then reads as follows.

\begin{theorem} \lb{RT5.1}
Assume Hypothesis \ref{h3.0}. Then the  minimally defined Dirac-type
operator $D_{\min}(q)$ associated with the Lax differential expression
\begin{equation}
	M(q)=i\begin{pmatrix} \frac{d}{dx} &
	-q\\-\bar{q}&-\frac{d}{dx}\end{pmatrix}
\end{equation} 
introduced in \eqref{3.2} is essentially $\cJ$-self-adjoint in
$L^2(\bbR)^2$, that is, 
\begin{equation}
\ol{D_{\min}(q)}=\cJ D_{\min}(q)^* \cJ, \lb{35.24}
\end{equation}
where $\cJ$ is the conjugation defined in \eqref{2.16}. Moreover,
\begin{equation}
\ol{D_{\min}(q)}=D_{\max}(q) \lb{35.25}
\end{equation}
and hence $D_{\max}(q)$ is $\cJ$-self-adjoint.
\end{theorem}
\begin{proof}
We first recall (cf.\ \eqref{3.3})
\begin{equation}
D_{\min}(q)^*=D_{\max}(-q) \lb{35.26}
\end{equation}
and also note 
\begin{equation}
\cJ D_{\max}(-q)\cJ=D_{\max}(q). \lb{35.27}
\end{equation}
Since $\ol{D_{\min}(q)}$ is closed and $\cJ$-symmetric 
(cf.\ \eqref{2.15}), its $\cJ$-self-adjointness is equivalent
to showing that (cf. \eqref{35.4})
\begin{equation}
\ker \big(D_{\min} (q)^* \cJ D_{\min}(q)^* \cJ+I\big)
=	\ker (D_{\max}(-q)D_{\max}(q)+I) =\{0\}. \lb{35.28}
\end{equation} 
Since by \eqref{35.10} $D_{\max}(-q)D_{\max}(q)$ is unitarily equivalent
to $(-d^2/dx^2 I_2)_{\max}\geq 0$, one concludes 
$D_{\max}(-q)D_{\max}(q)\geq 0$ and hence \eqref{35.28} obviously
holds.  The fact \eqref{35.25} then follows from \eqref{35.24} and
\eqref{35.26} since
\begin{equation}
	\ol{D_{\min}(q)}=\cJ D_{\min}(q)^*\cJ =\cJ D_{\max}(-q)\cJ 
	=D_{\max}(q).
\end{equation}
\end{proof}

As mentioned in the introductory paragraph to this section, Theorem
\ref{RT5.1} in the $\cJ$-self-adjoint context can be viewed as an
analog of the result of the corresponding (self-adjoint) Dirac operator
relevant in the defocusing $\NS_+$ case of the nonlinear Schr\"odinger
equation.

\section{Constructing $L^2(\bbR)^2$-Wave Functions for 
$\cJ$-Self-Adjoint \\
Dirac-Type Operators} \lb{s4}

In this section we discuss how to construct $L^2(\bbR)^2$-wave functions
for non-self-adjoint (but $\cJ$-self-adjoint) Dirac-type operators
associated with the Lax differential expression for the $\NS_-$ system.

By Remark \ref{YR2b}, in order to obtain a nonzero $z_1$-wave function
associated with $q^{(1)}_{z_1}$, we have to apply the transformation
matrix $\Gamma (z_1,\Psi(z_1),\cK \Psi(z_1))$ to a $z_1$-wave
function $\Phi(z_1)$  associated with $q$ that is linearly independent
with the original $z_1$-wave function $\Psi (z_1)$ associated with $q$.
Similarly, in order to obtain a nonzero $\bar{z_1}$-wave function
associated with $q^{(1)}_{z_1}$, we have to apply the transformation
matrix $\Gamma (\bar{z_1},\Psi(z_1),\cK \Psi(z_1))$ to a
$\bar{z_1}$-wave function $\cK \Phi(z_1)$ associated with $q$ that
is linearly independent with the original $\bar{z_1}$-wave function
$\cK \Psi(z_1)$ associated with $q$. The function $\Phi(z_1)$ is
constructed as follows.

Let $\Psi (z)=(\psi_1(z), \psi_2(z))^\top$, $z\in \mathbb{C}$, be a
$z$-wave function associated with  $q$ on $\bbR$ and
introduce 
\begin{equation}
\Omega_z=\{x\in\bbR \,|\, \psi_1(z,x)\psi_2(z,x)\neq 0\}.
\end{equation}
Next, consider
\begin{equation}
\Psi^\#(z,x)=(1/2)(\psi_2(z,x)^{-1},-\psi_1(z,x)^{-1})^\top, \quad  
x\in \Omega_z, 
\end{equation}
such that 
\begin{equation}
W(\Psi^\#(z,x) ,\Psi(z,x))=1. 
\end{equation}
Let $x_0,x\in\Omega_z$ such that $[x_0,x]\subseteq\Omega_z$ and 
define
\begin{align}
R(z,x,x_0)&=-\frac{1}{2}\int^x_{x_0}
dx'\bigg(\frac{\bar{q(x')}}{\psi_2(z,x')^2}
+\frac{q(x')}{\psi_1(z,x')^2}\bigg), \label{Y7.1} \\
\Phi (z,x)&=\Psi^\# (z,x)+R(z,x,x_0)\Psi (z,x), \quad
[x_0,x]\subseteq\Omega_z. \lb{3.5}
\end{align}
Using \eqref{Y1.2}, we have $(\Psi^\#)'=U (p,q)\Psi^\#-R'\Psi$. Thus,
$W(\Phi,\Psi)=1$ and $\Phi (z)$ is a $z$-wave function associated with 
$q$, since 
\begin{equation}
\Phi'=(\Psi^\#)'+R'\Psi+R\Psi'=U \Psi^\#-R'\Psi+R'\Psi+RU\Psi=U \Phi.
\end{equation}
We note that $\Psi^\bot \Psi^\#=1$ implies 
$\cK \Psi (z)\Psi (z)^\bot \Phi (z)=\cK \Psi (z)$. 

Thus, if $\Gamma (z,\Psi(z_1),\cK \Psi(z_1))$ is defined using
$\Psi(z_1)$ as in \eqref{Y5.4}, then the $z_1$-wave function 
$\Phi^{(1)}_{z_1}(z_1)$ associated with $q^{(1)}_{z_1}$, as prescribed in
Theorem~\ref{YT1}, is computed as follows  
\begin{align}
\Phi^{(1)}_{z_1}(z_1,x)
&=\Gamma (z_1,x,\Psi(z_1),\cK \Psi(z_1))\Phi(z_1,x) \no \\
&=\Im(z_1)\|\Psi(z_1,x)\|_{\bbC^2}^{-2}\cK \Psi (z_1,x).
\label{Y8.1}
\end{align} 
Moreover, by Remark~\ref{YR2a}, $\cK \Phi^{(1)}_{z_1}(z_1)$ is
computed as 
\begin{align}  
\cK \Phi^{(1)}_{z_1}(z_1,x)
&=\Gamma(\bar{z_1},x,\Psi(z_1),\cK \Psi (z_1))\cK \Phi (z_1,x) \no
\\ 
&=-\Im(z_1)\| \Psi (z_1,x)\|_{\bbC^2}^{-2}\Psi (z_1,x). \label{Y8.2}
\end{align}
By \eqref{Y8.1} and \eqref{Y5.4}, for each $z\in \mathbb{C}$, the
$z$-wave function $\Phi^{(1)}_{z_1}(z)$ associated with $q^{(1)}_{z_1}$
(constructed using the $z$-wave function $\Phi(z)$ associated with
$q$) is computed by  
\begin{align}
\Phi^{(1)}_{z_1}(z,x)
&=\Gamma (z,x,\Psi(z_1),\cK \Psi(z_1))\Phi (z,x) \no \\
&=-(i/2) (z-z_1)\Phi
(z,x)+\Phi^{(1)}_{z_1}(z_1,x)\Psi (z_1,x)^\bot \Phi (z,x) \no \\
&=-(i/2) (z-z_1)\Phi (z,x)-
W(\Psi (z_1,x),\Phi (z,x))\Phi^{(1)}_{z_1}(z_1,x). \label{Y8.3}
\end{align}
Formulas \eqref{Y5.2} and \eqref{Y8.1} now imply
\begin{equation} \label{Yq}
q^{(1)}_{z_1}(x)=q(x)+4\phi^{(1)}_{1,z_1}(z_1,x)\psi_1(z_1,x), 
\end{equation}
where $\Phi^{(1)}_{z_1}(z,x)=(\phi^{(1)}_{1,z_1}(z,x),
\phi^{(1)}_{2,z_1}(z,x))^\top$, $\Psi (z,x)=(\psi_1(z,x),
\psi_2(z,x))^\top$.

\begin{remark} \lb{r3.1}
We emphasize that while $R(z,x,x_0)$ in \eqref{Y7.1}, and hence
$\Phi(z,x)$ in \eqref{3.5}, in general, will have singularities on
$\bbR$, the formulas \eqref{Y8.1}--\eqref{Yq} are well-defined for all
$x\in\bbR$. 
\end{remark}

The next hypothesis will be crucial in our attempt to construct
$z_1$- and $\bar z_1$-wave functions in $L^2(\bbR)^2$ associated with the
Dirac-type differential expression $M\big(q^{(1)}_{z_1}\big)$.

\begin{hypothesis}\label{h3.2} 
Suppose $q\in L^1_{\loc}(\bbR)$, assume the $\NS_-$ case $p=-\bar q$,
and let $z_0\in\bbC$. Suppose $\Psi(z_0)$ to be a $z_0$-wave function
associated with $q$ that satisfies the condition 
$\|\Psi(z_0,\cdot)\|_{\bbC^2}^{-1}\in L^2(\bbR)$, that is,
\begin{equation}
\int^\infty_{-\infty} dx\,\|\Psi (z_0,x)\|_{\bbC^2}^{-2}<\infty. 
\lb{3.10}
\end{equation}
\end{hypothesis}

If a $z_0$-wave function $\Psi(z_0)$ associated with $q$ satisfies
condition \eqref{3.10}, we will henceforth say that $\Psi(z_0)$ satisfies 
Hypothesis~\ref{h3.2} at $z_0$.

\begin{remark} \lb{r3.3} Assume Hypothesis~\ref{h3.0} and suppose that 
$\Psi(z)$ satisfies Hypothesis~\ref{h3.2} at $z$. Then, \\
$(i)$ by H\"older's inequality, $\Psi (z)\not\in
L^2((-\infty,R])^2\cup L^2([R,\infty))^2$ for all $R\in\bbR$. \\ 
$(ii)$ $\cK \Psi (z)$ satisfies Hypothesis~\ref{h3.2} at
$\bar{z}$ by Lemma~\ref{YL1}\,$(iii)$.
\end{remark}

\begin{remark}\lb{RL4.0a}	Assume Hypothesis \ref{h3.0} and let 
$\lambda\in \bbR$. Then  all $\lambda$-wave functions $\Psi(\lambda)$
associated with $q$ satisfy $\|\Psi(\lambda,x)\|_{\bbC^2}=c(\lambda)$
$($independently of $x\in\bbR$$)$. In particular,
$\|\Psi(\lambda,\cdot)\|_{\bbC^2}, 
\|\Psi(\lambda,\cdot)\|_{\bbC^2}^{-1} \notin L^2([R,\pm\infty))$,
$R\in\bbR$, and hence there exists no $\lambda$-wave function associated
with $q$ that satisfies Hypothesis \ref{h3.2} at $\lambda\in\bbR$.
\end{remark}

The principal result of this section then reads as follows.

\begin{theorem} \label{YT2} Assume Hypothesis~\ref{h3.0}. Let
$z_1\in\bbC\backslash\bbR$ and suppose that the $z_1$-wave function
$\Psi (z_1)$ associated with $q$ satisfies Hypothesis~\ref{h3.2} at
$z_1$. Let $q^{(1)}_{z_1}$ be given by \eqref{Y5.2}. Then $z_1$ and
$\bar{z_1}$ are eigenvalues of the maximal Dirac-type operator
$D_{\max}\big(q^{(1)}_{z_1}\big)$ associated with 
$M\big(q^{(1)}_{z_1}\big)$ of geometric
multiplicity equal to one. The corresponding eigenfunctions 
$\Phi^{(1)}_{z_1}(z_1)$ and $\cK \Phi^{(1)}_{z_1}(z_1)$ are given by
\eqref{Y8.1} and \eqref{Y8.2}, respectively, that is, one has
\begin{align}
&\Phi^{(1)}_{z_1}(z_1), \cK \Phi^{(1)}_{z_1}(z_1)\in
\dom\big(D_{\max}\big(q^{(1)}_{z_1}\big)\big), \lb{3.12} \\
&D_{\max}\big(q^{(1)}_{z_1}\big)\Phi^{(1)}_{z_1}(z_1)
=z_1\Phi^{(1)}_{z_1}(z_1), 
\lb{3.13} \\
&D_{\max}\big(q^{(1)}_{z_1}\big)\cK \Phi^{(1)}_{z_1}(z_1)=\bar{z_1}
\cK \Phi^{(1)}_{z_1}(z_1). \lb{3.14} 
\end{align}
\end{theorem}
\begin{proof} Indeed, using \eqref{3.10} at $z_1$ one obtains
\begin{equation}
0<\big\|\Phi^{(1)}_{z_1}(z_1)\big\|^2_{L^2}= 
|\Im(z_1)|^2 \int^\infty_{-\infty} dx\,
\|\Psi (z_1,x)\|_{\bbC^2}^{-4}\|\Psi (z_1,x)\|_{\bbC^2}^2 <\infty. 
\end{equation} 
In order to show that $z_1$ has geometric multiplicity equal to one as an  
eigenvalue of $D_{\max}\big(q^{(1)}_{z_1}\big)$, we next assume that
$\Phi^{(1)}_{z_1}(z_1,\cdot)\in L^2(\bbR)^2$ and
$\ti\Phi^{(1)}_{z_1}(z_1,\cdot)\in L^2(\bbR)^2$ are linearly independent
$z_1$-wave functions associated with $q^{(1)}_{z_1}$. Then clearly 
$\Phi^{(1)}_{z_1}(z_1,\cdot), 
\ti\Phi^{(1)}_{z_1}(z_1,\cdot)\in 
\dom\big(D_{\max}\big(q^{(1)}_{z_1}\big)\big)$ 
and 
\begin{equation}
W\big(\Phi^{(1)}_{z_1}(z_1,\cdot), \ti\Phi^{(1)}_{z_1}(z_1,\cdot)\big)
\in L^1(\bbR). \lb{3.17}
\end{equation}  
However, since by \eqref{2.4}, $W(\Phi^{(1)}_{z_1}(z_1,x),
\ti\Phi^{(1)}_{z_1}(z_1,x))$ is constant with respect to $x\in\bbR$,
\eqref{3.17} represents a contradiction and hence $z_1$ has geometric
multiplicity equal to one. The analogous arguments apply to $\bar{z_1}$. 
\end{proof}

The argument that $D_{\max}\big(q^{(1)}_{z_1}\big)$ has only eigenvalues
of geometric multiplicity equal to one applies of course in complete
generality to any $D(q)$ with $q\in L^1_\loc(\bbR)$.   

Next we show that condition \eqref{3.10} is preserved under iterations, a
fact of great relevance in connection with the multi-soliton solutions
relative to arbitrary backgrounds discussed at the end of Section
\ref{s6}.

\begin{lemma}\label{YP1} Assume Hypothesis~\ref{h3.0} and let $z_1,z_2\in
\mathbb{C}$. Fix a $z_1$-wave function $\Psi(z_1)$ and a $z_2$-wave
function $\Phi (z_2)$ associated with $q$. Using $\Psi (z_1)$, construct
the $\NS_-$ potential $q^{(1)}_{z_1}$ by formula \eqref{Y5.2} and
consider the transformation matrix $\Gamma (z_2,\Psi(z_1),\cK \Psi(z_1))$ 
given by formula \eqref{Y5.4} for $z=z_2$. Let $\Phi^{(1)}_{z_1}(z_2)$ be
the $z_2$-wave function associated with $q^{(1)}_{z_1}$ as in
\eqref{Y8.3}. If $\Phi(z_2)$ satisfies Hypothesis~\ref{h3.2} at $z_2$,
then $\Phi^{(1)}_{z_1}(z_2)$ satisfies Hypothesis~\ref{h3.2} at $z_2$.
\end{lemma}
\begin{proof} Without loss of generality we may assume
$\Im(z_1)\Im(z_2)\geq 0$. Formulas \eqref{2.10}, \eqref{Y8.1}, and
\eqref{Y8.3} imply 
\begin{align} 
&\big\| \Phi^{(1)}_{z_1}(z_2)\big\|_{\bbC^2}^2 \no \\ 
&=\frac{1}{4}|z_2-z_1|^2\|\Phi(z_2)\|_{\bbC^2}^2 +
\frac{1}{4} |\bar{z_1}-z_1|^2\|\Psi (z_1)\|_{\bbC^2}^{-4}
\|\cK \Psi(z_1))\Psi (z_1)^\bot \Phi (z_2)\|_{\bbC^2}^2 \no \\
&\quad -\frac{1}{2}\Re \big(((z_2-z_1)\Phi (z_2),(\bar{z_1}-z_1)
\| \Psi (z_1)\|_{\bbC^2}^{-2}\cK \Psi(z_1)\Psi (z_1)^\bot 
\Phi (z_2))_{\bbC^2}\big) \no \\ 
&=\frac{1}{4}|z_2-z_1|^2\| \Phi (z_2)\|_{\bbC^2}^2
+\frac{1}{4}|\bar{z_1}-z_1|^2
\| \Psi (z_1)\|_{\bbC^2}^{-2}|\Psi (z_1)^\bot \Phi (z_2)|^2 \no \\
&\quad + \frac{1}{2}\| \Psi (z_1)\|_{\bbC^2}^{-2}|\Psi (z_1)^\bot 
\Phi (z_2)|^2\Re ((z_2-z_1)(\bar{z_1}-z_1)) \no \\
&=\frac{1}{4} |z_2-z_1|^2\| \Phi (z_2)\|_{\bbC^2}^2+
\|\Psi (z_1)\|_{\bbC^2}^{-2}|\Psi (z_1)^\bot \Phi (z_2)|^2
\Im (z_1)\Im(z_2) \no \\ 
&\geq \frac{1}{4} |z_2-z_1|^2\| \Phi (z_2)\|_{\bbC^2}^2.
\end{align}
Thus, if $\int^\infty_{-\infty} dx\,\| \Phi(z_2,x)\|_{\bbC^2}^{-2}
<\infty$, then 
$\int^\infty_{-\infty} dx\, \big\|
\Phi^{(1)}_{z_1}(z_2,x)\big\|_{\bbC^2}^{-2}<\infty$.  
\end{proof}

It will be shown in Remark \ref{r6.1} that for any
$z\in\rho\big(D_{\max}\big(q_{z_1}^{(1)}\big)\big)$ all but two
$z$-wave functions of $D_{\max}\big(q_{z_1}^{(1)}\big)$ satisfy
Hypothesis \ref{h3.2} at $z$.

\section{Some Spectral Properties and the Existence of
Weyl--Titchmarsh-type Solutions for $\cJ$-Self-Adjoint \\ Dirac-Type 
Operators} \lb{s5}

The principal purpose of this section is to establish the existence of
Weyl--Titchmarsh-type solutions for formally $\cJ$-self-adjoint Dirac
differetial expressions $M(q)$ associated with the focusing $\NS_-$ case.
The latter are well-known to exist in the case of self-adjoint Dirac
operators (in particular, they are well-known to exist in the context of
the defocusing $\NS_+$ equation) and are known to be a fundamental
ingredient in the spectral analysis in the self-adjoint context (cf., 
for instance, \cite[Chs.\ 3, 4,]{LS75}). As far as we know, no such
result appears to be known in the general $\cJ$-self-adjoint case studied
in this section. Along the way we also collect some results concerning
spectral properties of $D_{\max}(q)$.

Thus, assuming the $\NS_-$ case and hence the basic Hypothesis
\ref{h3.0} throughout this section, we adopt the simplified notation
(cf.\ Section
\ref{s3})
\begin{equation}
D(q)=\ol{D_{\min}(q)}=D_{\max}(q), \quad q\in L^1_{\loc}(\bbR). \lb{5.1}
\end{equation}
We also denote by $I_2$ the identity operator in $L^2(\bbR)^2$
(as well as in $\bbC^2$). Moreover, we find it convenient to introduce
the following notations,
\begin{align}
\begin{split}
   & L^2_{\loc}([-\infty,\infty))=\big\{f\colon \bbR\to\bbC 
	\text{ is measurable} \,\big|\, f\in L^2((-\infty,R]) \text{ for all
	$R\in\bbR$}\big\}, \\ 
   & L^2_{\loc}((-\infty,\infty])=\big\{f\colon \bbR\to\bbC 
	\text{ is measurable} \,\big|\, f\in L^2([R,\infty)) \text{ for all
	$R\in\bbR$}\big\}. \lb{5.2} 
\end{split}
\end{align}

We start with the following auxiliary result (the variation of parameters
formula).

\begin{lemma} \lb{l5.1}
Assume Hypothesis \ref{h3.0} and let $(z,x_0)\in\bbC\times\bbR$. 
Let $\Psi_1(z)$ and $\Psi_2(z)$ be linearly independent $z$-wave
functions for $M(q)$  defined on $[x_0,\infty) $ and denote by $\Xi
(z,x)=\big[\Psi_1(z,x),\Psi_2(z,x)]$ the $2\times 2$ fundamental
matrix  solution of $M(q)\Xi(z) =z\Xi(z)$. Assume the Wronskian
of $\Psi_1$ and $\Psi_2$ satisfies
$W(\Psi_1(z,x),\Psi_2(z,x))=\det(\Xi(z,x))=1$ for some
$($and hence for all\,$)$ $x\in [x_0,\infty)$. Moreover, suppose 
$B\in L^1_{\loc}([x_0,\infty))^{2\times 2}$. Then
$\Phi(z,\cdot)\in AC_{\loc}(\bbR)^2$ satisfies
$M(q)\Phi(z)= z\Phi(z) +B\Phi(z)$ if and only if
\begin{equation}
	\Phi(z,x)=\Xi (z,x)C+ \int_{x_0}^{x}
	dx'\, \Xi(z,x)\Xi(z,x')^{-1}B(x')\Phi(z,x'), \quad x\geq x_0
\lb{5.6}
\end{equation}
for some $C=(c_1,c_2)^{\top}\in \bbC^2$ independent of $x$. Moreover,
$\Phi(z,x)=0$ for all $x\geq x_0$ if and only if $C=0$.
\end{lemma}
\begin{proof}
The computation
\begin{align}
	M(q)&\big( \Phi(x)-\int_{x_0}^{x}
	dx'\, \Xi(x)\Xi(x')^{-1}B(x')\Phi(x')\big) \no\\
   &=z\Phi(x) +B(x)\Phi(x)-M(q)\bigg( \Xi(x) \int_{x_0}^{x}
	dx'\, \Xi(x')^{-1}B(x')\Phi(x') \bigg) \no\\
   &=z\bigg( \Phi(x)-\int_{x_0}^{x}
	dx'\, \Xi(x)\Xi(x')^{-1}B(x')\Phi(x')\bigg) \lb{5.7}
\end{align}	
shows that $\Phi$ satisfies
$M(q)\Phi(z)= z\Phi(z) +B\Phi(z)$ if and only if $\Phi$ satisfies
\eqref{5.6} since $\Xi$ is a fundamental matrix for the first-order
linear differential system $M(q)\Psi(z)=z\Psi(z)$. That $\Phi(z,x)=0$
for all $x\geq x_0$ if and only if $C=0$ follows upon iterating
the Volterra-type integral equation \eqref{5.6} in a standard
manner.
\end{proof}

Next, we find it convenient to recall a number of basic definitions and
well-known facts in connection with the spectral theory of 
non-self-adjoint operators (we refer to \cite[Chs.\ I, III, IX]{EE89},  
\cite[Sects.\ 1, 21--23]{Gl65}, and \cite[p.\ 178--179]{RS78} for
more details). Let $S$ be a densely defined closed operator in a complex
Hilbert space $\cH$. Denote by $\cB(\cH)$ the Banach space of all
bounded linear operators on $\cH$. The spectrum,
$\sigma(S)$, point spectrum (the set of eigenvalues), 
$\sigma_{\rm p}(S)$, continuous spectrum, $\sigma_{\rm c}(S)$, residual
spectrum, $\sigma_{\rm r}(S)$, approximate point spectrum, $\sigma_{\rm
a}(S)$, essential spectrum, $\sigma_{\rm e}(S)$, field of regularity,
$\pi(S)$, resolvent set, $\rho(S)$, and $\Delta (S)$ are defined by
\begin{align}
\sigma(S)&=\{\lambda\in\bbC\,|\, (S-\lambda I)^{-1}\in \cB(\cH)\}, 
\lb{5.8} \\
\sigma_{\rm p}(S)&=\{\lambda\in\bbC\,|\, \ker(S-\lambda I)\neq\{0\} \}, 
\lb{5.9} \\
\sigma_{\rm c}(S)&=\{\lambda\in\sigma (S)\,|\, \text{$\ker(S-\lambda
I)=\{0\}$ and $\ran(S-\lambda I)$ is dense in $\cH$} \no \\ &
\hspace*{2.25cm} \text{but not equal to
$\cH$}\}, \lb{5.10} \\
\sigma_{\rm r}(S)&=\{\lambda\in\bbC\,|\, \text{$\ker(S-\lambda I)=\{0\}$  
and $\ran(S-\lambda I)$ is not dense in $\cH$}\}, \lb{5.11} \\
\sigma_{\rm a}(S)&=\{\lambda \in \bbC \, |\, \text{there exists a
sequence $\{ f_n\}_{n\in\bbN}$ with $\| f_n \|_\cH=1$, $n\in\bbN$,} \no
\\ &\hspace*{1.8cm} \text{and $\lim_{n\to\infty} 
\|(S-\lambda I)f_n\|_\cH=0$}\}, \lb{5.12} \\
\sigma_{\rm e}(S)&=\{\lambda\in\bbC\,|\, \text{there exists a
sequence $\{ f_n\}_{n\in\bbN}$ s.t. $\{f_n\}_{n\in\bbN}$ contains no} \no \\
& \hspace*{-.6cm} \text{convergent subsequence, $\| f_n \|_\cH =1$,
$n\in\bbN$,  and $\lim_{n\to\infty} \|(S-\lambda I)f_n\|_\cH=0$}\},
\lb{5.13} \\
\pi(S)&=\{z \in\bbC \,|\, \text{there exists $k_z >0$ s.t. 
$\| (S- zI)u\|_\cH \ge k_z \| u\|_\cH$} \no \\
& \hspace*{1.75cm} \text{for all $u\in\dom(S)$}\}, \lb{5.14} \\
\rho(S)&=\bbC\backslash\sigma (S), \lb{5.15} \\
\Delta(S)&=\{z\in\bbC \,|\, \text{$\dim(\ker(S-z))<\infty$ and
$\ran(S-z)$ is closed}\}, \lb{5.16}
\end{align}
respectively. One then has 
\begin{align}
\sigma (S)&=\sigma_{\rm p}(S)\cup\sigma_{\rm{c}}(S)\cup
\sigma_{\rm r}(S) \quad \text{(disjoint union)} \lb{5.17} \\
&=\sigma_{\rm p}(S)\cup\sigma_{\rm{e}}(S)\cup\sigma_{\rm r}(S), 
\lb{5.18} \\
\sigma_{\rm c}(S)&\subseteq\sigma_{\rm e}(S)\backslash
(\sigma_{\rm p}(S)\cup\sigma_{\rm r}(S)), \lb{5.18a} \\
\sigma_{\rm r}(S)&=\sigma_{\rm p}(S^*)^* \backslash\sigma_{\rm p}(S), 
\lb{5.19} \\
\sigma_{\rm a}(S)&=\{\lambda\in\bbC \,|\, \text{for all $\varepsilon>0$,
there exists
$0\neq f_\varepsilon$} \no \\
& \hspace*{1.8cm} \text{s.t. $\|(S-\lambda)f_\varepsilon\|_\cH
\leq\varepsilon \|f_\varepsilon\|_\cH$}\} \lb{5.20} \\
&=\bbC\backslash\pi(S), \lb{5.21} \\
\sigma(S)\backslash&\sigma_{\rm a}(S)\subseteq\sigma_{\rm r}(S), \quad 
\sigma(S)\backslash\sigma_{\rm a}(S) \, \text{ is open}, \lb{5.21a} \\
\sigma_{\rm e}(S)&=\{\lambda\in\bbC\,|\, \text{there exists a
sequence $\{ g_n\}_{n\in\bbN}$ s.t. $\wlim_{n\to\infty} g_n=0$,} \no \\ 
& \hspace*{1.75cm} \text{$\| g_n \|_\cH =1$,
$n\in\bbN$,  and $\lim_{n\to\infty} \|(S-\lambda I)g_n\|_\cH=0$}\} 
\lb{5.22}\\
& = \bbC\backslash\Delta (S),\lb{5.22b}
\\
\sigma_{\rm e}(S)&\subseteq\sigma_{\rm a}(S)\subseteq\sigma(S) 
\, \text{ (all three sets are closed)}, \lb{5.23} \\
\rho(S)&\subseteq \pi(S) \subseteq \Delta (S) \, 
\text{ (all three sets are open).}
\lb{5.24} 
\end{align}
Here $\omega^*$ in the context of \eqref{5.19} denotes the complex 
conjugate of the set $\omega\subseteq\bbC$, that is,
\begin{equation}
\omega^*=\{\ol{\lambda}\in\bbC \,|\, \lambda\in\omega\}. \lb{5.25}
\end{equation}
For future reference we note that the sequence $\{f_n\}_{n\in\bbN}$ in
\eqref{5.13} (and the sequence $\{g_n\}_{n\in\bbN}$ in \eqref{5.22}) is
called a {\it singular $($or Weyl\,$)$ sequence} of $S$ corresponding to
$\lambda$. We also note that there are several  versions of the
concept of the essential spectrum in the non-self-adjoint context (cf.\
\cite[Ch.\ IX]{EE89}) but we will only use the one in \eqref{5.13}
(respectively,
\eqref{5.22}) in this paper.

In the special case where $S$ is $\cJ$-self-adjoint one obtains the
following  simplifications (cf.\ \cite[p.\ 118]{EE89}, \cite[p.\
76]{Gl65}):
\begin{align}
\sigma(S)&=\sigma_{\rm p}(S)\cup\sigma_{\rm c}(S) \lb{5.26} \\
&=\sigma_{\rm p}(S)\cup\sigma_{\rm e}(S), \lb{5.27} \\
\sigma_{\rm r}(S)&=\emptyset, \lb{5.28} \\
\sigma_{\rm p}(S)&=\sigma_{\rm p}(S^*)^*, \lb{5.28a} \\
\sigma_{\rm a}(S)&=\sigma(S), \lb{5.29} \\
\pi(S)&=\rho(S),  \lb{5.30} 
\end{align}
whenever $S$ is $\cJ$-self-adjoint. Note that $\pi(S)=\emptyset$ may
occur for $\cJ $-symmetric operators (see \cite{Ra85}) in sharp contrast
to the case of densely defined, closed, symmetric operators $T$, where
any nonreal number is in the field of regularity $\pi(T)$.

Returning to the $\NS_-$ case, we next recall an elementary but useful
consequence of \eqref{2.11} and \eqref{2.14}.

\begin{lemma} \lb{l5.4} Assume Hypothesis \ref{h3.0}. Then 
\begin{align}
\cK D(q)\cK&=-D(q), \lb{5.30a} \\
\sigma_3 D(q) \sigma_3 &=D(q)^*=D(-q). \lb{5.30b}
\end{align}
Consequently,
\begin{equation}
\sigma(D(q))^*=\sigma(D(q))=\sigma(D(-q)). \lb{5.30c}
\end{equation}
Moreover, 
\begin{align}
& \text{if }\, z_0\in\sigma_{\rm p}(D(q)) \, \text{ and } \, D(q)F=z_0F 
\, \text{ for some $F\in\dom(D(q))$,} \no \\
& \text{then } \, D(q)\cK F= \bar{z_0} \cK F. \lb{5.30d}
\end{align}
\end{lemma}
\begin{proof}
Relations \eqref{5.30a} and \eqref{5.30b} are clear from \eqref{2.11},
\eqref{2.14}, \eqref{3.1}, \eqref{3.3}, and \eqref{5.1}. Since 
$(D(q)^*-zI_2)^{-1}=[(D(q)-\bar{z}I_2)^{-1}]^*$, $\sigma(D(q)^*)=
\sigma(D(q))^*$ and hence \eqref{5.30c} follows from \eqref{5.30b}.
Finally, \eqref{5.30d} is clear from \eqref{5.30a} and $\cK^2=-I_2$.
\end{proof}

Next we introduce the basic hypothesis to be assumed for the remainder
of this section.

\begin{hypothesis} \lb{h4.2a} Suppose $q\in L^1_{\loc}(\bbR)$, assume
the $\NS_-$ case $p=-\bar q$, and suppose that the 
$($$\cJ$-self-adjoint\,$)$ operator $D(q)$ has nonempty resolvent set,
$\rho(D(q))\ne \emptyset$.  
\end{hypothesis}

Recalling the standard notation
\begin{align}
\nul(T)&=\dim(\ker(T)), \lb{5.31} \\
\text{def}(T)&= \dim(\ran(T)^\bot)=\dim(\ker(T^*))=\nul(T^*), \lb{5.32} 
\end{align} 
where $T$ denotes a densely defined closed operator in $\cH$, 
we can state the following fundamental result, establishing the
existence of Weyl--Titchmarsh-type solutions for $\cJ$-self-adjoint
Dirac-type operators relevant to the $\NS_-$ case.

\begin{theorem}\lb{RL4.1a} Assume Hypothesis \ref{h4.2a} and pick
$z\in \rho(D(q))$. Then there exist two unique $($up to constant
multiples$)$ linearly independent $z$-wave functions
$\Psi_{-}(z,\cdot)$ and $\Psi_{+}(z,\cdot)$ associated with $q$
satisfying
\begin{align}
&	\Psi_{-}(z,\cdot) \in
	L_{\loc}^2([-\infty,\infty)), \quad
	\Psi_{+}(z,\cdot) \in
	L_{\loc}^2((-\infty,\infty]), \lb{R4.1} \\
&	\|\Psi_{-}(z,\cdot)\|_{\bbC^2}^{-1}
	\in L_{\loc}^2((-\infty,\infty]),\quad 
	\|\Psi_{+}(z,\cdot)\|_{\bbC^2}^{-1}
	\in L_{\loc}^2([-\infty,\infty)), \lb{R4.2} \\
&\lim_{x\to\pm\infty}\|\Psi_\pm(z,x)\|_{\bbC^2}=
\lim_{x\to\mp\infty}\|\Psi_\pm(z,x)\|^{-1}_{\bbC^2}=0, \lb{5.40} \\
&	\sup_{r\in\bbR}
	\left[\left( \int_{-\infty}^r dx\,\|\Psi_{-}(z,x)\|^{2}
	_{\bbC^2} \right)
	\left( \int_r^{\infty} dx\,\|\Psi_{+}(z,x)\|^{2}_{\bbC^2}
	\right)\right] < \infty. \lb{R4.3}  
\end{align}
The corresponding $\bar z$-wave functions $\cK\Psi_\pm(z,x)$ associated
with $q$ satisfy \eqref{R4.1}--\eqref{R4.3} with $z$ replaced by $\bar
z$.
\end{theorem}
\begin{proof}
We prove the existence of the two $z$-wave functions following the
lines of \cite[Sect 10.4]{EE89}. To this end we introduce the
operators
\begin{align}
   &D_{\max}(q;-\infty)F=M(q)F, \no\\
	&F \in \dom(D_{\max}(q;-\infty)) = \big\{G\in
	L^2((-\infty,0]) \,|\, G\in	AC_{\loc}((-\infty,0]),  \\
& \hspace*{4.4cm} M(q)G \in	L^2((-\infty,0]) \big\}, \no\\
   & D_{\min}(q;-\infty)F=M(q)F, \no\\
	&F \in\dom(D_{\min}(q;-\infty)) = \{G\in
	\dom(D_{\max}(q;-\infty))\, \,| \, G(0)=0; \lb{R4.1a} \\
& \hspace*{4.35cm}  
	\supp (G) \subset (-\infty,0] \text{ is compact}\}, \no \\
  &D_{\max}(q;\infty)F=M(q)F, \no\\
	&F \in \dom(D_{\max}(q;\infty)) =
	\big\{G\in	L^2([0,\infty)) \,| \,	G\in AC_{\loc}([0,\infty)),  \\ 
& \hspace*{4.1cm} M(q)G \in	L^2([0,\infty))\big\}, \no\\
   &D_{\min}(q;\infty)F=M(q)F, \no\\
	&\dom(D_{\min}(q;\infty)) = \{G\in
	\dom(D_{\max}(q;\infty))
	\,|\, G (0)=0; \lb{R41.b} \\
& \hspace*{3.35cm} \supp (G) \subset [0,\infty) \text{ is compact}\}. \no
\end{align}

In close analogy to \cite[Theorem III.10.20]{EE89} one can prove that for
all $z\in \rho(D(q))$,
\begin{equation}
	\text{def}\big(\bar{D_{\min}(q)}-z I_2\big)=
\text{def}\big(\bar{D_{\min}(q;-\infty)}-z
	I_2\big)+\text{def}\big(\bar{D_{\min}(q;\infty)}-z I_2\big)-2.  
\end{equation}
Since $D(q)=\ol{D_{\min}(q)}$ is $\cJ$-self-adjoint and $z\in
\rho(D(q))$, we necessarily have (see, e.g., 
\cite[Theorem III.5.5]{EE89}) that
\begin{equation}
\text{def}\big(\bar{D_{\min}(q)}-z I_2\big)=\text{def}(D(q)-z
I_2)=0. 
\end{equation}
Thus, 
\begin{equation}
\text{def}\big(\bar{D_{\min}(q;-\infty)}-z I_2\big)+\text{def}
\big(\bar{D_{\min}(q;\infty)}-z I_2\big)=2. 
\end{equation}
We claim that the only possibility is
\begin{equation}
	\text{def}\big(\bar{D_{\min}(q;-\infty)}-zI_2\big)=
	\text{def}\big(\bar{D_{\min}(q;\infty)}-z I_2\big)=1. \lb{5.43}
\end{equation}
Indeed, arguing by contradiction, we assume, for instance, that
\begin{equation} 
\text{def}\big(\bar{D_{\min}(q;-\infty)}-z_0 I_2\big)=0. 
\end{equation}
This implies
\begin{equation}
\text{nul} \big(D_{\max}(-q;\infty)-\bar{z_0}I_2\big)=2 \lb{5.51a}
\end{equation}
since $(D_{\min}(q);\infty)^*=D_{\max}(-q;\infty)$. Thus all
$\bar{z_0}$-wave functions for the $\NS_-$ potential $-q$ are in
$L^2_{\loc}((-\infty,\infty])^2$. According to Remark \ref{RL4.0a},
this is clearly impossible for $\bar{z_0}\in \bbR\cap\rho(D(-q))$.
Next we show that this is impossible  also for $\bar{z_0}\in
(\bbC\backslash\bbR)\cap\rho(D(-q))$. Using Lemma \ref{YL1}\,(iii), we
can simplify notations and replace $\bar{z_0}$ by $z_0$ without loss of
generality. Moreover, to avoid confusion with the change $q\to -q$ and
the corresponding change for $q_{z_0}^{(1)}$, we simply use $q$ instead
of $-q$ in the proof of \eqref{R4.1} below.

To this end, we fix $z_0\in\rho(D(q))$ with $\Im(z_0)\ne 0$
and $\Psi_1(z_0), \Psi_2(z_0)$ two linearly independent
$z_0$-wave functions associated with the background potential $q$. The
latter are in $L^2([0,\infty))^2$ by hypothesis \eqref{5.51a}. Then for
any $z_0$-wave function $\Psi(z_0)=(\psi_1(z_0),\psi_2(z_0))^{\top}$
associated with the $\NS_-$ potential $q$, one infers (cf.\ \eqref{Y8.1}
and \eqref{Yq}) that
$\Phi^{(1)}_{z_0}(z_0)=\Im(z_0)\cK
\Psi(z_0) 
\| \Psi(z_0)\|_{\bbC^2}^{-2}$ is a
$z_0$-wave function associated with the $\NS_-$ potential 
\begin{equation}
q^{(1)}_{z_0}=q+4\Im(z_0)\psi_1(z_0)\bar{\psi_2(z_0)} 
\|\Psi(z_0)\|_{\bbC^2}^{-2}. 
\end{equation}
Thus $\Phi^{(1)}_{z_0}(z_0)$ satisfies
\begin{equation}
z_0\Phi^{(1)}_{z_0}(z_0)=M(q^{(1)})\Phi^{(1)}_{z_0}(z_0)
=M(q)\Phi^{(1)}_{z_0}(z_0)-B(z_0)\Phi^{(1)}_{z_0}(z_0),
\end{equation}
where
\begin{equation}
	B(z_0,x)=4i\Im(z_0)\|\Psi(z_0,x)\|_{\bbC^2}^{-2}\begin{pmatrix} 0&
\psi_1(z_0,x)\bar{\psi_2(z_0,x)} \\
	\bar{\psi_1(z_0,x)}\psi_2(z_0,x) & 0	\end{pmatrix}
\end{equation}
belongs to $L^{\infty}(\bbR)^{2\times 2}$. In particular,  
\begin{equation}
\text{ess\,sup}_{x\in\bbR}
\|B(z_0,x)\|_{\bbC^{2\times 2}}\le 2|\Im(z_0)|. \lb{5.48}
\end{equation}
Since no confusion can arise we occasionally suppress
the explicit $z_0$-dependence in the calculations below.
The variation of parameters formula \eqref{5.6} then yields the
following for the fundamental system of solutions $\Xi =
[\Psi_1,\Psi_2]$, $W(\Psi_1,\Psi_2)=1$, associated with $q$ and $z_0$: 
\begin{align}
   \Phi^{(1)}_{z_0}(x)&=\Xi(x)C+\int_{a}^{x}
	dx'\,\Xi(x)\Xi(x')^{-1}B(x')\Phi^{(1)}_{z_0}(x')
\lb{5.49} \\
   &=c_1\Psi_1(x)+c_2\Psi_2(x)+\Xi(x)\int_{a}^{x}
	dx'\,\Xi(x')^{-1}B(x')\Phi^{(1)}_{z_0}(x'), \quad x\geq a>0, \no
\end{align}
where $C=(c_1,c_2)^{\top}\neq 0$ depends on $a$ and we assume 
$W(\Psi_1,\Psi_2)=1$ according to Lemma \ref{l5.1}. Hence one 
obtains 
\begin{equation}
\|\Xi(x)\|_{\bbC^{2\times 2}} \leq
2^{1/2}\max\big(\|\Psi_1(x)\|_{\bbC^2},
\|\Psi_2(x)\|_{\bbC^2}\big). \lb{5.50}
\end{equation}
Moreover, writing $\Psi_j=(\psi_{1,j},\psi_{2,j})^\top$, $j=1,2$,
one infers 
\begin{equation}
\|\psi_{k,j}(z_0,\cdot)\|_{L^2([a,\infty))}\le K(a), \quad
j,k=1,2 \lb{5.51}
\end{equation}
for some constant $K(a)>0$ with 
\begin{equation}
\lim_{a\uparrow\infty}K(a)=0. 
\end{equation}
Hence,
\begin{align}
\int_a^x dx' \, \|\Xi(x')^{-1}\|^2_{\bbC^{2\times 2}} &\leq 
2\int_a^x dx' \, \max\bigg(\bigg\|\begin{pmatrix}
\psi_{2,2}(x')\\ -\psi_{2,1}(x')\end{pmatrix}\bigg\|^2_{\bbC^2},
\bigg\|\begin{pmatrix} -\psi_{1,2}(x')\\
\psi_{1,1}(x')\end{pmatrix}\bigg\|^2_{\bbC^2}\bigg) \no \\
& \leq 4 K(a)^2. \lb{5.52}
\end{align}
Thus, \eqref{5.48}--\eqref{5.52} yield for $x\ge a$,
\begin{align}
	&\big\|\Phi^{(1)}_{z_0}(x)\big\|_{\bbC^2}\le
|c_1|\|\Psi_1(x)\|_{\bbC^2}
	+|c_2|\|\Psi_2(x)\|_{\bbC^2} \lb{5.53} \\
	& \quad +2^{1/2}4|\Im(z_0)|K(a) \max\big(
\|\Psi_1(x)\|_{\bbC^2},\|\Psi_2(x)\|_{\bbC^2}\big)
	\bigg( \int_a^{x} dx' \, \big\|
\Phi^{(1)}_{z_0}(x')\big\|^2_{\bbC^2}\bigg)^{1/2}. \no
\end{align}
Squaring \eqref{5.53} and integrating the result from $a$ to $x$, 
one estimates 
\begin{align}
& \int_a^{x} dx' \, \big\|\Phi^{(1)}_{z_0}(x')\big\|^2_{\bbC^2} \no \\
& \leq 3|c_1|^2 \int_a^x dx'\,\|\Psi_1(x')\|^2_{\bbC^2}
+3|c_2|^2 \int_a^x dx'\,\|\Psi_2(x')\|^2_{\bbC^2} \no \\
& \quad +96|\Im(z_0)|^2 K(a)^2\int_a^x dx'\int_a^{x'} dx''\, 
\max\big(\|\Psi_1(x')\|_{\bbC^2},
\|\Psi_2(x')\|_{\bbC^2}\big)
\big\|\Phi^{(1)}_{z_0}(x'')\big\|^2_{\bbC^2} \no \\ & \leq
6K(a)^2\big(|c_1|^2+|c_2|^2\big) \no \\ & \quad + 96|\Im(z_0)|^2
K(a)^2\int_a^x dx''\int_{x''}^{x} dx'\, 
\max\big(\|\Psi_1(x')\|_{\bbC^2},
\|\Psi_2(x')\|_{\bbC^2}\big)
\big\|\Phi^{(1)}_{z_0}(x'')\big\|^2_{\bbC^2} \no \\ & \leq
6K(a)^2\big(|c_1|^2+|c_2|^2\big) \no \\ & \quad + 96|\Im(z_0)|^2
K(a)^2\int_a^x dx''\,
\big\|\Phi^{(1)}_{z_0}(x'')\big\|^2_{\bbC^2}
\int_{a}^{x} dx'\, \max\big(\|\Psi_1(x')\|_{\bbC^2},
\|\Psi_2(x')\|_{\bbC^2}\big)  \no \\
& \leq 6K(a)^2\big(|c_1|^2+|c_2|^2\big) + 192 |\Im(z_0)|^2
K(a)^4 \int_a^{x} dx'' \,
\big\|\Phi^{(1)}_{z_0}(x'')\big\|^2_{\bbC^2}. \lb{5.54}
\end{align}
Here we applied the Fubini--Tonelli theorem to the integrand 
\begin{equation}
\max\big(\|\Psi_1(x')\|_{\bbC^2},\|\Psi_2(x')\|_{\bbC^2}\big)
\big\|\Phi^{(1)}_{z_0}(x'')\big\|^2_{\bbC^2}\chi_{[a,x']}(x'')\geq 0
\end{equation}
($\chi_{\Lambda}$ the characteristic function of the set
$\Lambda\subset\bbR$) to prove equality of the iterated integrals
$\int_a^x dx' \int_a^{x'} dx'' \,\cdots$ and $\int_a^x dx'' \int_{x''}^x
dx'\,\cdots$ in \eqref{5.54}. Hence,
if one chooses $a$ large enough (such that
$192|\Im(z_0)|^2K(a)^4<1$), then
\begin{equation}
\big[1-192|\Im(z_0)|^2K(a)^4\big] \int_a^{x} dx' \,
\big\|\Phi^{(1)}_{z_0}(z_0,x')\big\|^2_{\bbC^2}
\le 6K(a)^2\big(|c_1|^2+|c_2|^2\big),
\end{equation}
and thus, $\Phi^{(1)}_{z_0}(z_0,\cdot)\in
L^2([a,\infty))^2$. Since
$\big\|\Phi^{(1)}_{z_0}(z_0,\cdot)\big\|_{\bbC^2}=
\|\Psi(z_0,\cdot)\|_{\bbC^2}^{-1}$ this contradicts the assumption
$\Psi(z_0,\cdot) \in L^2([0,\infty))^2$. This proves \eqref{5.43}.  

Finally, if $\Psi_-(z)$ and $\Psi_+(z)$ satisfying \eqref{R4.1} 
were linearly dependent then it would follow that
$z\in\sigma_p(D(q))$ by
\eqref{R4.1}, contradicting the initial assumption $z\in\rho(D(q))$.
Summing up, \eqref{5.43} implies existence and uniqueness (up to
constant multiples) of $\Psi_\pm(z)$ satisfying \eqref{R4.1}.

To prove \eqref{R4.2} we assume without loss of generality that
\begin{equation}
W(\Psi_-(z,x),\Psi_+(z,x))=1, \quad x\in\bbR. \lb{5.56}
\end{equation}
Then one computes
\begin{equation}
1+|(\Psi_-(z,x), \Psi_+(z,x))_{\bbC^2}|^2=
\| \Psi_+(z,x)\|^2_{\bbC^2}\| \Psi_-(z,x)\|^2_{\bbC^2}
\geq 1 \lb{R4.3AA}
\end{equation}
and hence,
\begin{equation}
	\|\Psi_{\mp}(z,x)\|_{\bbC^2}^{-1}\le \|
	\Psi_{\pm}(z,x)\|_{\bbC^2}, \quad x\in\bbR. \lb{R4.2b}
\end{equation}
Thus, $\|\Psi_{-}(z,\cdot)\|_{\bbC^2}^{-1}\in
L^2_{\loc}((-\infty, \infty])$. The fact that 
$\|\Psi_{+}(z,\cdot)\|_{\bbC^2}^{-1}\in
L^2_{\loc}([-\infty, \infty))$ in \eqref{R4.2} is proved analogously.

By \eqref{R4.1} and \eqref{R4.2} one infers
\begin{equation}
\liminf_{x\to\pm\infty}\|\Psi_\pm(z,x)\|_{\bbC^2}=
\liminf_{x\to\pm\infty}\|\Psi_\mp(z,x)\|^{-1}_{\bbC^2}=0. \lb{5.80}
\end{equation}
To prove \eqref{5.40} one first integrates \eqref{2.13} to obtain
\begin{equation}
\|\Psi_\pm(z,x_2)\|^2_{\bbC^2}-\|\Psi_\pm(z,x_1)\|^2_{\bbC^2}=2\Im(z) 
\int_{x_1}^{x_2} dx\, \big[|\psi_{1,\pm}(z,x)|^2- |\psi_{2,\pm}(z,x)|^2
\big]. 
\end{equation}
Thus, $\lim_{x\to\pm\infty}\|\Psi_\pm(z,x)\|_{\bbC^2}$
exist and hence equal $0$ by \eqref{5.80}. By \eqref{R4.2b} one then
infers
\begin{equation}
\lim_{x\to\pm\infty}\|\Psi_\mp(z,x)\|^{-1}_{\bbC^2}=0.
\end{equation}

What remains to be proved is \eqref{R4.3}. To this end consider the
Green's function for the Dirac-type operator $D(q)$. It can be written in
terms of the $z$-wave functions
\begin{equation}
	\Psi_{-}(z,x)=(\psi_{1,-}(z,x), \psi_{2,-}(z,x))^\top, \quad
	\Psi_{+}(z,x)=(\psi_{1,+}(z,x), \psi_{2,+}(z,x))^\top,
\end{equation}
whose existence is guaranteed by \eqref{R4.1} and whose Wronskian
is still assumed to satisfy \eqref{5.56}. More precisely, define the
$2\times 2$ matrix-valued function on $\bbR^2$
\begin{equation}
	G(z,x,y)=-\begin{cases}\Psi_{+}(z,x)\Psi_{-}(z,y)^\top, \quad
x\le y, \\
\Psi_{-}(z,x)\Psi_{+}(z,y)^\top, \quad x\ge y, \end{cases} 
\quad z\in\rho(D(q)). \lb{5.61}
\end{equation}
Explicitly,
\begin{equation}
	G(z,x,y)=-\begin{cases}\begin{pmatrix}
		\psi_{1,+}(z,x)\psi_{1,-}(z,y) &
\psi_{1,+}(z,x)\psi_{2,-}(z,y)\\
		\psi_{2,+}(z,x)\psi_{1,-}(z,y) &
\psi_{2,+}(z,x)\psi_{2,-}(z,y)
\end{pmatrix},\quad x\le y,\\[5mm]
	\begin{pmatrix}
		\psi_{1,-}(z,x)\psi_{1,+}(z,y) &
\psi_{1,-}(z,x)\psi_{2,+}(z,y)\\
		\psi_{2,-}(z,x)\psi_{1,+}(z,y) &
\psi_{2,-}(z,x)\psi_{2,+}(z,y)
\end{pmatrix},\quad x\ge y.
\end{cases}
\end{equation}
To prove that \eqref{5.61} indeed represents the Green's function
associated with the $\cJ$-self-adjoint operator $D(q)$ one can argue
as follows. One introduces a densely defined operator $R(z)$ in
$L^2(\bbR)^2$ by
\begin{align}
&(R(z)F)(x)=\int_{\bbR} dx'\, G(z,x,x')F(x'), \quad
z\in\rho(D(q)), \; x\in\bbR, \\ 
&F\in\dom(R(z))=\big\{G\in L^2(\bbR))^2 \,|\, \text{$\supp(G)$
is compact}\big\}. 
\end{align}
By inspection one infers that  
\begin{equation}
R(z)F\in AC_{\loc}(\bbR)^2\cap L^2(\bbR)^2 \, \text{ and } \,
M(q)R(z)F\in L^2(\bbR)^2.
\end{equation}
Thus, $R(z)$ maps $L^2(\bbR)^2$-elements of compact support into
the domain of $D(q)$. Moreover, an explicit computation shows that 
\begin{align}
&(M(q)-z)\big[R(z)F-(D(q)-z)^{-1}F\big] \no \\
& \quad =(D(q)-z)\big[R(z)F-(D(q)-z)^{-1}F\big]=0, \quad
F\in\dom(R(z)). \lb{5.66}
\end{align}
Since by hypothesis, $z\in\rho(D(q))$, \eqref{5.66} implies 
\begin{equation}
R(z)F=(D(q)-z)^{-1}F, \quad F\in\dom(R(z)).
\end{equation}
Hence $R(z)$ extends boundedly to all of $L^2(\bbR)^2$ and its
closure coincides with the resolvent $(D(q)-z)^{-1}$ of $D(q)$. 
Thus, $z\in \rho(D(q))$ is equivalent to the boundedness of the
operator
\begin{align}
   L^2(\bbR)^2 \ni F(x) \longmapsto &\int_{\bbR} 
G(z,x,y)F(y)\,dy\\
   &=\Psi_{+}(z,x)\int_{-\infty}^{x} \Psi_{-}(z,y)^\top F(y)dy
\no \\
& \quad	+\Psi_{-}(z,x)\int_x^{\infty}\Psi_{+}(z,y)^\top
F(y)dy \no
\end{align}
in $L^2(\bbR)$.
Taking $z \in \rho(D(q))$ and $F=(f,0)^\top$ and
$F=(0,f)^\top$ it follows that the operators 
\begin{align}
   L^2(\bbR)\ni f(x)\mapsto &\, \psi_{j,+}(z,x)\int_{-\infty}^{x}
	\psi_{k,-}(z,y) f(y)dy\no\\
   &+\psi_{j,-}(z,x)\int_x^{\infty} \psi_{k,+}(z,y) f(y)dy,
\quad j,k=1,2 \lb{R4.3a1}
\end{align}
are bounded in $L^2(\bbR)$. The last statement implies the
relations (in fact, it is equivalent to them, cf.\ \cite{Mu72} and
Lemma \ref{l6.1}) 
\begin{equation}
\sup_{r\in \bbR}\left[\left( \int_{-\infty}^{r} |\psi_{k,-}(x)|^2\,
dx \right)
	\left( \int_r^{\infty} |\psi_{j,+}(x)|^2 \, dx \right)\right] <
	+\infty, \quad j,k=1,2. \lb{R4.3a0}
\end{equation}
Indeed, we will prove next that \eqref{R4.3a0} follows from
\eqref{R4.3a1}. For simplicity we consider the case $j=k=1$. (The
proof for the remaining combinations of indices $j,k$ proceeds
analogously, cf.\ also \cite{CE70}). Assuming
\eqref{R4.3a1} for $j=k=1$, there exists a constant $C>0$ such that
\begin{align}
	&\int_{\bbR} \bigg{|} \psi_{1,+}(x)\int_{-\infty}^x  
	\psi_{1,-}(y) f(y)\, dy + \psi_{1,-} (x)
	\int_x^{\infty}
	\psi_{1,+}(y) f(y) dy\bigg{|}^2 \, dx\no\\
	&\le C\int_{\bbR} |f(x)|^2 \, dx. \lb{RM1}
\end{align}
For fixed $r\in \bbR$ and $f\in L^2(\bbR)$ satisfying $f(x)=0$, 
for all $x > r$, the inequality \eqref{RM1} implies (restricting the
interval of integration)
\begin{equation}
	\left( \int_{r}^{\infty}|\psi_{1,+}(x)|^2\, dx \right) \left|
	\int_{-\infty}^r \psi_{1,-}(y)f(y)\, dy \right|^2\le C
	\int_{-\infty}^r |f(x)|^2\, dx.
\end{equation}
Thus, choosing $f(x)=\bar{\psi_{1,-}(x)}$, for $x\le r$ and $f(x)=0$
otherwise (then clearly $f\in L^2(\bbR)$) one obtains \eqref{R4.3a0} with
$j=k=1$. Since
$\|\Psi_\pm(x)\|^2_{\bbC^2}=|\psi_{1,\pm}(x)|^2+|\psi_{2,\pm}(x)|^2$,
\eqref{R4.3a0} yields \eqref{R4.3}.

Finally, $\|\cK\Psi(x)\|_{\bbC^2}=\|\Psi(x)\|_{\bbC^2}$ proves the
statement about the $\bar z$-wave functions associated with $q$.
\end{proof}

The solutions $\Psi_\pm (z,x)$ in \eqref{R4.1} are analogs of
the familiar Weyl--Titchmarsh solutions in the context of self-adjoint
Dirac-type operators (cf.\ \cite[Ch.\ 3]{LS75}). 

The following is a consequence of \eqref{R4.1} and Remark \ref{RL4.0a}.

\begin{corollary} \lb{c5.5}
Assume $q\in L^1_{\loc}(\bbR)$ and suppose the $\NS_-$ case $p=-\bar q$.
Then 
\begin{align}
\sigma_{\rm c}(D(q))&\supseteq \bbR, \lb{5.73} \\
\sigma_{\rm e}(D(q))&\supseteq \bbR, \lb{5.73a} \\
\sigma_{\rm p}(D(q))&\cap\bbR=\emptyset. \lb{5.74}
\end{align}
\end{corollary}
\begin{proof}
Relation \eqref{5.74} holds by Remark \ref{RL4.0a} which excludes
the existence of an $L^2(\bbR)^2$ solution $F$ of $M(q)F=\lambda
F$ near $\pm\infty$ for all $\lambda\in\bbR$. To prove
\eqref{5.73} we can restrict ourselves to the case in which
$\rho(D(q))\ne\emptyset$. Pick $\lambda_0\in \bbR$. Then Remark
\ref{RL4.0a}, \eqref{R4.1}, and \eqref{5.74} imply that
$\lambda_0\notin\rho(D(q))\cup\sigma_{\rm p}(D(q))$.  Since 
$\sigma_{\rm r} (D(q))=\emptyset$ by \eqref{5.28}, it follows that
$\lambda_0\in\sigma_{\rm c}(D(q))$ by \eqref{5.17}. Relation
\eqref{5.73a} is then obvious from \eqref{5.18a}. 
\end{proof}

As a consequence of \eqref{5.73}, our frequent assumption
$z\in\rho(D(q))$ (especially in the next Section \ref{s6})
automatically implies $z\in\bbC\backslash\bbR$.

Interesting restrictions on the permissible location of eigenvalues of
$\cJ$-self-adjoint Dirac-type operators $D(q)$ under strong additional
constraints on $q$ were recently derived in \cite{KS02}. 

\begin{remark}
Given normalized Weyl--Titchmarsh-type solutions $\Psi_\pm(z,x,x_0)$ of
$M(q)\Psi(z,x)=z\Psi(z,x)$ for $z\in\rho(D(q))$ satisfying 
\begin{equation}
\psi_{1,\pm}(z,x_0,x_0)=1, \quad z\in\rho(D(q)) \lb{589}
\end{equation}
for some $x_0\in\bbR$, one can formally introduce associated
Weyl--Titchmarsh $m$-functions as follows. Denote by
$\Xi(z,x,x_0)$ a normalized $2\times 2$ fundamental system of solutions
of 
\begin{equation}
M(q)\Psi(z,x)=z\Psi(z,x), \quad z\in\bbC \lb{5.89a}
\end{equation}
at some $x_0\in\bbR$, that is,
$\Xi(z,x,x_0)$ satisfies \eqref{5.89a} for a.e.\ $x\in\bbR$ and
\begin{equation} 
\Xi(z,x_0,x_0)=I_2, \quad z\in\bbC. \lb{5.90}
\end{equation}
One then partitions $\Xi(z,x,x_0)$ as 
\begin{equation}
\Xi(z,x,x_0)=(\Theta(z,x,x_0)\; \Phi(z,x,x_0)) 
=\begin{pmatrix}\theta_1(z,x,x_0) & \phi_1(z,x,x_0)\\
\theta_2(z,x,x_0)& \phi_2(z,x,x_0) \end{pmatrix}, \lb{5.91}
\end{equation}
where $\theta_j(z,x,x_0)$ and $\phi_j(z,x,x_0)$, $j=1,2$, are entire with
respect to $z\in\bbC$ and normalized according to \eqref{5.90}. Then the
normalized Weyl--Titchmarsh solutions $\Psi_\pm(z,x,x_0)$ can be
expressed in terms of the basis $(\Theta(z,x,x_0)\;\Phi(z,x,x_0))$ as 
\begin{equation}
\Psi_{\pm}(z,x,x_0)=\Theta(z,x,x_0)+m_\pm(z,x_0)\Phi(z,x,x_0), \quad 
z\in\rho(D(q)) 
\end{equation}
for some coefficients $m_\pm(z,x_0)$. Clearly, $m_\pm(z,x_0)$ are 
analytic on $\rho(D(q))$ and they are the obvious analogs of the
half-line Weyl--Titchmarsh coefficients, familiar from
$($second-order scalar and first-order $2\times 2$$)$ self-adjoint
differential and difference operators (cf., e.g., 
\cite[\S~VII.1]{Be68},
\cite[Chs.\ 2, 3]{LS75}). It is tempting to conjecture that appropriate
boundary values of
$m_\pm(z,x_0)$ as $z$ approaches
$\sigma(D(q))$ encode the spectral information on $D(q)$, but this is
left for future investigations.
\end{remark}

\section{Transformation Operators for $\cJ$-Self-Adjoint \\ 
Dirac-Type Operators} \lb{s6}

The principal goal in this section is to construct transformation
operators in $L^2(\mathbb{R})^2$ that intertwine the 
$\cJ$-self-adjoint operators $D(q)$ and $D\big(q^{(1)}_{z_1}\big)$
corresponding to the Lax differential expressions $M(q)$ and
$M\big(q^{(1)}_{z_1}\big)$ in the focusing $\NS_-$-case and to use these
transformation operators to relate the spectra of $D(q)$ and
$D\big(q^{(1)}_{z_1}\big)$, the principal goal of this paper. 

In the following we always assume Hypothesis \ref{h4.2a} and freely use
the notation established in Section \ref{s5} for $D(q)$ as the maximally
defined $\cJ$-self-adjoint Dirac operator in the focusing $\NS_-$
case and the Weyl--Titchmarsh-type solutions $\Psi_\pm(z,x)$,
$z\in\rho(D(q))$,  established in Theorem \ref{RL4.1a}. 

We start with an elementary but important observation.

\begin{remark} \lb{r6.1}
Since the two $z$-wave functions $\Psi_-(z)$ and $\Psi_+(z)$ of $D(q)$
are linearly independent,
\begin{equation}
	W(\Psi_{-}(z,x),\Psi_{+}(z,x))=c(z)\ne 0, 
\lb{R4.4}
\end{equation}
all other $($nontrivial\,$)$ $z$-wave functions $\Psi$ satisfy
\begin{equation}
\Psi(z)=\alpha \Psi_{-}(z)+\beta\Psi_{+}(z) \text{ for some     
$\alpha, \beta \in \bbC\backslash \{0\}$.}  \lb{6.2}
\end{equation}
In addition, as we will prove next, 
\begin{equation}
	\| \Psi(z,\cdot)\|^{-1}_{\bbC^2} \in L^2(\bbR). \lb{6.3}
\end{equation}
Indeed, $\Psi(z)=\alpha\Psi_-(z)+\beta\Psi_+(z)$ and hence
\begin{equation} 
\|\Psi(z)\|_{\bbC^2}^{-1}\leq \big||\alpha| \|\Psi_+(z)\|_{\bbC^2}-
|\beta|\|\Psi_-(z)\|_{\bbC^2}\big|^{-1}, 
\end{equation}
\eqref{5.40}, and the fact that
$\Psi_\pm(z,\cdot)\in AC_\loc(\bbR)^2$, yield the existence of
constants $C_\pm >0$ such that
\begin{align}
\|\Psi(z,x)\|_{\bbC^2}^{-1} 
&= \big(\|\Psi_\pm(z,x)\|_{\bbC^2} \|\Psi(z,x)\|_{\bbC^2}^{-1}\big) 
\|\Psi_\pm(z,x)\|_{\bbC^2}^{-1} \no \\
& \leq C_\pm
\|\Psi_\pm(z,x)\|_{\bbC^2}^{-1}, \quad x\in\bbR. \lb{6.4}
\end{align}
By \eqref{R4.2} this implies that all $z$-wave functions $\Psi(z)$
associated with $q$, except $\Psi_{\pm}(z)$, satisfy \eqref{3.10}. Hence,
Hypothesis \ref{h4.2a} guarantees the existence of $z$-wave functions
$\Psi(z)$ satisfying Hypothesis \ref{h3.2} for all $z\in \rho(D(q))$. In
particular, all but two $z$-wave functions of $D(q)$ $($viz.,
$\Psi_\pm(z)$$)$ satisfy Hypothesis \ref{h3.2} at $z\in\rho(D(q))$.
\end{remark}

Without loss of generality we will restrict our considerations to the
special case $\alpha =\beta =1$ in \eqref{6.2} for the remainder of this
section up to \eqref{6.124}, that is, we choose 
\begin{equation}
	\Psi (z)=\Psi_{-}(z)+\Psi_{+}(z) \lb{R4.5}
\end{equation}
in the following.

Next, we pick some fixed $z_1\in \rho(D(q))$. We take $\Psi (z_1)$ as
in \eqref{R4.5}, $\Psi(z_1)=\Psi_{-}(z_1)+\Psi_{+}(z_1)$,  where
$\Psi_\pm(z_1)$ satisfy \eqref{R4.1}--\eqref{R4.3} with $z$ replaced by
$z_1$, and let $\cK \Psi(z_1)$ be the corresponding $\bar{z_1}$-wave
function associated with $q$. By Theorem \ref{RL4.1a},  
$\cK \Psi_{\pm}(z_1,\cdot)$ satisfy \eqref{R4.1}--\eqref{R4.3} with $z$
replaced by $\bar{z_1}$. Moreover, 
\begin{equation}
\cK \Psi (z_1)=\cK \Psi_{-}(z_1)+ \cK \Psi_{+}(z_1).
\end{equation}

Define the $\NS$ potential $q^{(1)}_{z_1}$ as in \eqref{Y5.2} and
consider the $z_1$-wave function $\Phi^{(1)}_{z_1}(z_1)
=(\phi^{(1)}_{1}(z_1),\phi^{(1)}_{2}(z_1))^\top$ associated
with $q^{(1)}_{z_1}$ as defined in \eqref{Y8.1},
\begin{equation}
	\Phi^{(1)}_{z_1}(z_1,x)=\Im(z_1)\|\Psi(z_1,x)\|_{\bbC^2}^{-2}
	\cK \Psi(z_1,x), \label{4.14a}
\end{equation}
and the $\bar{z_1}$-wave function $\cK \Phi^{(1)}_{z_1}(z_1)$ 
associated with $q^{(1)}_{z_1}$ as defined in \eqref{Y8.2},
\begin{equation}
\cK \Phi^{(1)}_{z_1}(z_1,x)=-\Im(z_1)\| \Psi
(z_1,x)\|_{\bbC^2}^{-2}\Psi (z_1,x). \lb{4.15a}
\end{equation}
We recall that according to \eqref{Yq} the new $\NS_-$ potential is 
then given by 
\begin{equation}
   q^{(1)}_{z_1}(x)=q(x)+4\phi^{(1)}_1
	(z_1,x)\psi_1(z_1,x). \lb{Yq.1} 
\end{equation}
Of course, both $D(q)$ and $D\big(q^{(1)}_{z_1}\big)$ (associated with
the differential expressions $M(q)$ and $M\big(q^{(1)}_{z_1}\big)$,
respectively) are $\cJ $-self-adjoint by Theorem \ref{RT5.1} and
Corollary \ref{YCor1}. 

In order to motivate the introduction of transformation operators one 
can argue as follows. Since
\begin{equation}
	\Gamma(z,\Psi(z_1),\cK \Psi (z_1))=-(i/2)(z-z_1)I_2 +
	\Phi^{(1)}_{z_1}(z_1)\Psi (z_1)^\bot, 
\end{equation}
one computes, for every $z$-wave function $\Upsilon(z,\cdot)$
associated with $q$, 
\begin{align}
   &\Upsilon^{(1)}_{z_1}(z,x)=\Gamma(z,x,\Psi(z_1),
	\cK \Psi (z_1))\Upsilon(z,x)\no \\
   &=-(i/2)(z-z_1)\Upsilon(z,x)+ \Phi^{(1)}_{z_1}(z_1,x)\Psi(z_1,x)^\bot
	\Upsilon(z,x) \no \\ 
   &=-(i/2)(z-z_1)\Upsilon(z,x) -\Phi^{(1)}_{z_1}(z_1,x)
	W(\Psi(z_1,x),\Upsilon(z,x))  \no \\
   &=-(i/2)(z-z_1)\Upsilon(z,x)- \Phi^{(1)}_{z_1}(z_1,x)
   W(\Psi_{-}(z_1,x)+\Psi_{+}(z_1,x),\Upsilon(z,x)) \no \\
   &=-(i/2)(z-z_1)\bigg\{\Upsilon(z,x)+2  \Phi^{(1)}_{z_1}(z_1,x) \no\\
   &\quad \times \bigg[\int_{-a}^x dx'\,
	\Psi_{-}(z_1,x')^\top\sigma_1 \Upsilon(z,x')-\int_x^{a} dx'\,
	\Psi_{+}(z_1,x')^\top\sigma_1 \Upsilon(z,x') \bigg] 
\bigg\}  \lb{4.10} \\
& \quad - \Phi^{(1)}_{z_1}(z_1,x)[W(\Psi_{-}(z_1,-a),\Upsilon(z,-a))+
W(\Psi_{+}(z_1,a),\Upsilon(z,a))], \quad a>0. \no
\end{align}
To arrive at \eqref{4.10}, we used the integrated form of \eqref{2.6}. 
Replacing $\Upsilon(z)$ by $F\in L^2(\bbR)^2$ in
\eqref{4.10}, noting that 
\begin{equation}
	\liminf_{a\uparrow \infty}|W(\Psi_{\pm}(z_1,\pm a),F(\pm a))|=0, \quad 
F\in L^2(\bbR)^2, 
\end{equation}
and repeating the same argument with $\Upsilon(z)$ replaced by 
$\cK\Upsilon(z)$, then leads to the introduction of the following
transformation operators $T_{z_1}$ and $\wti T_{\bar{z_1}}$ in
$L^2(\bbR)^2$,
\begin{align} 
L^2(\bbR&)^2 \ni F(x) \mapsto 
  (T_{z_1}F)(x)=F(x)+2\Phi^{(1)}_{z_1}(z_1,x) \lb{Y11.3} \\
   &\times \bigg[\int^x_{-\infty} dx' \, \Psi_{-} (z_1,x')^\top
   \sigma_1F(x')-\int_x^{\infty} dx' \, \Psi_{+} (z_1,x')^\top
   \sigma_1F(x')\bigg],  \no\\
L^2(\bbR&)^2 \ni F(x) \mapsto 
  (\wti T_{\bar{z_1}}F)(x)=F(x)+2\cK \Phi^{(1)}_{z_1}(z_1,x)  
\lb{Y11.4}\\
   &\times\bigg[\int^x_{-\infty} dx' \, (\cK \Psi_{-} (z_1,x'))^{\top} 
   \sigma_1 F(x') - \int_x^{\infty} dx' \,(\cK \Psi_{+} (z_1,x'))^{\top} 
   \sigma_1 F(x')  \bigg], \no \\ 
   & \hspace*{9cm} z_1\in\rho(D(q)). \no
\end{align}

That $T_{z_1}$ and $\wti T_{\bar{z_1}}$ are in fact bounded operators
in $L^2(\bbR)^2$ follows from Lemmas \ref{l6.1} and \ref{RL4.2} below. 

\begin{lemma} [Talenti \cite{Ta69}, Tomaselli \cite{To69} (see also
\cite{CE70}, \cite{Mu72}] \lb{l6.1}  ${}$ \\
Let $f\in L^2(\bbR)$, $U\in L^2((-\infty,R])$, $V\in L^2([R,\infty))$
for all $R\in\bbR$. Then the following assertions $(i)$--$(iii)$ are
equivalent: \\
$(i)$ There exists a finite constant $C>0$ such that
\begin{equation}
\int_\bbR dx \, \bigg|U(x)\int_x^\infty dx' \,V(x')f(x')
\bigg|^2 \leq C \int_\bbR dx\, |f(x)|^2. \lb{6.13a}
\end{equation}
$(ii)$ There exists a finite constant $D>0$ such that
\begin{equation}
\int_\bbR dx \, \bigg|V(x)\int_{-\infty}^x dx' \,U(x')f(x')
\bigg|^2 \leq D \int_\bbR dx\, |f(x)|^2. \lb{6.13b}
\end{equation}
$(iii)$ 
\begin{equation}
\sup_{r\in\bbR}\Bigg[\bigg(\int_{-\infty}^r dx\, |U(x)|^2\bigg)
\bigg(\int_r^\infty dx \, |V(x)|^2\bigg)\Bigg] <\infty.
\lb{6.14}
\end{equation}
\end{lemma}

\begin{lemma} \lb{RL4.2} 
Assume Hypothesis \ref{h4.2a} and $z_1\in\rho(D(q))$. Then the operators
\begin{align}
  & L^2(\bbR) \ni f(x) \mapsto
	\frac{1}{\|\Psi_{\pm}(z_1,x)\|_{\bbC^2}}\int_{\pm\infty}^{x} dx'\,
	\|\Psi_{\pm}(z_1,x')\|_{\bbC^2}f(x'), \lb{6.12} \\
  & L^2(\bbR) \ni f(x) \mapsto
	\|\Psi_{\pm}(z_1,x)\|_{\bbC^2}\int_{\mp\infty}^{x} dx'\,
	\frac{1}{\|\Psi_{\pm}(z_1,x')\|_{\bbC^2}}f(x') \lb{6.13}
\end{align}
are bounded in $L^2(\bbR)$.
\end{lemma}
\begin{proof}
Using \eqref{R4.1}, \eqref{R4.2}, \eqref{R4.3}, and \eqref{R4.2b} one
obtains
\begin{equation}
\sup_{r\in\bbR}
	\left( \int_{-\infty}^r dx\, \|\Psi_{-}(z_1,x)\|^{2}
	_{\bbC^2}\right) 
	\left( \int_r^{\infty} dx\, \|\Psi_{-}(z_1,x)\|^{-2}_{\bbC^2}
	\right) < \infty \lb{R4.3a}
\end{equation}
and  
\begin{equation}
\sup_{r\in\bbR}
	\left( \int_{-\infty}^r dx\, \|\Psi_{+}(z_1,x)\|^{-2}
	_{\bbC^2}\right) 
	\left( \int_r^{\infty} dx\, \|\Psi_{+}(z_1,x)\|^{2}_{\bbC^2}
	\right) < \infty. \lb{R4.3b}
\end{equation}
By Lemma \ref{l6.1}, \eqref{R4.3a} is a  necessary and
sufficient condition for the operators associated with
$\Psi_-(z_1,\cdot)$ in \eqref{6.12} and \eqref{6.13} to be bounded in
$L^2(\bbR)$. Similarly, \eqref{R4.3b} is a  necessary and
sufficient condition for the operators associated with
$\Psi_+(z_1,\cdot)$ in \eqref{6.12} and \eqref{6.13} to be bounded in
$L^2(\bbR)$.
\end{proof}

Thus, \eqref{6.12}, \eqref{Y8.1}, and \eqref{Y8.2} imply the following
result.

\begin{corollary} \lb{c6.2}
Assume Hypothesis \ref{h4.2a} and $z_1\in\rho(D(q))$. Then the
operators $T_{z_1}$ and $\wti T_{\bar{z_1}}$ defined in \eqref{Y11.3}
and \eqref{Y11.4} are bounded operators in $L^2(\bbR)^2$.
\end{corollary}
\begin{proof}
By \eqref{4.14a} one obtains
$\big\|\Phi^{(1)}_{z_1}(z_1,x)\big\|_{\bbC^2} =
|\Im(z_1)|\|\Psi(z_1,x)\|_{\bbC^2}^{-1}$. Applying \eqref{6.4} 
then yields the existence of constants $C_\pm >0$ such that
\begin{align}
\big\|\Phi^{(1)}_{z_1}(z_1,x)\big\|_{\bbC^2} & =
|\Im(z_1)| \|\Psi(z_1,x)\|_{\bbC^2}^{-1} \no \\
& \leq |\Im(z_1)| C_\pm
\|\Psi_\pm(z_1,x)\|_{\bbC^2}^{-1}, \quad x\in\bbR. \lb{6.20}
\end{align}
At this point an application of Lemma \ref{RL4.2} proves the
boundedness of $T_{z_1}$ in $L^2(\bbR)^2$. Since $\|\cK G\|_{\bbC^2}=
\|G\|_{\bbC^2}$ for all $G\in\bbC^2$, the same arguments prove 
boundedness of $\wti T_{\bar{z_1}}$ in $L^2(\bbR)^2$.
\end{proof}

In order to motivate the introduction of the inverse transformation
operators, one inverts the matrix
$\Gamma(z,\Psi(z_1),\cK \Psi(z_1))$ to obtain
\begin{align}
\Gamma(z,\Psi(z_1),\cK \Psi (z_1))^{-1}
&=c(z,z_1)\big[-(i/2)(z-\bar{z_1})I_2
+  \cK \Psi(z_1)\cK \Phi^{(1)}_{z_1} (z_1)^\bot \big] ,  
\end{align}
where 
\begin{equation}
c(z,z_1)=-4(z-z_1)^{-1}(z-\bar z_1)^{-1}.
\end{equation}
Thus, for $z$-wave functions $\Upsilon^{(1)}_{z_1}(z,x)$ associated
with $q^{(1)}_{z_1}$,
\begin{align}
&\Upsilon(z,x)=\Gamma(z,x,\Psi(z_1),\cK \Psi (z_1))^{-1}
\Upsilon^{(1)}_{z_1}(z,x) \no \\
&=c(z,z_1)\big[-(i/2)(z-\bar{z_1})\Upsilon^{(1)}_{z_1}(z,x) +
\cK \Psi(z_1,x)\cK \Phi^{(1)}_{z_1}(z_1,x)^\bot
\Upsilon^{(1)}_{z_1}(z,x)\big] \no \\
&=c(z,z_1)\big[-(i/2)(z-\bar{z_1})\Upsilon^{(1)}_{z_1}(z,x) -
\cK \Psi(z_1,x)
W\big(\cK\Phi^{(1)}_{z_1}(z_1,x),\Upsilon^{(1)}_{z_1}(z,x)\big)\big] \no
\\ &=c(z,z_1)\big\{-(i/2)(z-\bar{z_1})\Upsilon^{(1)}_{z_1}(z,x) \no \\
& \hspace*{1.75cm} -[\cK \Psi_-(z_1,x)+\cK \Psi_+(z_1,x)]
W\big(\cK\Phi^{(1)}_{z_1}(z_1,x),\Upsilon^{(1)}_{z_1}(z,x)\big)\big\}
\no \\ 
&=2i(z-z_1)^{-1}\bigg\{\Upsilon^{(1)}_{z_1}(z,x)+2\cK\Psi_{+}(z_1,x)
\int_{-a}^x dx'\, \big(\cK \Phi^{(1)}_{z_1}(z_1,x')\big)^\top \sigma_1 
\Upsilon^{(1)}_{z_1}(z,x')\no \\
&\hspace{4cm} - 2\cK\Psi_{-}(z_1,x) \int_x^{a} dx'\, 
\big(\cK \Phi^{(1)}_{z_1}(z_1,x')\big)^\top \sigma_1
\Upsilon^{(1)}_{z_1}(z,x')
\bigg\} \no \\
& \quad - c(z,z_1) \big[\cK \Psi_-(z_1,x) 
W\big(\cK\Phi^{(1)}_{z_1}(z_1,-a),\Upsilon^{(1)}_{z_1}(z,-a)\big) \no \\
& \hspace*{1.95cm} + \cK \Psi_+(z_1,x) 
W\big(\cK\Phi^{(1)}_{z_1}(z_1,a),\Upsilon^{(1)}_{z_1}(z,a)\big)\big],
\quad a>0. \lb{4.16}
\end{align}
Replacing $\Upsilon^{(1)}_{z_1}(z)$ by $F\in L^2(\bbR)^2$
in \eqref{4.16}, and repeating the same argument for 
$\cK \Upsilon^{(1)}_{z_1}(z)$ instead of $\Upsilon^{(1)}_{z_1}(z)$ , 
then leads to the introduction of the following (inverse) 
transformation operators $\hatt S_{\bar{z_1}}$ and $\wti S_{z_1}$ in
$L^2(\bbR)^2$,
\begin{align} 
L^2(\bbR)^2\ni F(x) \mapsto &(\hatt S_{\bar{z_1}}F)(x)= F(x) \no \\
&\quad +2\cK\Psi_{+} (z_1,x) \int^x_{-\infty}
dx'\, \big(\cK \Phi^{(1)}_{z_1}(z_1,x')\big)^\top \sigma_1 F(x')
\no\\ & \quad -2\cK \Psi_{-}(z_1,x) \int_x^{\infty} dx' \,
\big(\cK \Phi^{(1)}_{z_1}(z_1,x')\big)^\top \sigma_1 F(x'),
\lb{4.19} \\ 
L^2(\bbR)^2\ni F(x) \mapsto &(\wti S_{z_1}F)(x)=F(x) \no \\
& \quad +2\Psi_{+}(z_1,x) \int^x_{-\infty} dx' \,
\Phi^{(1)}_{z_1}(z_1,x')^\top \sigma_1 F(x') \lb{4.18}  \\
& \quad -2\Psi_{-}(z_1,x) \int_x^{\infty} dx' \,
\Phi^{(1)}_{z_1}(z_1,x')^\top \sigma_1 F(x'), \quad 
 z_1\in\rho(D(q)). \no
\end{align}

Equations \eqref{6.13}, \eqref{Y8.1}, and \eqref{Y8.2} then imply the
following result.

\begin{corollary} \lb{c6.3}
Assume Hypothesis \ref{h4.2a} and $z_1\in\rho(D(q))$. Then the
operators $\hatt S_{\bar{z_1}}$ and $\wti S_{z_1}$ defined in
\eqref{4.19} and \eqref{4.18} are bounded operators in $L^2(\bbR)^2$.
\end{corollary}
\begin{proof}
Using again the estimate \eqref{6.20}, one can follow the arguments
in the proof of Corollary \ref{c6.2}.
\end{proof}

We define two vectors $F,G\in L^2(\bbR)^2$ to be $\cJ $-orthogonal if 
\begin{equation}
(F, \cJ G)_{L^2}=0, 
\end{equation}
and we then write 
\begin{equation}
F \bot_{\cJ} G.
\end{equation}
Since 
\begin{align}
&(F(x),\cJ G(x))_{\bbC^2}=(\cJ G(x))^* F(x)= G(x)^\top \sigma_1 F(x),
\lb{6.24} \\  
& F \bot_{\cJ} G \text{ is equivalent to }  
(F, \cJ G)_{L^2}=\int_\bbR dx\, G(x)^\top \sigma_1 F(x)=0, \quad 
F,G\in L^2(\bbR)^2. \no 
\end{align}
We also introduce the following notation of the $\cJ$-orthogonal
complement $\cV^{\bot_\cJ }$ to a subset $\cV \subset L^2(\bbR)^2$,
\begin{equation}
\cV^{\bot_\cJ }=\{F\in L^2(\bbR)^2 \,|\, \text{$F\bot_\cJ G$ 
for all $G\in \cV$}\}.
\end{equation}
In analogy to the orthogonality property of eigenvectors corresponding
to different (necessarily real) eigenvalues of a symmetric operator in
some complex Hilbert space $\cH$, one infers the $\cJ$-orthogonality of
eigenvectors corresponding to different eigenvalues of a $\cJ$-symmetric
operator $S$ in $\cH$. Indeed, $Sf_j=z_j f_j$, $j=1,2$, with $z_1\neq
z_2$ implies
\begin{equation}
\bar{z_2}(\cJ f_1,f_2)_{\cH}=(\cJ f_1,Sf_2)_{\cH}=(\cJ Sf_1,f_2)_{\cH}
=(\cJ z_1 f_1,f_2)_{\cH}=\bar{z_1}(\cJ f_1,f_2)_{\cH} \lb{6.26}
\end{equation}
and hence
\begin{equation}
(z_1-z_2)(f_1,\cJ f_2)_{\cH}=0, \text{ implying } (f_1,\cJ
f_2)_{\cH}=0. \lb{6.27}
\end{equation}

Next, let $\sigma_0$ be an isolated  subset of $\sigma
\big(D\big(q^{(1)}_{z_1}\big)\big)$ in the sense that $\sigma_0$ can be
surrounded by a positively oriented, rectifiable, simple, closed
countour 
$\gamma_{\sigma_0}\subset \rho\big(D\big(q^{(1)}_{z_1}\big)\big)$
separating $\sigma_0$ from the remaining spectrum
$\sigma\big(D\big(q^{(1)}_{z_1}\big)\big)\backslash \sigma_0$. Then the
Riesz projection onto the spectral subspace corresponding to
$\sigma_0$ is given by
\begin{equation}
	P_{z_1}^{(1)}(\sigma_0)=-\frac{1}{2\pi i}\int_{\gamma_{\sigma_0}} dz \,
\big(D\big(q^{(1)}_{z_1}\big)-z I_2\big)\big)^{-1}. 
\end{equation}
The spectral subspace $\Sigma_{z_1}^{(1)}(\sigma_0)$ corresponding to
$\sigma_0$ is then the range of the Riesz projection, 
$\Sigma_{z_1}^{(1)}(\sigma_0)=P_{z_1}^{(1)}(\sigma_0)L^2(\bbR)^2$.

We record the following result.

\begin{lemma}\lb{RL4.1} Assume Hypothesis \ref{h4.2a}. \\
$(i)$ Let $z_0, \ti z_0 \in \sigma_p\big(D\big(q^{(1)}_{z_1}\big)\big)$,
$z_0\ne
\ti z_0$. Then the $L^2(\bbR)^2$-eigenfunctions $\Phi^{(1)}_{z_1}(z_0)$,
$\Phi^{(1)}_{z_1}(\ti z_0)$ corresponding to $z_0$ and $\ti z_0$, 
respectively, are $\cJ$-orthogonal,
\begin{equation}
	\Phi^{(1)}_{z_1}(z_0)\bot_\cJ \Phi^{(1)}_{z_1}(\ti z_0). \lb{6.29}
\end{equation}
$(ii)$ Let $z_0 \in \sigma_p\big(D\big(q^{(1)}_{z_1}\big)\big)$ and
$\sigma_0$ an isolated subset of
$\sigma\big(D\big(q^{(1)}_{z_1}\big)\big)$ which does not contain
$z_0$. Then the $L^2(\bbR)^2$-eigenfunction $\Phi^{(1)}_{z_1}(z_0)$
corresponding to $z_0$ is $\cJ$-orthogonal to the spectral subspace
$\Sigma_{z_1}^{(1)}(\sigma_0)$ corresponding to $\sigma_0$,
\begin{equation}
	\Phi^{(1)}_{z_1}(z_0)\bot_\cJ  \Sigma_{z_1}^{(1)}(\sigma_0).
\end{equation}
\end{lemma}
\begin{proof}
Assertion $(i)$ is clear from \eqref{6.26} and \eqref{6.27}. 
To prove $(ii)$, one chooses a positively oriented, rectifiable, simple, 
closed contour $\gamma_{\sigma_0}$ which separates $\sigma_0$  and
$\{z_0\}$. For $F\in L^2(\bbR)^2$ one then computes
\begin{align}
   \big(P_{z_1}^{(1)}(\sigma_0)F, \cJ
\Phi^{(1)}_{z_1}(z_0)\big)_{L^2}&=-\frac{1}{2\pi
i}\int_{\gamma_{\sigma_0}}
	dz\, \big(D\big(q^{(1)}_{z_1}\big)-z I_2\big)^{-1}F, \cJ
\Phi^{(1)}_{z_1}(z_0)\big)_{L^2} \no\\
&=-\frac{1}{2\pi i}\int_{\gamma_{\sigma_0}} dz\, \big(F,\cJ
\big(D\big(q^{(1)}_{z_1}\big)-z I_2\big)^{-1}
	\Phi^{(1)}_{z_1}(z_0)\big)_{L^2} \no\\
   &=-\frac{1}{2\pi i}\int_{\gamma_{\sigma_0}} dz\, (z_0 - z)^{-1}
	\big(F,\cJ \Phi^{(1)}_{z_1}(z_0)\big)_{L^2}=0. \no
\end{align} 
Here we used the fact that since $D\big(q^{(1)}_{z_1}\big)$ is 
$\cJ $-self-adjoint, so is $\big(D\big(q^{(1)}_{z_1}\big)-z
I_2\big)^{-1}$. 
\end{proof}

\begin{remark}
{}From Lemma \ref{RL4.1} and Theorem \ref{YT2} one concludes that 
\begin{equation}
\Phi^{(1)}_{z_1}(z_1)\bot_\cJ \cK \Phi^{(1)}_{z_1}(z_1) \lb{6.36a}
\end{equation}
since $z_1$ is a nonreal eigenvalue. Thus,
\begin{equation}
	\Phi^{(1)}_{z_1}(z_1)\in \big\{ \cK 
	\Phi^{(1)}_{z_1}(z_1)\big\}^{\bot_\cJ }.\lb{4.21}
\end{equation}
\end{remark}

\begin{remark}
By Corollary \ref{c6.3}, the operator $\hatt S_{\bar{z_1}}$,
$z_1\in\rho(D(q))$, is well-defined and bounded in
$L^2(\bbR)^2$. However, for future considerations it is more
appropriate to restrict it to the closed subspace 
$\big\{\cK \Phi^{(1)}_{z_1}(z_1)\big\}^{\bot_\cJ }$. Hence from now on,
we will denote by $S_{\bar{z_1}}$ the restriction of $\hatt
S_{\bar{z_1}}$ to $\big\{\cK \Phi^{(1)}_{z_1}(z_1)\big\}^{\bot_\cJ }$,
\begin{equation}
S_{\bar{z_1}}=\hatt S_{\bar{z_1}}\big|_{\big\{\cK
\Phi^{(1)}_{z_1}(z_1)\big\}^{\bot_\cJ }}, \quad z_1\in\rho(D(q)).
\lb{6.37}
\end{equation}
An elementary computation $($based on \eqref{6.24}$)$ then reveals that 
\begin{align}
&	(S_{\bar{z_1}}G)(x)=(\hatt S_{\bar{z_1}}G)(x)=
G(x)+2\cK \Psi(z_1,x)\int_{-\infty}^x
	dx'\, \big(\cK \Phi^{(1)}_{z_1}(z_1,x')\big)^\top\sigma_1 G(x'), \no \\
& \hspace*{8cm} G\in \big\{\cK \Phi^{(1)}_{z_1}(z_1)\big\}^{\bot_\cJ }.
\lb{4.29}
\end{align}

\end{remark}

Next, we prove several results leading up to the principal theorems of
this section.

\begin{lemma} \lb{RL4.3a} 
Assume Hypothesis \ref{h4.2a} and $z_1\in\rho(D(q))$. Then 
\begin{align} 
& \ran(T_{z_1})\subseteq\big\{\cK\Phi^{(1)}_{z_1}(z_1)\big\}^{\bot_\cJ},
\quad \Phi^{(1)}_{z_1}(z_1)\in\ker(S_{\bar{z_1}}), \lb{6.35} \\ 
& \big(\Phi^{(1)}_{z_1}(z_1),\cJ \Phi^{(1)}_{z_1}(z_1)\big)_{L^2}=
\Im(z_1) (2W(\Psi_+(z_1),\Psi_-(z_1)))^{-1}\ne 0, \lb{6.36} \\ 
& T_{z_1}S_{\bar{z_1}}\, G=G-\frac{\big(G,\cJ 
	\Phi^{(1)}_{z_1}(z_1)\big)_{L^2}}{\big(\Phi^{(1)}_{z_1}(z_1),
 	\cJ \Phi^{(1)}_{z_1}(z_1)\big)_{L^2}}\Phi^{(1)}_{z_1}(z_1), \quad 
G\in \big\{ \cK \Phi^{(1)}_{z_1}(z_1)\big\}^{\bot_\cJ}, \lb{4.28} \\
&	S_{\bar{z_1}}T_{z_1}F=F, \quad F\in L^2(\bbR)^2. \lb{4.28b}
\end{align}
\end{lemma}
\begin{proof}
For brevity, we introduce
\begin{align}
&	\xi(x;F)=\int_{-\infty}^x dx' \, \Psi_{-}(z_1,x')^\top\sigma_1F(x')
	-\int_x^{\infty} dx' \, \Psi_{+}(z_1,x')^\top\sigma_1F(x'), 
\lb{4.28a} \\
& \hspace*{7.6cm} F\in L^2(\bbR)^2, \; x\in\bbR, \no 
\end{align} 
such that
\begin{equation}
(T_{z_1}F)(x)=F(x)+2\Phi^{(1)}_{z_1}(z_1,x)\xi(x;F), \quad F\in
L^2(\bbR)^2. \lb{6.40}
\end{equation}
Using 
\begin{equation}
\xi _x(x;F)=\Psi(z_1,x)^\top\sigma_1F(x) \text{ for a.e.\ $x\in\bbR$},
\lb{6.41}
\end{equation}
one computes
\begin{align}
   &\big(T_{z_1}F,\cJ \cK \Phi^{(1)}_{z_1}(z_1)\big)_{L^2}=
	\big(F,\cJ \cK \Phi^{(1)}_{z_1}(z_1)\big)_{L^2}
+2\big(\Phi^{(1)}_{z_1}(z_1)\xi(\cdot;F),\cJ\cK
\Phi^{(1)}_{z_1}(z_1)\big)_{L^2} \no \\
   & =\big(F,\cJ \cK \Phi^{(1)}_{z_1}(z_1)\big)_{L^2}-(\Im (z_1))^{-1}
\int_\bbR dx\,
	\big(\big\|\Phi^{(1)}_{z_1}(z_1,x)\big\|^2_{\bbC^2}\big)_x 
\xi(x;F) \no \\
& =\big(F,\cJ \cK \Phi^{(1)}_{z_1}(z_1)\big)_{L^2}+(\Im
	z_1)^{-1}\int_\bbR dx\, \big\|\Phi^{(1)}_{z_1}(z_1,x)\big\|^2_{\bbC^2}
	\Psi(z_1,x)^\top\sigma_1 F(x) \no \\
& = \big(F,\cJ \cK \Phi^{(1)}_{z_1}(z_1)\big)_{L^2}+\bar{\Im(z_1)}
\int_\bbR dx\,\|\Psi(z_1,x)\|^{-2}_{\bbC} \Psi(z_1,x)^\top \sigma_1 F(x)
\no \\
& = 0, \quad F\in L^2(\bbR)^2. \lb{6.45a}
\end{align}
To justify the integration by parts step in \eqref{6.45a} one can argue
as follows. By \eqref{6.20}, the estimate
\begin{align}
& \big\|\Phi^{(1)}_{z_1}(z_1,x)\big\|^2_{\bbC^2}|\xi (x;F)| \no \\
& \leq |\Im(z_1)|^2 \|\Psi(z_1,x)\|^{-2}_{\bbC^2} 
\bigg[ \int_{-\infty}^x dx' \, \|\Psi_-(z_1,x')\|_{\bbC^2}
\|F(x')\|_{\bbC^2} \no \\
& \hspace*{3.9cm} + \int^{\infty}_x dx' \, \|\Psi_+(z_1,x')\|_{\bbC^2}
\|F(x')\|_{\bbC^2}\bigg] \no \\
& \leq |\Im(z_1)|^2 \|\Psi(z_1,x)\|^{-1}_{\bbC^2} \bigg[C_- 
\|\Psi_-(z_1,x)\|^{-1}_{\bbC^2} \int_{-\infty}^x dx' \, 
\|\Psi_-(z_1,x')\|_{\bbC^2} \|F(x')\|_{\bbC^2} \no \\
& \hspace*{2.8cm} + C_+ 
\|\Psi_+(z_1,x)\|^{-1}_{\bbC^2} \int^{\infty}_x dx' \, 
\|\Psi_+(z_1,x')\|_{\bbC^2} \|F(x')\|_{\bbC^2}\bigg] \lb{6.46a}
\end{align}
yields
\begin{equation}
\big\|\Phi^{(1)}_{z_1}(z_1,\cdot)\big\|^2_{\bbC^2}|\xi (\cdot;F)|\in 
L^1(\bbR), \lb{6.47}
\end{equation}
using \eqref{6.3} and \eqref{6.12}. Thus, 
\begin{equation}
\liminf_{x\to\pm\infty}
\big\|\Phi^{(1)}_{z_1}(z_1,x)\big\|^2_{\bbC^2}|\xi (x;F)|=0 \lb{6.42} 
\end{equation}
(actually,  
$\lim_{x\to \pm\infty}|\cdots|=0$ in \eqref{6.42} since all Lebesgue
integrals involved in \eqref{6.45a} are finite), which was to be proven.
Hence, one concludes 
$\cK\Phi^{(1)}_{z_1}(z_1)\bot_{\cJ}\ran(T_{z_1})$. 

Next, using \eqref{2.6}, Lemma
\ref{YL1}\,(iii), \eqref{4.14a}, \eqref{4.21}, and \eqref{4.29}, one
computes
\begin{align}
&\big(S_{\bar{z_1}}\Phi^{(1)}_{z_1}(z_1)\big)(x)=\Phi^{(1)}_{z_1}(z_1,x)
+2\cK \Psi(z_1,x)\int^x_{-\infty} dx' \,
\big(\cK\Phi^{(1)}_{z_1}(z_1,x') \big)^\top \sigma_1
\Phi^{(1)}(z_1,x') \no \\ 
& = \Phi^{(1)}_{z_1}(z_1,x)-2\cK \Psi(z_1,x)(2\Im
(z_1))^{-1}	W\big(\cK \Phi^{(1)}_{z_1}(z_1,x),
\Phi^{(1)}_{z_1}(z_1,x)\big) \no \\
& =\Phi^{(1)}_{z_1}(z_1,x)-\cK\Psi(z_1,x)(\Im(z_1))^{-1} 
\big\|\Phi_{z_1}^{(1)}(z_1,x)\big\|^2_{\bbC^2} \no \\
& = \Phi^{(1)}_{z_1}(z_1,x)-\Im(z_1)\|\Psi(z_1,x)\|^{-2}_{\bbC^2}
\cK\Psi(z_1,x) \no \\
& =0 \lb{6.43}
\end{align}
and hence $\Phi^{(1)}_{z_1}(z_1)\in\ker(S_{\bar{z_1}})$.

For the proof of \eqref{6.36} we next assume $G\in \big\{ \cK
\Phi^{(1)}_{z_1}(z_1)\big\}^{\bot_\cJ }$. Using
\begin{equation}
2\Im(z)\Psi_\mp(z_1,x)^\top\sigma_1\cK\Psi(z_1,x)=W(\Psi_\mp(z_1,x),
\cK\Psi(z,x))_x
\end{equation}
(cf.\ \eqref{2.6}), one then computes  
\begin{align}
   &(T_{z_1}S_{\bar{z_1}}\, G)(x)=(S_{\bar{z_1}}G)(x)+
	2\Phi^{(1)}_{z_1}(z_1,x)\xi(x;S_{\bar{z_1}}G)\no\\
   &=G(x)+2\cK \Psi (z_1,x)\int_{-\infty}^x dx'\,
	\big(\cK \Phi^{(1)}_{z_1}(z_1,x')\big)^\top\sigma_1
	G(x')+2\Phi^{(1)}_{z_1}(z_1,x)\xi(x;G)\no \\
   &\quad +2(\Im (z_1))^{-1}\Phi^{(1)}_{z_1}(z_1,x) \\
   &\quad\quad\times\bigg[ \int_{-\infty}^x dx' \,
	W(\Psi_-(z_1,x'),\cK \Psi (z_1,x'))_{x'}\int_{-\infty}^{x'} 
	dx'' \, \big(\cK \Phi^{(1)}_{z_1}(z_1,x'')\big)^\top\sigma_1
G(x'')\no\\
&\quad\quad\quad\; - \int_x^{\infty} dx' \, 
	W(\Psi_+(z_1,x'),\cK \Psi (z_1,x'))_{x'}\int_{-\infty}^{x'} 
	dx'' \,\big(\cK \Phi^{(1)}_{z_1}(z_1,x'')\big)^\top\sigma_1
G(x'')\bigg]. \no
\end{align}
By \eqref{6.20} one estimates
\begin{align}
&\bigg|W(\Psi_{\mp}(z_1,x),\cK \Psi (z_1,x))
	\int_{\mp\infty}^{x} dx' \,
\big(\cK \Phi^{(1)}_{z_1}(z_1,x')\big)^\top \sigma_1 G(x')\bigg| \no \\
&\leq \|\Psi_\mp(z_1,x)\|_{\bbC^2} \|\Psi(z_1,x)\|_{\bbC^2} 
\int_{\mp\infty}^{x} dx' \,
\big\|\Phi^{(1)}_{z_1}(z_1,x')\big\|_{\bbC^2} \|G(x')\|_{\bbC^2}
\no \\ 
& \leq \|\Psi_\mp(z_1,x)\|_{\bbC^2} \|\Psi(z_1,x)\|_{\bbC^2} |\Im(z_1)|
\int_{\mp\infty}^{x} dx' \,
\big\|\Psi(z_1,x')\big\|^{-1}_{\bbC^2} \|G(x')\|_{\bbC^2} \no \\
& \leq |\Im(z_1)| C_\pm \|\Psi_\mp(z_1,x)\|_{\bbC^2}
[\|\Psi_-(z_1,x)\|_{\bbC^2} + \|\Psi_+(z_1,x)\|_{\bbC^2} ] \no \\
& \quad \times \int_{\mp\infty}^{x} dx' \,
\big\|\Psi_\pm(z_1,x')\big\|^{-1}_{\bbC^2} \|G(x')\|_{\bbC^2}, 
\lb{6.50}
\end{align}
and hence the left-hand side of \eqref{6.50} is in
$L^1((\mp\infty,R])$ for all $R\in\bbR$ by Lemma \ref{RL4.2}. Thus, 
\begin{equation}
	\liminf_{x\to\mp\infty}\bigg|W(\Psi_{\mp}(z_1,x),\cK \Psi (z_1,x))
	\int_{\mp\infty}^{x} dx' \,
\big(\cK \Phi^{(1)}_{z_1}(z_1,x')\big)^\top \sigma_1 G(x')\bigg|=0.  
\lb{6.46}
\end{equation}
An integration by parts, using \eqref{6.46} and 
\begin{equation}
(\Im(z_1))^{-1}\Phi^{(1)}_{z_1}(z_1,x) W(\Psi(z_1,x),\cK\Psi(z_1,x))
=-\cK\Psi(z_1,x),
\end{equation}
then yields
\begin{align}
   &(T_{z_1}S_{\bar{z_1}}G)(x)=G(x)+2\Phi^{(1)}_{z_1}(z_1,x)\xi(x;G)
	-2(\Im (z_1))^{-1}\Phi^{(1)}_{z_1}(z_1,x) \no \\
   & \quad\times \bigg[ \int_{-\infty}^x dx'\,
	W(\Psi_-(z_1,x'),\cK \Psi (z_1,x'))
	\big(\cK \Phi^{(1)}_{z_1}(z_1,x')\big)^\top\sigma_1
	G(x') \no\\
   &\quad\quad\quad-\int_{x}^{\infty}dx'\,
	W(\Psi_+(z_1,x'),\cK \Psi (z_1,x'))\big(\cK 
	\Phi^{(1)}_{z_1}(z_1,x')\big)^\top\sigma_1 G(x')\bigg]\no\\
   & =G(x)+2\Phi^{(1)}_{z_1}(z_1,x)\no\\
&\quad\times \bigg[\int_{-\infty}^x dx'\,
\Theta_{-}(z_1,x')^\top\sigma_1 G(x')-\int_x^{\infty} dx'\,
\Theta_{+}(z_1,x')^\top\sigma_1 G(x')\bigg]. \lb{4.30}
\end{align}
(Again,  $\lim_{x\to \pm\infty}|\cdots|=0$ in \eqref{6.46} since all
Lebesgue integrals involved are finite). Here we introduced the
abbreviation
\begin{equation}
\Theta_{\pm}(z_1,x)=\Psi_{\pm}(z_1,x)+
\Psi(z_1,x)\|\Psi(z_1,x)\|_{\bbC^2}^{-2}
W(\Psi_{\pm}(z_1,x),\cK \Psi(z_1,x)).  
\end{equation}
Actually, using the Jacobi identity
\begin{equation}
	AW(B,C)+BW(C,A)+CW(A,B)=0 , \quad A,B,C\in \bbC^2, 
\end{equation}
one infers
\begin{equation}
\Psi_{\pm}(z_1)W(\cK \Psi (z_1),\Psi(z_1))+
\Psi(z_1)W(\Psi_{\pm}(z_1),\cK \Psi (z_1))=\cK \Psi (z_1)
W(\Psi_{\pm}(z_1),\Psi(z_1)), 
\end{equation} 
which in turn implies  
\begin{equation}
\Theta_{\pm}(z_1,x)=(\Im (z_1))^{-1}W(\Psi_{\pm}(z_1,x),
\Psi_{\mp}(z_1,x)) \Phi^{(1)}_{z_1}(z_1,x). \lb{4.31}
\end{equation}
Combining \eqref{4.30} and \eqref{4.31} results in 
\begin{align}
   (T_{z_1}S_{\bar{z_1}}G)(x)&=G(x)+2(\Im (z_1))^{-1}
	W(\Psi_{-}(z_1,x),\Psi_{+}(z_1,x))\Phi^{(1)}_{z_1}(z_1,x) \no\\
   &\quad \times \int_{\bbR} dx'\, 
	\Phi^{(1)}_{z_1}(z_1,x')^\top\sigma_1 G(x').
\end{align}
Thus,
\begin{align}
&	(T_{z_1}S_{\bar{z_1}}G)(x)=G(x) \no \\
&\quad +2 (\Im (z_1))^{-1} W(\Psi_{-}(z_1,x),\Psi_{+}(z_1,x)) 
\big(G,\cJ \Phi^{(1)}_{z_1}(z_1)\big)_{L^2} \Phi^{(1)}_{z_1}(z_1,x).
\lb{4.32}
\end{align}
Applying \eqref{4.32} to $G=\Phi^{(1)}_{z_1}(z_1) 
\in \big\{\cK \Phi^{(1)}_{z_1}(z_1)\big\}^{\bot_\cJ }$, one
obtains the relation 
\begin{equation}
0=1+2(\Im (z_1))^{-1} W(\Psi_{-}(z_1,x),\Psi_{+}(z_1,x)) 
\big(\Phi^{(1)}_{z_1}(z_1),\cJ \Phi^{(1)}_{z_1}(z_1)\big)_{L^2},
\lb{6.55} 
\end{equation}
which proves \eqref{6.36}. Insertion of \eqref{6.55} into \eqref{4.32}
proves \eqref{4.28}.

Finally, let $F\in L^2(\bbR)^2$. Then $T_{z_1}F\in \big\{ \cK
\Phi^{(1)}_{z_1}(z_1)\big\}^{\bot_\cJ }$ by \eqref{6.35} and hence
\eqref{4.29} yields 
\begin{align}
&(S_{\bar{z_1}} T_{z_1}F)(x)
 =F(x)+2\Phi^{(1)}_{z_1}(z_1,x)\xi(x;F)\no\\
   & \quad\quad\quad +2\cK \Psi (z_1,x)\int_{-\infty}^x dx'\,
\big(\cK \Phi^{(1)}_{z_1}(z_1,x')\big)^\top\sigma_1
	\big[F(x')+2\Phi^{(1)}(z_1,x')\xi(x';F) \big]\no\\
   & = F(x)+ 2\Phi^{(1)}_{z_1}(z_1,x)\xi(x;F)+2\cK 
	\Psi (z_1,x)\int_{-\infty}^x dx'\,
   	\big(\cK \Phi^{(1)}_{z_1}(z_1,x')\big)^\top\sigma_1 F(x')\no\\
   & \quad -2(\Im (z_1))^{-1}\cK \Psi (z_1,x)
	\int_{-\infty}^x dx' \, \big(W\big(\cK \Phi^{(1)}_{z_1} (z_1,x'),
	\Phi^{(1)}(z_1,x')\big)\big)_{x'}\xi(x',F), \lb{6.56} 
\end{align}
using
\begin{equation}
-2\Im(z_1)\big(\cK\Phi^{(1)}_{z_1}(z_1,x)\big)^\top\sigma
\Phi^{(1)}_{z_1}(z_1,x)=W\big(\cK\Phi^{(1)}_{z_1}(z_1,x), 
\Phi^{(1)}_{z_1}(z_1,x)\big)_x.
\end{equation}
The estimate
\begin{align}
& \big\|\cK\Psi(z_1,x)
W\big(\cK\Phi^{(1)}_{z_1}(z_1,x),\Phi^{(1)}_{z_1}(z_1,x)\big)\xi(x;F)
\big\|_{\bbC^2} \no \\
& \leq \|\Psi(z_1,x)\|_{\bbC^2} 2|\Im(z_1)|
\big\|\Phi^{(1)}_{z_1}(z_1,x)\big)\|_{\bbC^2}^2 |\xi(x;F)| \no \\
& \leq 2|\Im(z_1)|^3 \|\Psi(z_1,x)\|^{-1}_{\bbC^2} |\xi(x;F)| \no \\
& = 2|\Im(z_1)|^3 \|\Psi(z_1,x)\|^{-1}_{\bbC^2} 
\bigg[ \int_{-\infty}^x dx' \, \|\Psi_-(z_1,x')\|_{\bbC^2}
\|F(x')\|_{\bbC^2} \no \\
& \hspace*{4.1cm} + \int^{\infty}_x dx' \, \|\Psi_+(z_1,x')\|_{\bbC^2}
\|F(x')\|_{\bbC^2}\bigg] \no \\
& \leq 2|\Im(z_1)|^3 \bigg[C_- 
\|\Psi_-(z_1,x)\|^{-1}_{\bbC^2} \int_{-\infty}^x dx' \, 
\|\Psi_-(z_1,x')\|_{\bbC^2} \|F(x')\|_{\bbC^2} \no \\
& \hspace*{2.1cm} + C_+ 
\|\Psi_+(z_1,x)\|^{-1}_{\bbC^2} \int^{\infty}_x dx' \, 
\|\Psi_+(z_1,x')\|_{\bbC^2} \|F(x')\|_{\bbC^2}\bigg] \lb{6.57}
\end{align}
then proves
\begin{equation}
\big\|\cK\Psi(z_1)
W\big(\cK\Phi^{(1)}_{z_1}(z_1),\Phi^{(1)}_{z_1}(z_1)\big)\xi(\cdot;F)
\big\|_{\bbC^2} \in L^2(\bbR) \lb{6.58}
\end{equation}
by \eqref{6.12}. An integration by parts in the last term of 
\eqref{6.56}, using \eqref{4.14a}, \eqref{4.15a}, and 
\begin{align}
&\liminf_{x\downarrow -\infty}\big\|\cK\Psi(z_1,x)
W\big(\cK\Phi^{(1)}_{z_1}(z_1,x),\Phi^{(1)}_{z_1}(z_1,x)\big)\xi(x;F)
\big\|_{\bbC^2} \no \\
&\leq|2\Im(z_1)| \liminf_{x\downarrow -\infty} \|\Psi(z_1,x)\|_{\bbC^2} 
\big\|\Phi^{(1)}_{z_1}(z_1,x)\big\|^2_{\bbC^2}|\xi(x;F)|=0 \lb{6.60}
\end{align}
by \eqref{6.58} (in fact, $\lim_{x\downarrow -\infty}|\cdots|=0$ in
\eqref{6.60} since all Lebesgue integrals involved are finite), then
proves $S_{\bar{z_1}} T_{z_1}F=F$ and hence \eqref{4.28b}. 
\end{proof}

Next, we further restrict $S_{\bar{z_1}}$ and define the operator
$S_{z_1,\bar{z_1}}$ by
\begin{equation}
S_{z_1,\bar{z_1}}=
S_{\bar{z_1}}\big{|}_{\{\Phi^{(1)}_{z_1}(z_1),
\cK \Phi^{(1)}_{z_1}(z_1)\}^{\bot_\cJ }}
=\hatt S_{\bar{z_1}}\big{|}_{\{\Phi^{(1)}_{z_1}(z_1),
\cK \Phi^{(1)}_{z_1}(z_1)\}^{\bot_\cJ }}. \lb{6.61a}
\end{equation} 

\begin{lemma} \lb{RL4.3} 
Assume Hypothesis \ref{h4.2a} and $z_1\in\rho(D(q))$. Then 
\begin{align} 
& \ker(T_{z_1})=\{0\}, \quad \ran  (T_{z_1}) =
\big\{\Phi^{(1)}_{z_1}(z_1), 
\cK \Phi^{(1)}_{z_1}(z_1)\big\}^{\bot_\cJ }, \lb{6.61} \\
& \ker (S_{\bar{z_1}}) = \Span \{\Phi^{(1)}_{z_1}(z_1)\}, \quad 
\ran (S_{\bar{z_1}})=L^2(\bbR)^2.  \lb{6.62} 
\end{align}
Moreover, $S_{z_1, \bar{z_1}}$ is the inverse of $T_{z_1}$, that is,
\begin{align}
&\ker(S_{z_1, \bar{z_1}})=\{0\}, \quad \ran (S_{z_1, \bar{z_1}})=
L^2(\bbR)^2, \lb{6.63} \\
&S_{z_1, \bar{z_1}}T_{z_1}=I_2 \,\text{ on } \, L^2(\bbR)^2, \lb{4.24}
\\
&T_{z_1}S_{z_1, \bar{z_1}}=I_2 \,\text{ on } \, \big\{
\Phi^{(1)}_{z_1}(z_1),
	\cK \Phi^{(1)}_{z_1}(z_1)\big\}^{\bot_\cJ }. \lb{4.25}
\end{align}
\end{lemma}
\begin{proof}
Suppose $T_{z_1}F=0$ for some $F\in L^2(\bbR)^2$. Then \eqref{4.28b}
yields $0=S_{\bar{z_1}}T_{z_1}F=F$ and hence $\ker(T_{z_1})=\{0\}$. The 
assertion $\ran(S_{\bar{z_1}})=L^2(\bbR)^2$ in \eqref{6.62} is also 
clear from \eqref{4.28b}. Next, assume $S_{\bar{z_1}}G=0$ for some
$G\in \dom(S_{\bar{z_1}})=\big\{ \cK
\Phi^{(1)}_{z_1}(z_1)\big\}^{\bot_\cJ }$ (cf.\ \eqref{6.37}). Then
\eqref{4.28} implies
\begin{equation}
0=T_{z_1}S_{\bar{z_1}}\, G=G-\frac{\big(G,\cJ 
	\Phi^{(1)}_{z_1}(z_1)\big)_{L^2}}{\big(\Phi^{(1)}_{z_1}(z_1),
 	\cJ \Phi^{(1)}_{z_1}(z_1)\big)_{L^2}}\Phi^{(1)}_{z_1}(z_1), \quad 
G\in \big\{ \cK \Phi^{(1)}_{z_1}(z_1)\big\}^{\bot_\cJ}, \lb{6.72}
\end{equation}
and hence $G=c\Phi^{(1)}_{z_1}(z_1)$ for some $c\in\bbC$. This proves
\eqref{6.63}.

Next, suppose $G\in\big\{
	\Phi^{(1)}_{z_1}(z_1),\cK \Phi^{(1)}_{z_1}(z_1)\big\}^{\bot_\cJ }$.
Then $\big(G,\cJ\Phi^{(1)}_{z_1}\big)_{L^2}=0$ and \eqref{4.28} imply
$T_{z_1}S_{\bar{z_1}}G=G$ and hence
\begin{equation}
\big\{\Phi^{(1)}_{z_1}(z_1),\cK \Phi^{(1)}_{z_1}(z_1)\big\}^{\bot_\cJ }
\subseteq \ran(T_{z_1}) \subseteq 
\big\{\cK \Phi^{(1)}_{z_1}(z_1)\big\}^{\bot_\cJ }, \lb{6.74}
\end{equation}
taking
$\ran(T_{z_1})\subseteq\big\{\cK\Phi^{(1)}_{z_1}(z_1)\big\}^{\bot_\cJ}$
in \eqref{6.35} into account. The computation 
\begin{equation}
\big(\Phi^{(1)}_{z_1}(z_1),\cJ T_{z_1}S_{\bar{z_1}}G\big)_{L^2}=
\big(\Phi^{(1)}_{z_1}(z_1),\cJ G\big)_{L^2} -
\big(G,\cJ \Phi^{(1)}_{z_1}(z_1)\big)_{L^2}=0
\end{equation}
then proves $\Phi^{(1)}_{z_1}(z_1) \bot_\cJ \ran(T_{z_1})$ and 
\begin{equation}
\ran(T_{z_1})=\big\{\Phi^{(1)}_{z_1}(z_1),\cK
\Phi^{(1)}_{z_1}(z_1)\big\}^{\bot_\cJ }
\end{equation}
since $\ran(S_{\bar{z_1}})=L^2(\bbR)^2$.

Finally, \eqref{4.24} and \eqref{4.25} are clear from
\eqref{4.28}, \eqref{4.28b}, \eqref{6.61}, and \eqref{6.62}.
\end{proof}

Next we state an auxiliary result.

\begin{lemma}\label{YL.11.1} Assume Hypothesis~\ref{h3.0} and let
$\xi\in AC_{\loc}(\bbR)$, $\Phi=(\phi_1,\phi_2)^\top \in
AC_{\loc}(\bbR)^2$, $F=(f_1,f_2)^\top \in \dom(D(q))$, and  
\begin{align}
&A_-\in \big\{ G\in AC_{\loc}(\bbR)^2 \,\big|\, 
 G, M(q)G\in L^2_{\loc}([-\infty,\infty))^2	\big\}, \lb{5.4} \\
&A_+\in \big\{G\in AC_{\loc}(\bbR)^2 \,\big|\, 
 G, M(q)G \in L^2_{\loc}((-\infty,\infty])^2	\big\}. \lb{5.5} 
\end{align}
Then,
\begin{equation}
	(M(q)\xi \Phi)(x)=i\xi'(x)(\phi_1(x),-\phi_2(x))^\top+\xi(x)(M(q)
	\Phi)(x) \text{ for a.e.\ $x\in\bbR$} \lb{4.1a}
\end{equation}
and
\begin{align}
 \int^x_{-\infty} dx' \, A_-(x')^\top \sigma_1
	(M(q)F)(x')&=\int^x_{-\infty} dx'\,(M(q)A_-)(x')^\top \sigma_1
	F(x')\no\\
   & \quad +iA_-(x)^\bot F(x), \lb{4.2a}\\
\int_x^{\infty} dx' \, A_+(x')^\top \sigma_1
	(M(q)F)(x')&=\int_x^{\infty} dx'\,(M(q)A_+)(x')^\top \sigma_1
	F(x')\no\\
   & \quad -iA_+(x)^\bot F(x) \lb{4.3a}
\end{align}
for all $x\in \bbR$.
\end{lemma}
\begin{proof} Assertion \eqref{4.1a} follows directly from the 
definition of $M(q)$.  To prove assertion \eqref{4.2a} one
integrates by parts and obtains 
\begin{align} 
& \int^x_{-\infty} dx' \, A_{-}(x')^\top\sigma_1 (M(q)F)(x')
	=i[a_{2,-}(x)f_1(x)-a_{1,-}(x)f_2(x)] \no \\ 
   & \quad + i\int^x_{-\infty} dx' \, (-a_{1,-}\bar{q}
	f_1-a_{2,-}qf_{2,-}+a'_{1,-}f_2-a'_{2,-}f_1)(x') \no \\
& =\int_{-\infty}^x dx' \, ((M(q)A_-)(x'))^\top \sigma_1 F(x')
+iA_-(x)^\bot F(x), 
\end{align}
using $\liminf_{x\downarrow -\infty} |A_{-}(x)^{\bot} F(x)| =0$
(actually, $\lim_{x\downarrow -\infty} |\cdots|=0$). Relation 
\eqref{4.3a} is proved analogously.
\end{proof}

The following result shows that $T_{z_1}$ and $S_{\bar{z_1}}$
intertwine $D\big(q^{(1)}_{z_1}\big)$ and $D(q)$.

\begin{lemma} \label{YL12.1} 
Assume Hypothesis~\ref{h4.2a} and $z_1\in\rho(D(q))$. Then, 
\begin{align}
D\big(q^{(1)}_{z_1}\big) T_{z_1}&= T_{z_1}D(q), \lb{4.6a} \\
S_{\bar{z_1}}D\big(q^{(1)}_{z_1}\big)&=D(q) S_{\bar{z_1}}. \lb{4.7a}
\end{align}
\end{lemma}
\begin{proof} 
Using formulas \eqref{Yq} one infers 
\begin{equation}
	M\big(q^{(1)}_{z_1}\big)-M(q)=4i\begin{pmatrix}0 &
	-\phi^{(1)}_{1,z_1}(z_1)\psi_{1}(z_1)\\ \phi^{(1)}_{2,z_1}
	(z_1)\psi_{2}(z_1) & 0 \end{pmatrix}.
\end{equation}
Relation \eqref{4.1a} applied to $M(q^{(1)})$ with
$\Phi=\Phi^{(1)}_{z_1}(z_1)$ and (cf.\ \eqref{4.28a}) 
\begin{align}
&\xi(x)=\xi(x;F)=\int^{x}_{-\infty} dx' \,\Psi_-(z_1,x')^\top \sigma_1
F(x')-\int_{x}^{\infty} dx' \,\Psi_+(z_1,x')^\top \sigma_1 F(x'), \no \\
& \hspace*{8.8cm} F\in\dom(D(q)) 
\end{align}
then yields
\begin{align}   
\big(M\big(q^{(1)}_{z_1}\big)\big(\xi(\cdot;F)\,\Phi^{(1)}_{z_1}(z_1)
\big)\big)(x) &=i\Psi(z_1,x)^\top \sigma_1
	F(x)(\phi^{(1)}_{1,z_1}(z_1,x),-\phi^{(1)}_{2,z_1}(z_1,x))^\top \no \\
   & \quad +z_1\Phi^{(1)}_{z_1}(z_1,x)\xi(x;F)
\end{align}
since
$M\big(q^{(1)}_{z_1}\big)\Phi^{(1)}_{z_1}(z_1)=z_1\Phi^{(1)}_{z_1}(z_1)$.
By \eqref{4.2a}, with $A_-=\Psi_-(z_1)$ and $A_+=\Psi_+(z_1)$, one
infers
\begin{align}
&\int^x_{\pm\infty} dx' \, \Psi_{\pm} (z_1,x')^\top \sigma_1 (M(q)F)(x')
=z_1\int^x_{\pm\infty} dx' \, \Psi_{\pm} (z_1,x')^\top 
\sigma_1 F(x')\no\\
& \hspace{5cm} + i\Psi_{\pm} (z_1,x)^\bot F(x), \quad F\in\dom(D(q))
\end{align}
since $(M(q)\Psi_{\pm} (z_1))^\top =z_1\Psi_{\pm} (z_1)^\top$. Thus,
\begin{align}
&\big(M\big(q^{(1)}_{z_1}\big)T_{z_1}-T_{z_1}M(q)\big)F 
=\big(M\big(q^{(1)}_{z_1}\big)-M(q)\big)F \no \\
& \quad +2 M\big(q^{(1)}_{z_1}\big)\big(\Phi^{(1)}(z_1)\xi(\cdot;F)\big)
 -2\Phi^{(1)}_{z_1}(z_1)\xi(\cdot;M(q)F) \no\\
&= 4i\big(-\phi^{(1)}_{1,z_1}(z_1)\psi_{1}(z_1)f_2, 
 \phi^{(1)}_{2,z_1}(z_1)\psi_{2}(z_1)f_1\big)^\top \no \\
&\quad +2i \big[\Psi(z_1)^\top\sigma_1 F\big]
\big(\phi_1^{(1)}(z_1),  -\phi_2^{(1)}(z_1)\big)^\top
+2z_1\Phi^{(1)}_{z_1}(z_1)\xi(\cdot;F) \no \\  
&\quad -2z_1\Phi^{(1)}_{z_1}(z_1)\xi(\cdot;F)
 -2i \big[\Psi(z_1)^\bot F\big] \big(\phi^{(1)}_{1,z_1}(z_1),
\phi^{(1)}_{2,z_1}(z_1)\big)^\top \no \\ 
&=0, \quad F\in\dom(D(q)). \lb{6.86}
\end{align}
This computation also proves that 
\begin{equation}
T_{z_1}\dom(D(q))\subseteq \dom\big(D(q^{(1)}_{z_1})\big)
\text{ and hence }
T_{z_1}D(q)\subseteq D\big(q^{(1)}_{z_1}\big)T_{z_1}. \lb{6.87}
\end{equation}
Next, choosing $G\in\dom(D(q)S_{\bar{z_1}})$ (i.e., $G\in L^2(\bbR)^2$
such that $S_{\bar{z_1}}G\in\dom(D(q))$), \eqref{6.87} implies
\begin{align}
&S_{\bar{z_1}}D(q^{(1)}_{z_1})T_{z_1}S_{\bar{z_1}}G=
S_{\bar{z_1}}D(q^{(1)}_{z_1})\big[G-\big(\Phi^{(1)}_{z_1}(z_1),
\cJ\Phi^{(1)}_{z_1}\big)_{L^2}^{-1}
\big(G,\cJ\Phi^{(1)}_{z_1}(z_1)\big)_{L^2}\big]
\no \\
&=S_{\bar{z_1}}D(q^{(1)}_{z_1})G=S_{\bar{z_1}}T_{z_1}D(q)S_{\bar{z_1}}G 
=D(q)S_{\bar{z_1}}G, \quad G\in\dom(D(q)S_{\bar{z_1}}). \lb{6.88}
\end{align}
Here we successively used \eqref{4.28},
$S_{\bar{z_1}}\Phi^{(1)}_{z_1}(z_1)=0$ (cf.\ \eqref{6.35}), and
\eqref{4.28b}. Thus, one concludes
\begin{equation}
\dom\big(D(q^{(1)}_{z_1})\big)\subseteq \dom(D(q)S_{\bar{z_1}}) 
\text{ and hence } 
S_{\bar{z_1}}D\big(q^{(1)}_{z_1}\big)\subseteq D(q)S_{\bar{z_1}}.
\lb{6.89}
\end{equation}
Hence,
\begin{equation}
\dom\big(D(q^{(1)}_{z_1})T_{z_1}\big)=\big\{F\in L^2(\bbR)^2\,\big|\,
T_{z_1}F\in\dom\big(D(q^{(1)}_{z_1})\big)\big\}\subseteq \dom(D(q))
\lb{6.90}
\end{equation}
since $S_{\bar{z_1}}(T_{z_1}F)=F\in\dom(D(q))$ for 
$F\in\dom\big(D\big(q^{(1)}_{z_1}\big)T_{z_1}\big)$ by \eqref{6.89}.
Combining \eqref{6.87} and \eqref{6.90} then proves \eqref{4.6a}.
Equations \eqref{4.6a}, \eqref{4.28}, and \eqref{4.28b} in turn imply 
\begin{align}
&S_{\bar{z_1}}D(q^{(1)}_{z_1})T_{z_1}S_{\bar{z_1}}=
S_{\bar{z_1}}D(q^{(1)}_{z_1})\big[I-\big(\Phi^{(1)}_{z_1}(z_1),
\cJ\Phi^{(1)}_{z_1}\big)_{L^2}^{-1}
\big(\cdot,\cJ\Phi^{(1)}_{z_1}(z_1)\big)_{L^2}\big] \no \\
&=S_{\bar{z_1}}D(q^{(1)}_{z_1})=S_{\bar{z_1}}T_{z_1}D(q)S_{\bar{z_1}} 
=D(q)S_{\bar{z_1}} \lb{6.92}
\end{align}
and hence \eqref{4.7a}.
\end{proof}

\begin{remark}
One can prove, similarly to Lemma \ref{YL12.1}, that 
\begin{align}
D\big(q^{(1)}_{z_1}\big)\wti T_{\bar{z_1}}&=\wti T_{\bar{z_1}}D(q)
\text{ and } \ker (\wti T_{\bar{z_1}})=\{ 0\},\no\\
\wti S_{z_1}D\big(q^{(1)}_{z_1}\big)&=D(q)\wti S_{z_1} \text{ and } 
\ker(\wti S_{z_1})=\Span\big\{\cK \Phi^{(1)}_{z_1}(z_1)\big\} 
\end{align}
and that $\wti T_{\bar{z_1}}$ and
$\wti S_{z_1}\big{|}_{\big\{\Phi^{(1)}_{z_1}(z_1),
\cK \Phi^{(1)}_{z_1}(z_1)\big\}^{\bot_\cJ }}$ are transformation
operators $($inverse to each other\,$)$.
\end{remark}

Summarizing the results of Corollaries \ref{c6.2} and \ref{c6.3} and
Lemmas \ref{RL4.3a}--\ref{YL12.1} obtained thus far, we are now in
position to state one of the principal results of this section.
\begin{theorem}\lb{RT4.1}
Assume Hypothesis \ref{h4.2a} and $z_1\in\rho(D(q))$. Then $T_{z_1}$ and
$S_{\bar{z_1}}$ are bounded linear  operators in $L^2(\bbR)^2$ which
intertwine the operators $D(q)$ and $D\big(q^{(1)}_{z_1}\big)$,  
\begin{align}
D\big(q^{(1)}_{z_1}\big)T_{z_1}&=T_{z_1}D(q), \lb{4.22} \\
S_{\bar{z_1}}D\big(q^{(1)}_{z_1}\big)&=D(q)S_{\bar{z_1}}. \lb{4.23} 
\end{align}
Moreover,
\begin{align}
& \ker (T_{z_1})=\{ 0\}, \quad \ran 
(T_{z_1})=\big\{\Phi^{(1)}_{z_1}(z_1),\cK
\Phi^{(1)}_{z_1}(z_1)\big\}^{\bot_\cJ }, \lb{6.99} \\
& \ker (S_{\bar{z_1}})=\Span \big\{\Phi^{(1)}_{z_1}(z_1)\big\}, 
\quad \ran(S_{\bar{z_1}})=L^2(\bbR)^2,  \lb{6.100}
\end{align}
and $S_{z_1, \bar{z_1}}$ $($cf.\ \eqref{6.61a}$)$ is the inverse of
$T_{z_1}$, that is,
\begin{align}
   &S_{z_1, \bar{z_1}}T_{z_1}=I_2 \,\text{ on } \,
L^2(\bbR)^2,\lb{4.24A} \\
   &T_{z_1}S_{z_1, \bar{z_1}}=I_2 \,\text{ on } \, \big\{
\Phi^{(1)}_{z_1}(z_1),
	\cK \Phi^{(1)}_{z_1}(z_1)\big\}^{\bot_\cJ }.\lb{4.25A}
\end{align}
\end{theorem}

An analogous statement can be formulated when $z_1$ is replaced by 
$\bar{z_1}$, that is, for the operators $T_{\bar{z_1}}$ and
$\wti S_{z_1}$,  but we chose not to dwell on it here due to the
symmetry of the  arguments.

We conclude with the principal spectral theoretic result of this paper.

\begin{theorem}\lb{RT4.2}
Assume Hypothesis \ref{h4.2a} and\,\footnote{We recall that
$z\in\rho(D(q))$ implies $z\in\bbC\backslash\bbR$ by Corollary
\ref{c5.5}.}
$z_1\in\rho(D(q))$. Then 
\begin{align}
\sigma (D\big(q^{(1)}_{z_1}\big))&=\sigma (D(q))\cup \{
z_1,\bar{z_1}\}, \lb{4.26} \\
\sigma_{\rm p}\big(D\big(q^{(1)}_{z_1}\big)\big)
&=\sigma_{\rm p}(D(q))
\cup \{z_1,\bar{z_1}\}, \lb{b4.1}\\
\sigma_{\rm e}\big(D\big(q^{(1)}_{z_1}\big)\big)
&=\sigma_{\rm e}(D(q)). \lb{b4.2}  
\end{align}
In other words, constructing the new $\NS_-$ potential
$q^{(1)}_{z_1}$ amounts to inserting a pair of complex conjugate 
$($nonreal\,$)$ $L^2(\bbR)^2$-eigenvalues, $z_1$ and $\bar{z_1}$, into
the spectrum of the  background operator $D(q)$, leaving the rest of its 
spectrum invariant. 
\end{theorem}
\begin{proof}
We will prove \eqref{b4.1} and \eqref{b4.2} from which \eqref{4.26}
follows by \eqref{5.27}. By Theorem \ref{YT2} one has 
$z_1,\bar{z_1}\in \sigma_{\rm p} \big(D\big(q^{(1)}_{z_1}\big)\big)$. 
Thus, \eqref{b4.1} is equivalent to
\begin{equation}
	\sigma_{\rm p} \big(D\big(q^{(1)}_{z_1}\big)\big)\backslash 
\{z_1,\bar{z_1}\}=\sigma_{\rm p} (D(q)). \lb{4.27}
\end{equation}
Next we denote 
\begin{equation}
X^{(1)}=\big\{\Phi^{(1)}_{z_1}(z_1),\cK
\Phi^{(1)}_{z_1}(z_1)\big\}^{\bot_\cJ }. \lb{6.108} 
\end{equation}
Since $D(q^{(1)}_{z_1})$ is $\cJ$-self-adjoint, $X^{(1)}$ is a closed,
$D\big(q^{(1)}_{z_1}\big)$-invariant subspace. In addition, we denote by
$D\big(q^{(1)}_{z_1}\big)\big{|}_{X^{(1)}}$ the part of
$D\big(q^{(1)}_{z_1}\big)$ in $X^{(1)}$ with
\begin{equation}
\dom\big(D\big(q^{(1)}_{z_1}\big)\big{|}_{X^{(1)}}\big)=X^{(1)}\cap 
\dom\big(D\big(q^{(1)}_{z_1}\big)\big).
\end{equation}
{}From \eqref{4.22} and
Theorem \ref{RT4.1} it follows that
\begin{equation}
	D\big(q^{(1)}_{z_1}\big)\big{|}_{X^{(1)}}=TD(q)T^{-1} 
\text{ on }X^{(1)} \lb{6.109} 
\end{equation}
and 
\begin{equation}
	D(q)=T^{-1}D\big(q^{(1)}_{z_1}\big)\big{|}_{X^{(1)}} T 
\text{ on }L^2(\bbR)^2, \lb{b4.4}
\end{equation}
with $T=T_{z_1}$ and
$T^{-1}=\hatt S_{\bar{z_1}}\big|_{\big\{\Phi^{(1)}_{z_1}(z_1),
\cK\Phi^{(1)}_{z_1}(z_1)\big\}^{\bot_\cJ }}$ (cf.\ Lemma \ref{RL4.3}).
Assuming $\mu\in\sigma_{\rm p}(D(q))$ and 
\begin{equation}
D(q)F(\mu)=\mu F(\mu), \quad 0\neq F(\mu)\in\dom(D(q)),
\end{equation}
then \eqref{6.109} implies
\begin{equation}
D\big(q^{(1)}_{z_1}\big)\big|_{X^{(1)}}TF(\mu)=\mu TF(\mu).
\end{equation}
Since $\ker(T)=\{0\}$ (cf.\ \eqref{6.99}), this implies
\begin{equation}
\sigma_{\rm p} (D(q))\subseteq 	\sigma_{\rm p} 
\big(D\big(q^{(1)}_{z_1}\big)\big) \text{ and hence }
\sigma_{\rm p} (D(q))\subseteq 	\sigma_{\rm p} 
\big(D\big(q^{(1)}_{z_1}\big)\big)\backslash \{z_1,\bar{z_1}\}.
\end{equation}
Conversely, assuming $\nu\in \sigma_{\rm p} 
\big(D\big(q^{(1)}_{z_1}\big)\big)\backslash \{z_1,\bar{z_1}\}$ and 
\begin{equation}
D\big(q^{(1)}_{z_1})\big|_{X^{(1)}}F^{(1)}(\nu)=\nu F^{(1)}(\nu), \quad 
0\neq F^{(1)}(\nu)\in\dom\big(D\big(q^{(1)}_{z_1}\big)\big),
\end{equation}
one infers from \eqref{b4.4} that
\begin{equation}
D(q)T^{-1}F^{(1)}(\nu)=\nu T^{-1}F^{(1)}(\nu). 
\end{equation}
Since $\ker(T^{-1})=\{0\}$ (cf.\ \eqref{6.63}), one concludes
$T^{-1}F^{(1)}(\nu)\neq 0$ and hence 
\begin{equation}
\sigma_{\rm p} (D(q))\supseteq 	\sigma_{\rm p} 
\big(D\big(q^{(1)}_{z_1}\big)\big)\backslash \{z_1,\bar{z_1}\},
\end{equation}
implying \eqref{b4.1}.

Next, one observes that \eqref{b4.4} also implies
\begin{equation}
\sigma_{\rm e}(D(q))=
\sigma_{\rm e}\big(D\big(q^{(1)}_{z_1}\big)\big{|}_{X^{(1)}}\big).
\lb{b4.7}
\end{equation} 
To prove \eqref{b4.2} we note that \eqref{b4.7} implies 
\begin{equation}
\sigma_{\rm e}(D(q))
=\sigma_{\rm e}\big(D\big(q^{(1)}_{z_1}\big)\big{|}_{X^{(1)}}\big) 
\subseteq \sigma_{\rm e}\big(D\big(q^{(1)}_{z_1}\big)\big). 
\end{equation}
Conversely,
let $\lambda\in\sigma_{\rm e}\big(D\big(q^{(1)}_{z_1}\big)\big)$ and 
$\{ G_n\}_{n\in\bbN}$ be a singular sequence of 
$D\big(q^{(1)}_{z_1}\big)$
corresponding to $\lambda$, that is, a bounded sequence in
$\dom\big(D\big(q^{(1)}_{z_1}\big)\big)$ without any convergent
subsequence such that
$\lim_{n\to\infty}\big(D\big(q^{(1)}_{z_1}\big)-\lambda I\big)G_n=0$. 
Since $L^2(\bbR)^2$ permits the direct sum decomposition (not to be
confused with an orthogonal direct sum decomposition)
\begin{equation}
L^2(\bbR)^2= X^{(1)} \dot + 
\Span \big\{\Phi^{(1)}_{z_1}(z_1), \cK  \Phi^{(1)}_{z_1}(z_1)\big\}
\end{equation}
(here one uses the fact that $\Phi^{(1)}_{z_1}(z_1)\bot_{\cJ}\cK 
\Phi^{(1)}_{z_1}(z_1)$, cf.\ \eqref{6.36a}), one can write 
\begin{equation}
	G_n=F_n+\frac{\big(G_n, \cJ \Phi^{(1)}_{z_1}(z_1)\big)_{L^2}}
	{\big(\Phi^{(1)}_{z_1}(z_1),\cJ \Phi^{(1)}_{z_1}(z_1)\big)_{L^2}} 
	\Phi^{(1)}_{z_1}(z_1)+ \frac{\big(G_n, \cJ
\cK\Phi^{(1)}_{z_1}(z_1)\big)_{L^2}}
	{\big(\cK \Phi^{(1)}_{z_1}(z_1),\cJ \cK
\Phi^{(1)}_{z_1}(z_1)\big)_{L^2}} 
	\cK \Phi^{(1)}_{z_1}(z_1), \lb{b4.5}
\end{equation}
where $F_n\in X^{(1)}\cap\dom\big(D\big(q^{(1)}_{z_1}\big)\big)$ is the
projection of $G_n$ onto the space $X^{(1)}$, the projection being
parallel to the subspace
$\Span \big\{\Phi^{(1)}_{z_1}(z_1), \cK  \Phi^{(1)}_{z_1}(z_1)\big\}$.
Since the coefficients of
$\Phi^{(1)}_{z_1}(z_1)$ and $ \cK \Phi^{(1)}_{z_1}(z_1)$ in
\eqref{b4.5} are bounded with respect to $n\in\bbN$, one can assume that
they are convergent by restricting to a subsequence. Thus, for some real
numbers $c_1$ and $c_2$, one has 
\begin{equation}
	\big(D\big(q^{(1)}_{z_1}\big)-\lambda I\big) F_n
\underset{n\to\infty}{\rightarrow} c_1
	\Phi^{(1)}_{z_1}(z_1)+ c_2 \cK \Phi^{(1)}_{z_1}(z_1). 
\end{equation}
Since $X^{(1)}$ is $D\big(q^{(1)}_{z_1}\big)$-invariant, $F_n\in X^{(1)}$
implies $\big(D\big(q^{(1)}_{z_1}\big)-\lambda I\big) F_n\in
X^{(1)}$, and since $X^{(1)}$ is closed, the limit $c_1
	\Phi^{(1)}_{z_1}(z_1)+ c_2 \cK \Phi^{(1)}_{z_1}(z_1)$ also belongs to 
$X^{(1)}$. But this implies $c_1=c_2=0$, because
$\big(\Phi^{(1)}_{z_1}(z_1), \cJ \cK \Phi^{(1)}_{z_1}(z_1)\big)_{L^2}=0$
(cf.\ \eqref{6.36a}) and $\big(\Phi^{(1)}_{z_1}(z_1), \cJ
\Phi^{(1)}_{z_1}(z_1)\big)_{L^2}\ne 0$, $\big(\cK\Phi^{(1)}_{z_1}(z_1),
\cJ\cK \Phi^{(1)}_{z_1}(z_1)\big)_{L^2}\ne 0$ (cf.\ \eqref{6.36}). Thus,
one concludes that $\{ F_n \}_{n\in\bbN} \subset X^{(1)}$ is a singular
sequence of $D\big(q^{(1)}_{z_1}\big)\big|_{X^{(1)}}$ corresponding to
$\lambda$ since it is bounded and has no convergent subsequence.
(Otherwise, by \eqref{b4.5},
$\{G_n\}_{n\in\bbN}$ would have a convergent subsequence,  contradicting 
the assumption that it is a singular sequence). It follows that
$\lambda\in\sigma_{\rm e}\big(D\big(q^{(1)}_{z_1}\big)
\big{|}_{X^{(1)}}\big)\,
=\sigma_{\rm e}(D(q))$ and hence $\sigma_{\rm e}(D(q))
=\sigma_{\rm e}\big(D\big(q^{(1)}_{z_1}\big)\big{|}_{X^{(1)}}\big) 
\supseteq \sigma_{\rm e}\big(D\big(q^{(1)}_{z_1}\big)\big)$, proving
\eqref{b4.2}.
\end{proof}

Theorems \ref{RT4.1} and \ref{RT4.2} are new. We note that they are
proven under the optimal assumption $q\in L^1_{\loc}(\bbR)$ (but they
seem to be new under virtually any assumptions on $q$). 

In the special periodic case where the machinery of Floquet theory can
be applied, the issue of isospectral Darboux transformations is
briefly mentioned in \cite[Theorem\ 3]{LM94}. This excludes the insertion
of eigenvalues as in Theorem \ref{RT4.2}. Inserting eigenvalues into the
spectrum of a self-adjoint one-dimensional Dirac operator (not applicable
in the present $\NS_-$ context) was investigated by means of
transformation operators (along the lines of \cite{GT96}) in
\cite{Te98}.  

We conclude this section with a few facts on $N$-soliton $\NS_-$
potentials. As shown in Lemma \ref{YP1}, the insertion of pairs of 
complex
conjugate eigenvalues into the spectrum of the background operator $D(q)$
can be iterated. To fix the proper notation, we now slightly extend our
approach of Section \ref{s6} and we consider a more general linear
combination $\Psi_{\gamma_k}(z,x)$ of $\Psi_\pm(z,x)$: Assuming
$z_k\in\bbC\backslash\bbR$, $k=1,\dots,N$, one defines 
\begin{equation}
\Psi_{\gamma_k}(z_k,x)=\Psi_-(z_k,x)+\gamma_k \Psi_+(z_k,x), \quad
\gamma_k\in\bbC\backslash\{0\}, \; k=1\dots,N\lb{6.124} 
\end{equation} 
(as opposed to to our choice $\gamma=1$ in \eqref{R4.5}). In obvious
notation one then denotes the corresponding $N$th iteration of the
construction of $q^{(1)}_{z_1}(x)$ presented in Sections
\ref{s2} and \ref{s4} (cf.\ \eqref{Y5.2}, \eqref{Yq}), 
identifying $\Psi(z_1,x)$ and $\Psi_{\gamma_1}(z_1,x)$, by
$q^{(N)}_{z_1,\ldots z_N,\gamma_1,\ldots,\gamma_N}(x)$. 

In order to describe a well-known explicit formula for 
$q^{(N)}_{z_1,\ldots z_N,\gamma_1,\ldots,\gamma_N}(x)$ (cf., e.g., 
\cite[Sect.\ 4.2]{MS91}, \cite{NM84}, \cite{St97}) one introduces the
quantities
\begin{equation}
	\varphi_k(x)=
\frac{\psi_{2,-}(z_k,x)+\gamma_k \psi_{2,+}
(z_k,x)}{\psi_{1,-}(z_k,x)+\gamma_k\psi_{1,+}	(z_k,x)}, \lb{6.125}
\end{equation}
and the $2N\times 2N$ Vandermonde-type matrices
\begin{equation}
V_{2N}(x)=\begin{pmatrix}
	1&z_1&\hdots&z_1^{N-1}&\varphi_1&z_1\varphi_1
	&\hdots&z_1^{N-1}\varphi_1 \\
	\vdots&\vdots& &\vdots&\vdots&\vdots & &\vdots \\
	1&z_N&\hdots&z_N^{N-1}&\varphi_N&z_N\varphi_N
	&\hdots&z_N^{N-1}\varphi_N \\
	\\
	1&z_{N+1}&\hdots&z_{N+1}^{N-1}&\varphi_{N+1}&
	z_{N+1}\varphi_{N+1}&\hdots&z_{N+1}^{N-1}\varphi_{N+1}\\
	\vdots&\vdots& &\vdots&\vdots&\vdots & &\vdots\\
	1&z_{2N}&\hdots&z_{2N}^{N-1}&\varphi_{2N}&
	z_{2N}\varphi_{2N}&\hdots& z_{2N}^{N-1}\varphi_{2N} \\
	\end{pmatrix} \lb{6.126}
\end{equation}
and 
\begin{equation}
\wti V_{2N}(x)=\begin{pmatrix}
	1&z_1&\hdots&z_1^{N-1}&\varphi_1&z_1\varphi_1
	&\hdots&z_1^{N}\\
	\vdots&\vdots& &\vdots&\vdots&\vdots & &\vdots\\
	1&z_N&\hdots&z_N^{N-1}&\varphi_N&z_N\varphi_N
	&\hdots&z_N^{N}\\
	\\
	1&z_{N+1}&\hdots&z_{N+1}^{N-1}&\varphi_{N+1}&
	z_{N+1}\varphi_{N+1}&\hdots&z_{N+1}^{N}\\
	\vdots&\vdots& &\vdots&\vdots&\vdots & &\vdots\\
	1&z_{2N}&\hdots&z_{2N}^{N-1}&\varphi_{2N}&
	z_{2N}\varphi_{2N}&\hdots& z_{2N}^{N}\\
	\end{pmatrix}. \lb{6.127}
\end{equation} 
The $N$th iteration $q^{(N)}_{z_1,\ldots
z_N,\gamma_1,\ldots,\gamma_N}(x)$ is then explicitly given by
\begin{equation}
q^{(N)}_{z_1,\ldots z_N,\gamma_1,\ldots,\gamma_N}(x)=q(x)-2i
\f{\det(\wti V_{2N}(x))}{\det(V_{2N}(x))}. \lb{6.128}
\end{equation} 
Denoting by $D(q^{(N)}_{z_1,\ldots z_N,\gamma_1,\ldots,\gamma_N})$
the associated $\cJ$-self-adjoint operator in $L^2(\bbR)^2$, repeated
application of Theorem \ref{RT4.2} then yields
\begin{align}
\sigma(D(q^{(N)}_{z_1,\ldots z_N,\gamma_1,\ldots,\gamma_N}))&=
	\sigma(D(q))\cup \{ z_1,\bar{z_1},\ldots, z_N,\bar{z_N}\}, \\
\sigma_{\rm p}(D(q^{(N)}_{z_1,\ldots z_N,\gamma_1,\ldots,\gamma_N}))&=
\sigma_{\rm p}(D(q))\cup \{ z_1,\bar{z_1},\ldots, z_N,\bar{z_N}\}, \\
\sigma_{\rm e}(D(q^{(N)}_{z_1,\ldots z_N,\gamma_1,\ldots,\gamma_N}))&=
\sigma_{\rm e}(D(q)). 
\end{align}

\noindent {\bf Acknowledgements.} We are indebted to Igor Verbitsky for
a crucial discussion on boundedness of transformation operators in
connection with  Lemma \ref{l6.1}. We also gratefully acknowledge
discussions and a critical reading of our manuscript by Konstantin
Makarov and Alexander Motovilov. 

F.\ G.\ gratefully acknowledges support from Simula Research
Laboratory. F.\ G.\ and H.\ H.\ were supported in part by the Research
Council of Norway. Y.\ L.\ was partially supported by the Twinning
Program of the National Academy of Sciences and National Science
Foundation, and by the Research Council and Research Board of the
University of Missouri.


\end{document}